\documentclass[final,3p,times]{elsarticle}

\usepackage{color}
\usepackage{graphicx}
\usepackage{subfigure}
\usepackage{amssymb}

\usepackage{lineno}
 \usepackage{float}
 \usepackage{dblfloatfix}
 \usepackage{fixltx2e}
 
\biboptions{sort&compress }

\newcommand{\nuebar}{$\overline{\nu}_{e}$}

\journal{Nuclear Instruments and Methods}

\begin{document}

\begin{frontmatter}

\title{The Detector System of 
\\the Daya Bay  Reactor Neutrino Experiment 
}

\def\ECUST{1}
\def\IHEP{2}
\def\Wisconsin{3}
\def\Yale{4}
\def\BNL{5}
\def\NTU{6}
\def\NUU{7}
\def\DEAD{8}
\def\Dubna{9}
\def\NJU{10}
\def\CalTech{11}
\def\CUHK{12}
\def\NCTU{13}
\def\SDU{14}
\def\TsingHua{15}
\def\NCEPU{16}
\def\SZU{17}
\def\Princeton{18}
\def\LBNL{19}
\def\Siena{20}
\def\IIT{21}
\def\UIUC{22}
\def\CDUT{23}
\def\UCLA{24}
\def\RPI{25}
\def\SJTU{26}
\def\BNU{27}
\def\UH{28}
\def\VirginiaTech{29}
\def\CIAE{30}
\def\NanKai{31}
\def\UC{32}
\def\DGUT{33}
\def\UCB{34}
\def\HKU{35}
\def\Charles{36}
\def\USTC{37}
\def\ZSU{38}
\def\WM{39}
\def\XJTU{40}
\def\TempleUniversity{41}
\def\CUC{42}
\def\CGNPG{43}
\def\NUDT{44}
\def\IowaState{45}
\def\CQU{46}
\fntext[DEAD]{Deceased.}
\author[\ECUST]{F.~P.~An}
\author[\IHEP]{J.~Z.~Bai}
\author[\Wisconsin]{A.~B.~Balantekin}
\author[\Yale]{H.~R.~Band}
\author[\BNL]{D.~Beavis}
\author[\BNL]{W.~Beriguete}
\author[\BNL]{M.~Bishai}
\author[\NTU,\NUU]{S.~Blyth}
\author[\BNL]{R.~L.~Brown\fnref{DEAD}}
\author[\Dubna]{I.~Butorov}
% ...


%

\begin{abstract}

The Daya Bay experiment was the first to report simultaneous  measurements of  reactor antineutrinos
at multiple baselines leading to the discovery of $\bar{\nu}_e$ oscillations 
over km-baselines. 
Subsequent data  
has provided the world's most precise measurement 
of   $\rm{sin}^22\theta_{13}$  and the  effective mass splitting $\Delta m_{ee}^2$. 
The experiment is located in Daya Bay, China where the cluster of six  nuclear reactors is
among the world's most prolific sources of electron antineutrinos. 
Multiple antineutrino detectors are deployed  in three underground  water pools at different distances 
from the reactor cores to search for deviations 
in the antineutrino  rate and energy spectrum due to neutrino mixing. 
Instrumented with photomultiplier tubes, the water pools serve
as shielding against natural radioactivity from the surrounding rock and provide efficient muon tagging. 
Arrays of resistive plate chambers over  the top of each pool provide additional muon detection. 
The antineutrino detectors were specifically designed for measurements of the antineutrino flux with minimal systematic uncertainty.
Relative detector efficiencies between the near and far detectors are known to better than 0.2\%.  
With the unblinding of the final two detectors' baselines and target masses, a complete description and comparison of the eight 
antineutrino detectors  can now be presented.
This paper describes the Daya Bay detector systems, consisting of eight antineutrino detectors in three
 instrumented water pools in three underground halls, and their operation through 
 the first year of eight detector data-taking.

\end{abstract}


\end{frontmatter}



\section{Introduction }

\label{introduction}

The Daya Bay experiment was the first to measure a nonzero $\theta_{13}$ 
with a significance of $\geq$ 5 standard deviations~\cite{prl_rate}  and continues to provide the 
most precise measurement of $\theta_{13}$~\cite{cpc_rate,prl_shape}. 
In the generally accepted  three-neutrino framework,
neutrino oscillations are described by three mixing angles ($\theta_{12}$, $\theta_{23}$, and $\theta_{13}$),
two mass-squared differences ($\Delta m^2_{31}$ and $\Delta m^2_{21}$),
and a phase  in the Pontecorvo-Maki-Nakagawa-Sakata matrix~\cite{pontecorvo,mns}.
Prior to Daya Bay, $ \theta_{13}$ was poorly known, constrained by a 
sin$^22\theta_{13} \leq 0.17$ limit from the Chooz experiment~\cite{chooz} and several low statistics measurements
by T2K~\cite{t2k}, MINOS~\cite{minosth13} and Double Chooz~\cite{dchooz}
that  indicated that $\theta_{13}$ could be non-zero.

For reactor-based experiments, $\theta_{13}$
can be extracted from the survival probability of the electron antineutrino \nuebar\ at distances a few km from the reactor:
\begin{eqnarray}
P_{\rm sur} =1 &-& \sin^2 2 \theta_{13}  \sin^2 \Delta_{ee}  \nonumber \\
                       &-& \cos^4 \theta_{13} \sin^2 2\theta_{12}  \sin^2 \Delta_{21} 
\label{eqn:Psurv}
\end{eqnarray}
where $\Delta_{ee}=\Delta_{32}\pm \phi/2$, with $\phi = \arctan\left(\frac{\sin2\Delta_{21}}{\cos2\Delta_{21}+\tan^2\theta_{12}}\right) $
for normal mass hierarchy (+) and inverted mass hierarchy (-),
$\Delta_{ji}\equiv 1.267\Delta m^2_{ji}$(eV$^2$)$L($m$)/E($MeV), 
 with $\Delta m^2_{ji}$  the difference between the mass-squares of the mass eigenstates $\nu_j$ and $\nu_i$,  
$E$ is the \nuebar\ energy and $L$ is the distance between
the \nuebar\ source and the detector (baseline).
The effective mass-squared difference $\Delta m_{ee}^2$ has
been measured in Daya Bay~\cite{prl_shape}.

Antineutrinos with energy $\ge 1.8$~MeV are detected  via the inverse beta-decay (IBD) reaction on ``free" protons in the liquid scintillating target:

\begin{equation}
\bar{\nu}_e + p \to e^+ + n\,. \label{eqn:IBD}
\end{equation}

The positron, carrying most of the antineutrino energy, deposits its energy and annihilates with an electron 
producing a prompt signal with energy ranging from 1 MeV to about 8 MeV.
The neutron, after thermalizing, captures on a gadolinium or hydrogen nucleus, producing a delayed signal with energy 
of  $\sim$8~MeV or 2.2 MeV, typically within $100~\mu s$. 
The correlation in time and space between the prompt and delayed signals provides a 
distinctive $\bar{\nu}_e$ signature, greatly suppressing backgrounds.

The number of detected
antineutrinos $N_{\rm det}$ is given by
\begin{eqnarray}
N_{\rm det}=\frac{N_p \epsilon}{4\pi L^2}\int{\sigma(E) P_{\rm sur}(E,L,\theta_{13})S(E) dE}
\label{eq:absolute}
\end{eqnarray}
where $N_p$ is the number of free protons in the target, 
$\epsilon$ is the
efficiency of detecting an IBD, $\sigma$ is the total cross
section of the IBD process, $P_{\rm sur}$ is the $\bar{\nu}_e \to \bar{\nu}_e$ survival
probability (Eq.~\ref{eqn:Psurv}), and $S$ is the differential
energy distribution of the antineutrinos.

Using Eq.~\ref{eq:absolute} to precisely measure $\theta_{13}$  from either a deficit in the observed antineutrino rate 
or a distortion in the energy spectrum requires that all systematic uncertainties  be well controlled.
Background corrections to the observed antineutrino spectrum can be reduced with overburden and detector shielding. 
Detector efficiency uncertainties in the  2-3\% range were achieved in previous experiments.
The reactor antineutrino flux and spectrum can also be calculated to 2-3\% precision ~\cite{reactorflux1, reactorflux2}. 

To eliminate the dependence on the neutrino flux,
Mikaelyan and Sinev proposed the use of two antineutrino detectors~\cite{Russian}. 
The first detector, located close to the reactor, measures the antineutrino flux as a function of energy
while a second detector further from the reactor is located close to the oscillation minimum of $P_{\rm sur}$. 
For this situation, one near and one far detector observing antineutrino events from a  single reactor, 
the ratio of  antineutrinos observed in a specific energy range at the two detectors is given by 

\begin{eqnarray}
\frac{N_{\rm f}}{N_{\rm n}}&=&
\left (\frac{N_{\rm p,f}}{N_{\rm p,n}}\right )
\left (\frac{L_{\rm n}}{L_{\rm f}}\right)^2
\left (\frac{\epsilon_{\rm f}}{\epsilon_{\rm n}}\right )
\left [\frac{P_{\rm sur}(E,L_{\rm f},\theta_{13})}{P_{\rm sur}(E,L_{\rm n},\theta_{13})}\right]
\label{eq:relative}
\end{eqnarray}

\noindent where  $N_{\rm p,f}$ and $ N_{\rm p,n}$ refer to the numbers of target protons at the far and near sites, 
and $L_{\rm f}$ and $L_{\rm n}$ are the distances of the far and near detectors from the reactor core.
The ratio of the  detector efficiencies ($\epsilon_{\rm f} / \epsilon_{\rm n}$) can be determined more precisely than the individual 
efficiency.  The near-far relative measurement cancels  nearly all reactor-related and   detector-related systematic errors.

The Daya Bay experiment uses eight antineutrino detectors (ADs) to monitor 6 nuclear reactor cores.
To minimize  efficiency differences between the near and far detectors,  
interchangeable  ADs of identical design were built 
and  assembled above ground using standardized procedures. Built in pairs, antineutrino detectors were
filled with liquid scintillator  within eight days of each other to further minimize  possible differences in performance.

\begin{figure}
    \centering
    \includegraphics[clip=true, trim=20mm 10mm 10mm 10mm,width=3.1in]{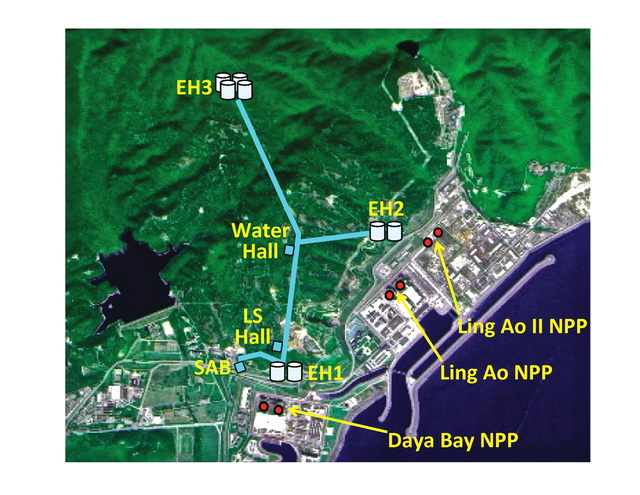}
    \caption{
Layout of the Daya Bay experiment. The dots represent reactor cores.
Each pair of cores forms a Nuclear Power Plant (NPP). The three NPPs, Daya Bay, Ling Ao, and Ling Ao-II 
are between the Daya Bay coast and 400-m-high mountains inland. Eight antineutrino detectors (ADs) were 
installed in three experimental halls (EHs). Additional halls were used for filling the detectors (LS Hall) or 
processing the water (Water Hall) for the experimental hall water pools. 
The ADs were assembled in the Surface Assembly Building (SAB) before being moved underground. 
Not shown are surface support buildings adjacent to the SAB containing the control room, 
electrical and ventilation utilities.
}
    \label{fig-layout}
\end{figure}

This paper  describes all of the Daya Bay detector systems. 
After a brief overview  of the experimental layout  in Section~\ref{sec-layout}, 
the construction, assembly and installation  of the detectors are presented in Section~\ref{sec-overview}. 
Sections~\ref{sec-ad} and ~\ref{sec-adcalib} describe the antineutrino detector and calibration systems.  
The muon detector system overview is given in Section~\ref{sec-mudet}, followed by  
a description of the data acquisition system in Section~\ref{sec-electronics-daq} and offline computing 
in Section~\ref{sec-offline}. 
Several years of calibration and antineutrino data measure detector performance 
over time and allow comparisons of the ADs in Section~\ref{sec-performance}.

\section{Experimental Layout  }

\label{sec-layout}

As depicted in Fig.~\ref{fig-layout}, the Daya Bay reactors are arranged in three pairs of reactor 
cores spread  over nearly 2 km of coastline in southern China. 
Each pair  of cores is designated a Nuclear Power Plant (NPP).
Experimental Hall 1 (EH1) measures the antineutrino rate and spectrum primarily from  the Daya Bay (DYB)  NPP.
Similarly, Experimental Hall 2 (EH2) measures the rate and spectrum primarily from  the Ling Ao and Ling Ao-II NPPs.  
These near sites are positioned  as close to the reactor cluster as possible, given constraints on the desired  overburden and rock condition. 
The location of the far hall (EH3) was  determined from a multi-parameter optimization of desired baselines for maximum oscillation
sensitivity given the mountain profile and rock quality~\cite{tdr}. 
The water equivalent overburden, simulated muon rate and average muon energy at each site are listed in Table~\ref{table-overburden}. 

\begin{table}
\caption{Vertical overburden in meters-water-equivalent  (mwe), estimated muon rate $R_\mu$ and average muon energy in 
the three experimental halls. }
\centering
\begin{tabular}{ cccc}
\hline
  Hall     &Overburden(mwe)       &  $R_\mu$(Hz/m$^2$)  & $\langle$ E $\rangle$(GeV) \\
\hline
EH1   &   250  & 1.27  & 57 \\
EH2 &  265 & 0.95 & 58 \\
EH3&   860  & 0.056 & 137 \\
\hline
\end{tabular}

\label{table-overburden}
\end{table}

Daya Bay deploys two ADs in each near hall and four ADs in the far hall.
Multiple, functionally identical, movable detectors are unique to   the Daya Bay experiment. 
Dividing the target mass in each hall into standard units, built above ground, allowed the construction and assembly
of the antineutrino detectors to proceed in parallel with preparation of  the experimental halls, considerably 
reducing the  total project time.
The large total target mass at the far hall enables the unprecedented statistical precision of the Daya Bay experiment.  
Side-by-side comparisons between detectors at  the near  sites cross check  calculations of  relative 
detector efficiencies to better than 0.2\%.

Daya Bay ADs are constructed from three-concentric-cylindrical tanks. 
The inner acrylic vessel (IAV) contains gadolinium doped liquid scintillator. 
A second acrylic vessel (OAV) containing un-doped liquid scintillator surrounds the IAV. 
The OAV is surrounded by a stainless steel vessel (SSV) containing mineral oil.  
The inner cylindrical wall of the SSV supports nearly two hundred photomultipliers (PMTs) pointed inward toward the OAV and IAV.
Reflectors above and below the OAV  redirect light to the PMTS, thereby reducing  the total number of PMTs  required.

\begin{figure}\centering
    \includegraphics[clip=true, trim=0mm 0mm 0mm 0mm,width=3.1in]{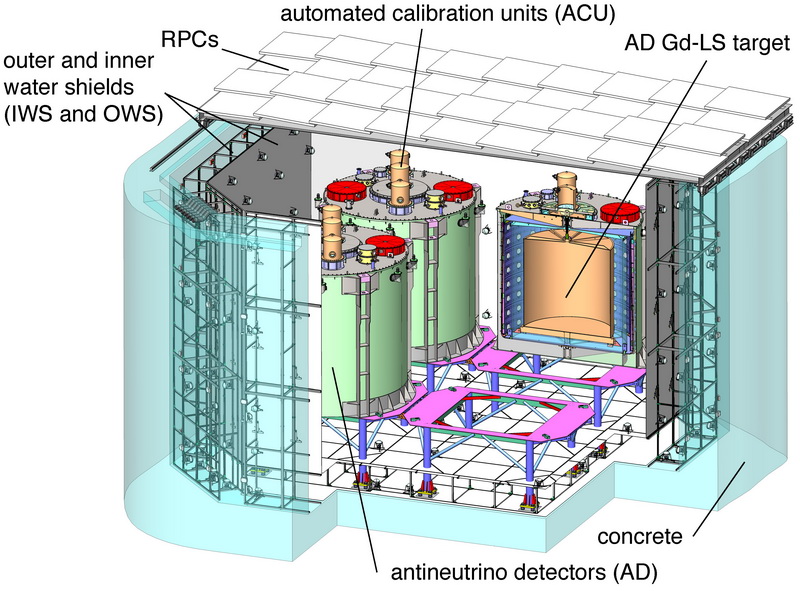}
    \caption{Far hall with 3 ADs installed. The water pools are segmented into two independent zones - an outer water shield
1m thick along the perimeter and over the floor and inner water shield containing the antineutrino detectors. 
Four layers of Resistive Plate Chambers over the water pool add further cosmic ray identification capability.} 
    \label{fig-farhall}
\end{figure}

The antineutrino detectors are placed in water pools.  
At least 2.5 meters of ultra pure water surround each antineutrino detector
and shield against background radiation. 
Instrumented with PMTs,  the pool  is segmented into two Cerenkov detector zones
for active identification of cosmic rays. 
The pool is covered by resistive plate chambers (RPCs) for additional cosmic ray muon identification.
Figure~\ref{fig-farhall} shows three  ADs in the far water pool.

\subsection{The reactor complex}

The Daya Bay nuclear power complex is located 
55 km northeast of Hong Kong and 45 km east of Shenzhen. 
The  six reactor cores are pressurized water reactors with a maximum of 2.9 GW thermal power~\cite{dayabay}. 
 The cores in each pair are separated by  88 m. 
 The Ling Ao and  Ling Ao-II NPPs are approximately 500 m apart, while 
the Daya Bay and Ling Ao NPPs are separated  by  $\sim$1100~m.  

The Daya Bay and Ling Ao NPPs
are based on the Framatone ANP French 990 MW$_e$ (gross electrical power) design. 
The Ling Ao-II NPPs have an updated design (CPR1000)  with 
1080  MW$_e$.  The Daya Bay reactors  utilize $\sim$4\% enriched $^{235}$U fuel.   
Each reactor core contains 157 3.7-m-long fuel rod assemblies. 
The centroid of the power production is known to about 2 cm horizontally and 20 cm vertically.  
Approximately 1/3 of the fuel assemblies of the Daya Bay NPP reactors are replaced during a refueling shut-down
every 18 months. 
In the  Ling Ao NPP reactors,  $\sim$1/4 of the fuel assemblies are replaced every 12-18 months.

The thermal power of each core is monitored continuously  by the power plant
with an uncertainty of 0.5\%~\cite{thermal_power1, thermal_power2,thermal_power3}. 
The fission fractions of the four main isotopes contributing to power (and antineutrino) production
are modeled as a function of time to follow the burn-up of the nuclear fuel~\cite{reactor}.  
Daily thermal power measurements and fission fraction estimates  are provided to the collaboration by the power plant.

\subsection{Underground Halls  }

The detectors reside in three large experimental halls (EH) excavated from granite bedrock. 
The halls are connected by large cross-section (6.2~m $\times$ 7.0~m) tunnels through which 
completed antineutrino detectors were moved from the filling hall to the experimental halls.  
The rock overburden is 93~m at EH1, 100~m at EH2 and 324~m at EH3. Core samples taken
throughout the area found the average rock density to be 2.6 g/cm$^3$.

\begin{figure}
\centering
\includegraphics[clip=true, trim=55mm 25mm 40mm 15mm,width=3.1in]{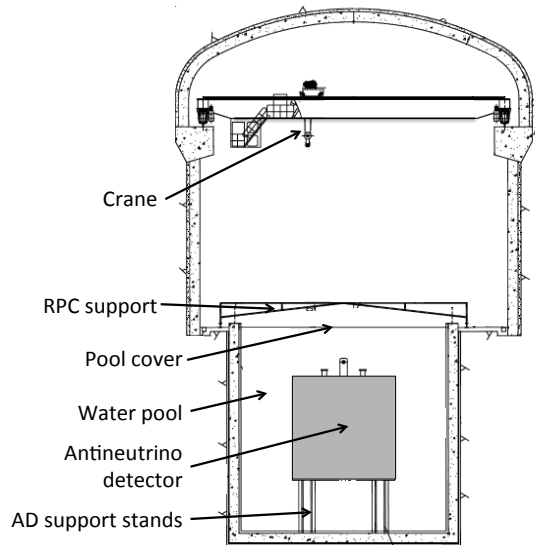}
\caption{Cross section of a near hall showing an antineutrino detector 
on a stand in the water pool beneath the RPC support structure and overhead crane.}
\label{fig-nearhall}
\end{figure}

The two near experimental halls are 15~m wide by 45-50~m long. As seen in Fig.~\ref{fig-nearhall} the roof 
of the 28~m long central section  of the hall is nearly 15~m above the hall floor to provide space 
for   a 120-ton  crane covering the water pool and  detector staging areas next to the pool.
An  area 18~m long  by 3~m high provides storage space for the RPC system when rolled back
from the water pool. 
The far experimental hall housing four ADs is 21~m wide by 62~m long.
The experimental halls are isolated from the tunnels by roll-up doors to
keep dust to a minimum. Fresh, filtered, and cooled air is delivered from above ground by a network of 
vents which run through the tunnels.  
Cleaning crews worked constantly to keep the halls clean during installation and commissioning.
Auxiliary rooms for the gas system, data taking, water purification and refuge are located next to each experimental hall. 
A secondary personnel egress tunnel links these auxiliary rooms to the main access tunnel.   

Ten-meter-deep water pools house the ADs in each hall~\cite{muon}. 
The water pools,  
10~m wide by 16~m long for near halls and 16~m by 16~m for the far hall, were
constructed in an  octagonal shape, as shown Fig.~\ref{fig-farhall}.
The water pools were excavated from the rock and lined with waterproof concrete strengthened by grounded rebar. 
A PermaFlex$^{TM}$ liner which had been extensively tested for compatibility with high purity water was sprayed over the concrete to seal the surface against radon penetration.  A  thick  black rubberized  sheet is suspended over the pool by cables for light
and gas isolation and is securely fastened to the edges of the pool.
Nitrogen from cryogenic liquid bottles is used as an inert cover gas between the cover and water surface. The measured average radon concentration in the water is less than 50 Bq/m$^3$ for the three halls, about
3-8 times lower than the radon concentration in the air above the pool cover.

Two special purpose underground halls were also built. A liquid scintillator (LS) hall located near EH1 was used to produce
and store AD liquids under clean room conditions. 
The LS hall contains two thermoplastic lined concrete storage tanks  of 200 ton 
capacity, five 40-ton capacity acrylic tanks, and related piping and pumping systems. 
After LS production the hall was used to fill  ADs.
Near the intersection point of the main tunnels  a water hall contains the central water purification station. 
Highly purified city water is  sent to the three experimental halls with pipes 
of about the same length to ensure equal water quality ~\cite{Water}. 
The LS and Water hall locations are shown in Fig.~\ref{fig-layout}.

\subsection{Baseline and position survey }

\begin{figure}
\centering
\includegraphics[width=3.1in]{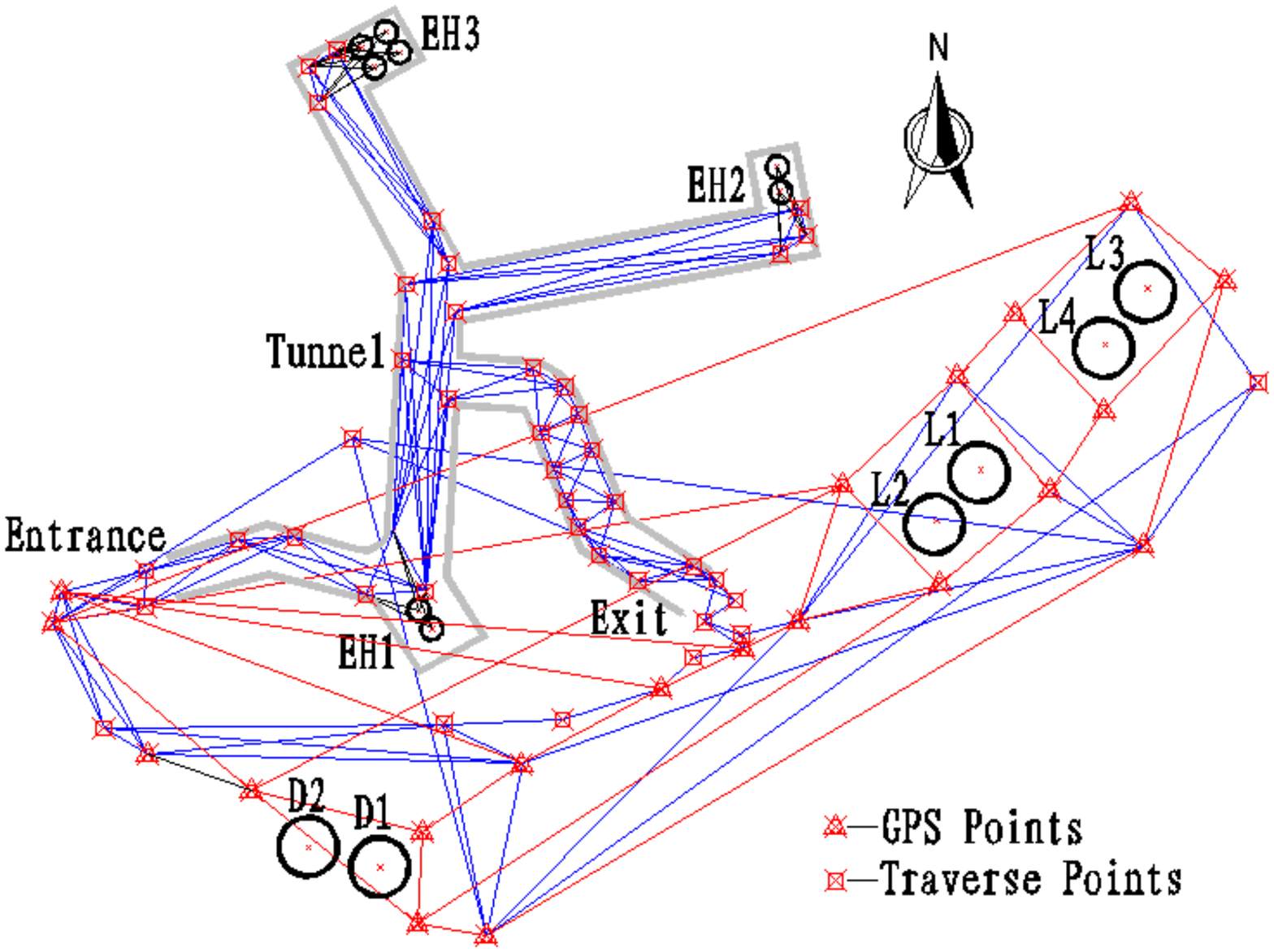}
\caption{Baseline survey. Bigger black circles are reactor cores, smaller black circles are ADs. Tunnels  are shown in grey.  Red lines represent the  GPS control network. Blue lines represent the Total Station transverse network.   }
\label{fig-baseline}
\end{figure}

The distances (baselines) from each AD to each reactor core are precisely measured via two independent surveys~\cite{prl_rate, cpc_rate,nim_sidebyside} using Global Positioning System (GPS) receivers and modern 
 theodolites.
 GPS control networks and Total Station traverse networks were installed to link
measurements of the reactor complex area to points outside each of the experimental halls.  
Further Total Station surveys linked each hall position to the external monuments on each AD which had 
been used as references for surveys of the AD interior during construction and assembly.
The GPS control points combined pre-exisiting power plant survey points with  newly installed points  to 
obtain a comprehensive, redundant network overlapping with the Total Station traverse network.   
Shown as red lines in Fig.~\ref{fig-baseline}, the GPS network gives  static synchronous observations  
based on a high-precision dual frequency GPS receiver.  
The Total Station transverse network shown as blue lines in Fig.~\ref{fig-baseline}  provides coordinates 
for all the transverse points using a Leica TDA5005 Total Station.

To minimize propagation errors inside the tunnels  several measures were adopted.
 Dual transverse networks were setup in each tunnel to increase reliability. 
 An additional transverse network loop through the main Daya bay tunnel and out the construction tunnel (Exit in Fig.~\ref{fig-baseline})
 checked for error buildup.
A  station was positioned on the top of the mountain to measure  points at the  tunnel entrance, exit,  and  reactors   to increase conversion precision between the two networks.  The overlap between GPS and transverse network   improved
the conversion precision between these two coordinate systems.

The measured  baselines are given in Table~\ref{table-baselines}. 
The  largest baseline difference between the two surveys is 4~mm and the uncertainty in the baselines is 
determined to be less than 18~mm. 
The largest contribution to the overall uncertainty was 
the  measurement error of 12.6~mm using the  GPS network.
Other contributions were a  9.1~mm error due to fitting uncertainties associated with the linking of the GPS 
and the Total Station networks,    
a measurement error of 6.5~mm using the Total Station transverse network, 
a 5.9~mm error coming from conversion 
from the reactor core to GPS control network, and an error of 2.8~mm coming from conversion between antineutrino detector 
center and the GPS control network.   When combined with the uncertainties of the effective center of fission~\cite{cpc_rate},
these  baseline uncertainties were found to make a negligible contribution to the  uncertainties of the oscillation parameters.

\begin{table*}
\centering
 \caption{Baselines from the antineutrino detector centers to the reactor core centers in the Daya Bay (D1, D2), Ling Ao (L1-2) and
 Ling Ao-II (L3-4) Nuclear Power Plants is given in meters. 
 The  total survey uncertainty is 18 mm.}
\centering
\begin{tabular}{|c|c|c|c|c|c|c|}
\hline
 AD$\backslash$ Core      &  D1 & D2 & L1 & L2 & L3 & L4 \\
       \hline
EH1-AD1 & 362.380 & 371.763 & 903.466 & 817.158 & 1353.618& 1265.315   \\
\hline
EH1-AD2 & 357.940 & 368.414 & 903.347 & 816.896 & 1354.229& 1265.886 \\
\hline
EH2-AD1 & 1332.479 &1358.148 & 467.574 & 489.577 & 557.579 &   499.207\\
\hline
EH2-AD2& 1337.429 & 1362.876 &  472.971 & 495.346  &  558.707  &   501.071\\
\hline
EH3-AD1& 1919.632 & 1894.337 & 1533.180 & 1533.628 & 1551.384 & 1524.940  \\
\hline
EH3-AD2 & 1917.519 & 1891.977 & 1534.919 & 1535.032 & 1554.767 & 1528.079 \\
\hline
EH3-AD3 & 1925.255& 1899.861& 1538.930& 1539.468 & 1556.344 & 1530.078  \\
\hline
EH3-AD4& 1923.149 & 1897.507 & 1540.667& 1540.872& 1559.721 & 1533.179\\
\hline

\end{tabular}
\label{table-baselines}
\end{table*}

\section{ Construction, Assembly and Installation Overview  }\label{sec-overview}

Civil construction started in October, 2007 with the beginning of tunnel excavation. 
Upon completion of the surface assembly building (SAB)  in 2009, test assembly of a full-size prototype antineutrino 
detector initiated onsite assembly activities.  As the tunnels and experimental halls were excavated and prepared for
beneficial occupancy,  AD assembly and preparation of the muon system PMTs and 
RPCs proceeded in parallel above ground in the SAB.
Well-coordinated underground and surface activities ensured the timely start of the experiment. 

\begin{figure}
\centering
\includegraphics[clip=true, width=3.1in]{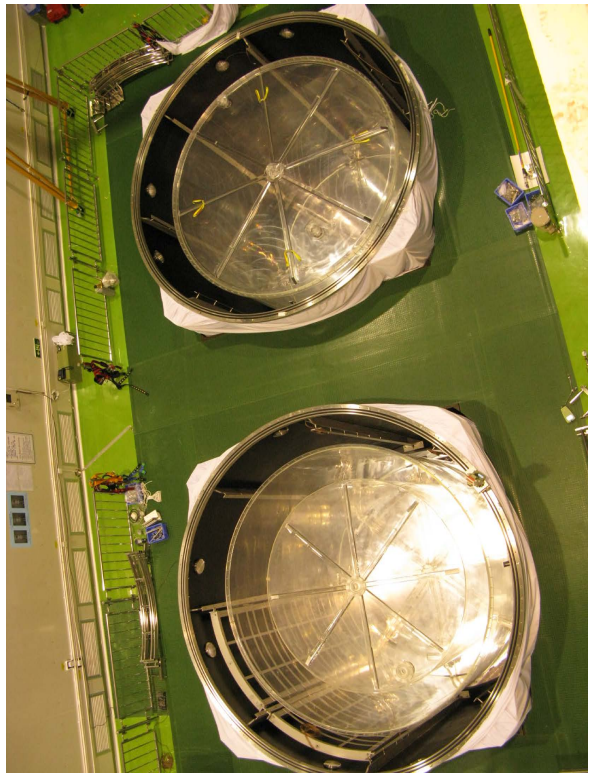}
\caption{Photograph of the SAB clean room showing two ADs under construction. 
Both construction pits contain a stainless steel vessel (with lid removed) and an outer acrylic vessel.
The  lower AD has two photomultiplier support ladders being test fit. The lid of the lower outer acrylic vessel
has been removed to allow insertion of an inner acrylic vessel.}
\label{fig-SAB2}
\end{figure}

All assembly of the antineutrino detectors except for filling and final cabling was  performed  in the  SAB.
The north side of the SAB, used primarily for testing muon system PMTs and storage,  was serviced with a 10-ton overhead crane.
The south side of the building was serviced by  a 40-ton overhead crane and contained a 11~m  wide by 33~m long by 11~m high
clean room and a smaller adjacent cleaning area. 
The clean room contained two four-meter deep pits to allow insertion of AD components into the Stainless Steel Vessel (SSV) at convenient working heights as shown in Fig.~\ref{fig-SAB2}. 
Specialized platforms provided safe access inside and around the perimeter of the AD. 
The clean room was designed to meet ISO7 cleanliness standards (ISO 14644-1). 
Cleaning crews were constantly at work in the clean room and cleaning area to keep the floors and tables dust and dirt free.
The weekly measured particle counts ($\geq 0.5$ micron) were typically  $\leq 10,000$ counts/m$^3$ which is better than the more rigorous ISO6 standard.
Radon levels in the clean room were measured to be  $147\pm13$ Bq/m$^3$.

A roll-up curtain wall separated the  cleaning area from the clean room. 
This semi-clean area was isolated from the rest of the SAB by a second roll-up curtain wall which kept particulate levels low. 
The cleaning area was equipped with a resin bed water filtration system which provided $>10$~megaohm-cm water 
for cleaning items that would enter the clean room. 

All eight ADs were constructed and assembled in the same clean room over the period of 2009-2012 using the 
procedures  developed during the prototype assembly. 
The time needed for AD assembly  in the SAB decreased from 13 months for the first pair to 3.5 months as the production 
rate of sub-components improved and the crew gained experience.  
 ADs 1 and 2 were moved underground, filled with liquids, installed and commissioned in EH1 in July - Sep.,  2011.  
 A comparison of 
detector performance of these ADs can be found in~\cite{nim_sidebyside}. AD3 was installed in EH2  the following October.  ADs 4-6 were then installed in EH3.  Six AD data collection started in Dec. 2011.
Installation of ADs 7 and 8 completed the planned deployment  in Aug.-Sep., 2012.  
All eight ADs have been collecting data since Oct. 2012.

The Liquid Scintillator hall was the first underground hall ready for occupation in 2010.   
Two large storage tanks and five smaller acrylic tanks were used to hold AD liquid raw materials and finished liquid scintillators.
Multiple tanks and pump  stations were installed  and kept in clean room conditions for AD liquid production and testing.   All
liquid scintillators were produced by early 2011 and monitored for nearly 1/2 year before first use in an AD.
ADs were brought into the LS hall one at a time for filling. A custom filling system precisely measured the liquid masses 
pumped into each AD while closely monitoring liquid levels to avoid damaging level mismatches.

The first near experimental hall was completed in 2011. The water pool was  lined with waterproof concrete, 
painted with three layers of PermaFlex$^{TM}$, and also lined with Tyvek $^{TM}$ sheets heat-welded 
together. Preassembled  frames were fastened to the wall and floor to support the muon PMTs.
An inner Tyvek layer was added to separate the inner and outer regions.

 The muon  detector assembly  proceeded  in parallel with AD assembly and preparation 
of the experimental halls. PMTs were tested and placed in mounting brackets and magnetic shields in the SAB.
After completion of the experimental halls, AD support stands were installed in the pools 
onto precisely surveyed studs protruding through the concrete floor and anchored to the underlying rock.
Muon PMTs were then installed onto the support structure and cabled to the electronics room.
Interlaced with the PMT installation, RPC support structures were installed on the rails on either side of the water pool.
RPC  modules containing four RPC layers were  installed onto the structural supports.
The entire RPC assembly was then rolled into its storage area for completion of the cabling and gas connections while
the ADs were installed.

A crucial element of the assembly plan was the custom automatic guided vehicle  transporter~\cite{AGV} (AGV) used to transport 
ADs and other large parts within the SAB, dry ADs down the 10\% grade from the SAB to LS Hall,
and filled ADs from the filling hall to the experimental halls. 
The AGV operated under diesel or electric power as required and could support up to 130 tons. 
When transporting filled ADs through the $\leq0.5$\% grades of the main Daya Bay tunnels, the AGV 
load bed was kept level using the auto-leveling feature to reduce  stress on AD components.
Eight sets of heavy duty omni-directional wheel modules provided 360 degree maneuverability  for precisely
positioning heavy loads.

ADs were brought into the experimental hall by the AGV and set down on a temporary support stand 
by the edge of  the water pool by lowering the AGV bed. 
The lifting fixture and video position monitoring system were mounted on top of the AD.
The AD was then lifted by the overhead crane into the pool onto the support stands. 
Access ladders and platforms were mounted  on top of the AD for removal of the lifting fixture and  video system. 
Cables from the AD PMTs were run through dry pipes to the edges of the pool before being routed to the electronics room.
Similarly, cables from the automated calibration units (ACU) and lid sensors were run through a large dry pipe to above 
the pool surface, and the cover gas lines were run to the gas room.
Function tests of all AD systems were performed before  final leak checks of all connections ensured that the detector
was leak tight.
A final survey of the  AD position within the hall was conducted before filling  the pool with water. 
The final installation steps were to seal  the pool with a gas-tight, opaque, black cover and to roll the  RPCs over the pool.   
More detailed descriptions of the assembly and installation processes can be found in
references \cite{muon,Assembly}.

\section{Antineutrino Detectors  }  
\label{sec-ad}

\begin{figure}
\centering
\includegraphics[clip=true, trim=0mm 10mm 0mm 10mm,width=3.3in]{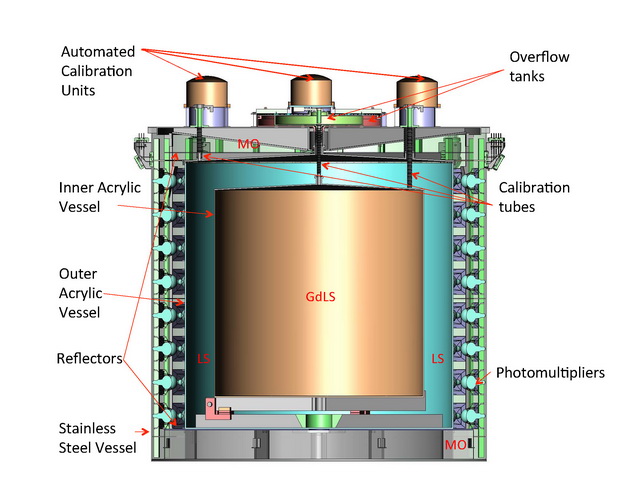}
\caption{Antineutrino detector. }
\label{fig-AD}
\end{figure}

The antineutrino detectors  are designed to detect IBDs while minimizing systematic errors. 
As shown in Fig.~\ref{fig-AD}, an  antineutrino detector is made from three concentric  vessels of cylindrical shape.
Cylinders are easier to construct than spheres and have a more uniform detector response than rectangular geometries.
The inner target volume of 
gadolinium doped liquid scintillator (GdLS)   is contained by an inner acrylic vessel (IAV) approximately 3 m high 
and 3 m in diameter.  Surrounding the IAV is an outer acrylic vessel (OAV) approximately 4 m high 
and 4 m in diameter filled with un-doped  liquid scintillator (LS).  
The LS improves the detection efficiency of gamma rays  from interactions in the GdLS target volume. 
The OAV is inside a SSV
approximately 5m high and 5m in diameter which supports ladders of photomultipliers (PMTs)
pointed at the GdLS and LS volumes and isolates the inner detectors from the water pool.   
The space between the SSV and OAV is filled with high purity mineral oil (MO) which suppresses 
radioactive backgrounds from the PMT glass and the surrounding environment.

Unlike similar experiments with cylindrical inner detectors~\cite{dchooz2,reno}, Daya Bay uses circular reflectors above and below the OAV
instead of PMTs~\cite{tdr}. 
The reflectors redirect scintillation light  towards the PMTs mounted on the perimeter and reduce the required number of PMTs by 50\%
without significantly degrading the energy or position resolution. 
The dimensions of the three vessels were chosen to maximize the target mass 
without increasing the systematic uncertainty due to backgrounds. 
A Daya Bay AD has a ratio of GdLS target mass to  total liquid mass of 26\%, higher
than  similar experiments, such as Double Chooz~\cite{dchooz2} or RENO~\cite{reno}.
Radioactive background rates are reduced sufficiently  to render systematics in the accidental background subtraction  negligible
in the near/far oscillation analysis.

Mounted on the SSV lid are three ACUs which can position
radioactive calibration sources or light emitting diode (LED) pulsers at different positions in the GdLS and LS volumes.
The ACUs and liquid volumes are
connected by calibration tubes. The central calibration tubes also connect the GdLS and LS volumes to concentric overflow
tanks on the SSV lid. All gas spaces above the overflow tanks were continuously purged with dry nitrogen 
from a cover gas system~\cite{Gas}.

The eight ADs were built above ground in pairs and then moved below ground for filling.
GdLS, LS, and MO were pumped simultaneously into an AD as needed while keeping the relative liquid levels 
within 10 cm of each other. The liquid levels were then topped up to fill the overflow tanks to about 1/3 capacity.

Although the nominal pair-wise production and assembly plan should have  resulted in an orderly association of the first SSV
with the first IAV and first OAV, this pattern and the orderly placement of ADs into the halls was quickly broken by practical schedule considerations. 
The first SSV needed extra work to obtain a suitably smooth finish in the O-ring grooves and was used in AD3. 
The second IAV pair hold-down tabs and calibration ports were mis-aligned, necessitating  complimentary changes in the last pair of OAVs.
IAV3 was damaged during an annealing accident and was replaced by IAV9. 
To obtain the earliest physics data, it was decided to run with only 6 ADs spread over the three experimental halls. 
The AD naming scheme and major components of each  AD are shown in Table~\ref{table-ADmap}.

\begin{table}
 \caption{Naming scheme for  antineutrino detector (AD) placement in the three experimental halls (EH). 
 The major components of the ADs: stainless steel vessel (SSV), 
 outer acrylic vessel (OAV), and inner acrylic vessel (IAV) were labeled by production order.  
 ADs were assembled and filled in pairs. Even though the components were 
 also typically produced in pairs, the order in which the components were used depended on special considerations described in the text. 
 }
 \begin{center}
\begin{tabular}{|c|c|c|c|c|}
\hline
 AD      &  \textbf{AD\#} & \textbf{SSV\#} & \textbf{OAV\#}&\textbf{IAV\#}  \\
\hline
\textbf{EH1-AD1} & AD1 & SSV2 & OAV1 & IAV2    \\
\hline
\textbf{EH1-AD2} & AD2 & SSV3 & OAV2 & IAV1  \\
\hline
\textbf{EH2-AD1} & AD3 &SSV1 & OAV3 & IAV5 \\
\hline
\textbf{EH2-AD2} & AD8 & SSV8 &  OAV7 & IAV4  \\
\hline
\textbf{EH3-AD1} & AD4 & SSV4 & OAV4 & IAV6   \\
\hline
\textbf{EH3-AD2} & AD5 & SSV5 & OAV5 & IAV8  \\
\hline
\textbf{EH3-AD3} & AD6& SSV6 & OAV6 & IAV7  \\
\hline
\textbf{EH3-AD4} & AD7 & SSV7 & OAV8 & IAV9\\
\hline

\end{tabular}
\end{center}
\label{table-ADmap}
\end{table}

Control of radioactive backgrounds was achieved by strict control of all materials used in the AD and of all
procedures used during assembly and construction. 
Details for  individual parts can be found in the related section.

\subsection{Stainless Steel Vessel }

\begin{figure}
\centering
\includegraphics[width=3.1in]{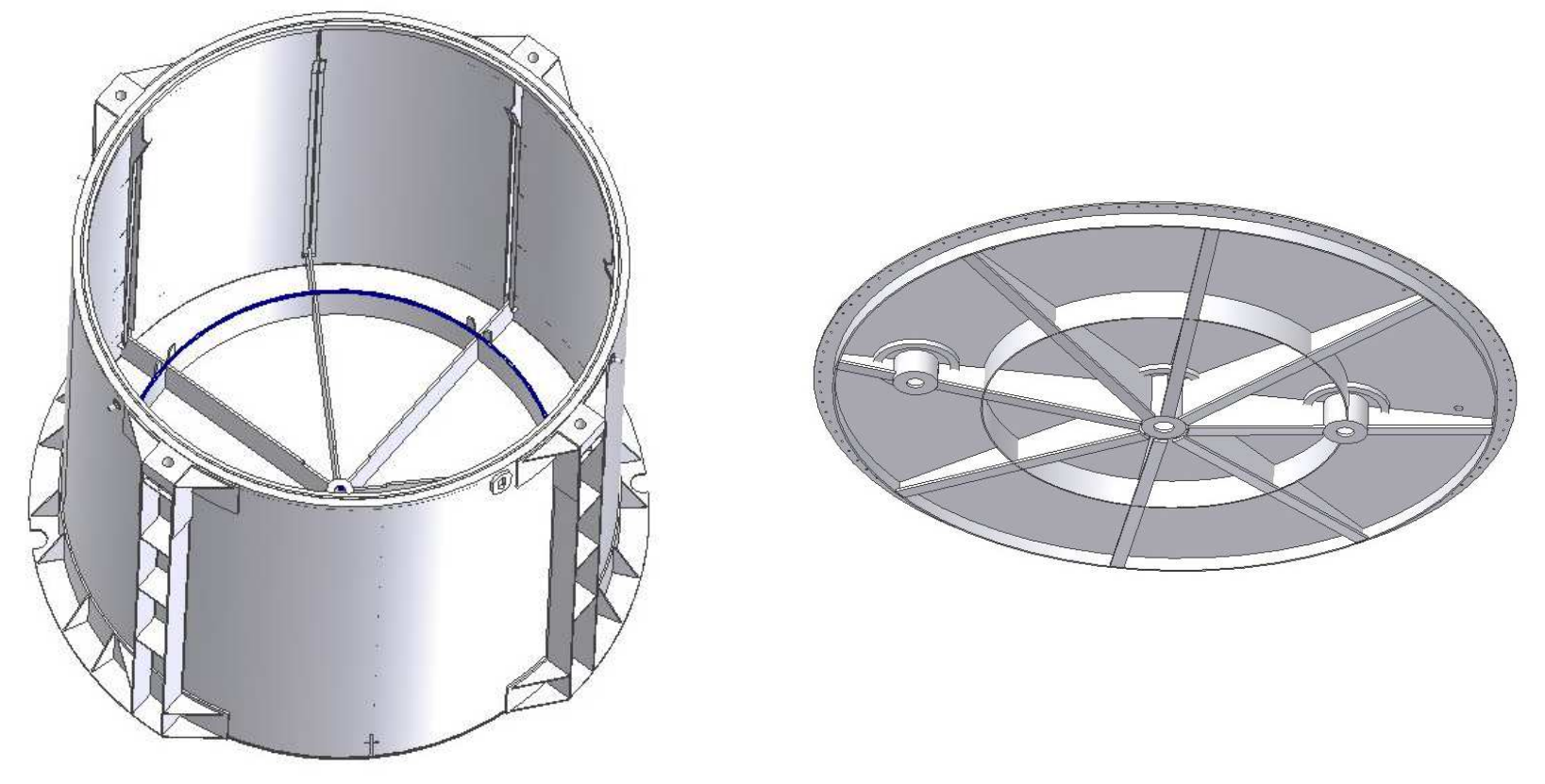}\hfil%
\caption{Stainless steel vessel (SSV). Left: barrel and bottom sections. Right: inner side of the tank lid.}
\label{fig-ssv}
\end{figure}

The stainless steel vessels (SSV) are  5000 mm high, 5000 mm diameter cylinders with 
12 mm thick low radioactivity stainless steel walls. 
The external walls are strengthened by 12 mm thick internal ribs. 
 Each SSV(dry) weighs  about 24 ton (with an inner 
volume of $\approx$95 m$^3$).  Figure~\ref{fig-ssv} shows the structure of the barrel and lid.  
When loaded with the acrylic vessels, PMTs, and liquids, an AD weighs $\approx$112 tons. 
Deformations and  mechanical strength for the SSV vessel were thoroughly studied using 
Finite Element Analysis (FEA) code  to insure that the design  had the required safety factor 
when a filled SSV was picked up and lowered onto its support stands or when 
the lid was submerged under  2.5 m of  the water. 
SSVs were constructed in pairs by the Guangdong Shanfeng Chemical Machinery Co. LTD.
A SSV barrel was welded together from  three subsections: a bottom section including 
bottom ribs, a top section  with a top flange, and a middle barrel section.  

\subsubsection{SSV O-ring and lid}

The removable SSV lid is sealed at the circumference by two 19 mm thick O-rings placed 
in  two concentric grooves 23 mm wide by 14 mm high. 
With the 112 lid bolts tightened to 120 N-m the O-ring is typically compressed by 16-26\%.
Similar, but smaller,  O-rings were used to seal the overflow tank lids to the SSV lid. 
The O-rings were  required to be compatible  with water, mineral oil and liquid scintillator,  
and to provide  effective sealing under pressure for more than five years.
 DuPont$^{TM}$ Fluorocarbon Rubber Viton$^{\textregistered}$ A supplied by Shenzhen-O Technology Co., Ltd was 
 chosen as the O-Ring material and  passed all requirements. 
O-rings were subjected to a 60 day  accelerated aging study at elevated temperature ( (110-170~$^o$C)
under the compression of 17\% at the Aerospace Materials and Technology Research Institute. 
Based on these studies,  permanent O-ring deformation  after five years at nominal  temperatures ($\leq 25^\circ$C) was
projected to be $\leq$~9\% in water.

The Viton-A  O-rings were made  using a segmented mold vulcanizing process, which achieved the same 
performance as a molded vulcanized O-ring seal without making a very big mold.  The
space between the O-rings was connected to an external port to enable leakage tests~\cite{OringLeaktest}
of the final O-ring seal to ensure that  the SSV would not leak when immersed in the water pool.

\subsubsection{Fluorocarbon painting inside SSV}

\begin{figure}
\centering
\includegraphics[clip=true, trim=40mm 37mm 20mm 30mm,width=3.1in]{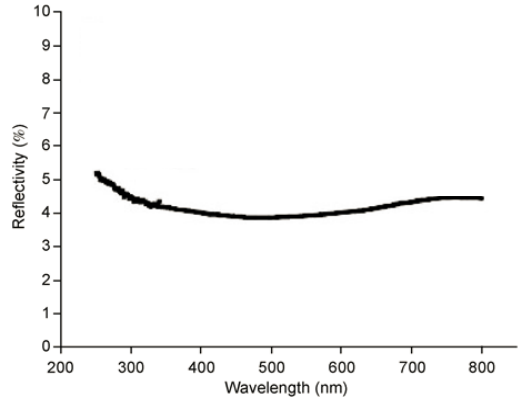}
\caption{Diffuse reflectivity of the black fluorocarbon paint developed for Daya Bay versus wavelength range. 
}
\label{fig-ssv-reflectivity}
\end{figure}

The inside of the SSVs are coated with a specially developed black fluorocarbon paint~\cite{ssvpaint} to minimize reflected photons.
Although common fluorocarbon paints are  compatible  with mineral oil,  low in radioactivity and strongly adhesive,
they typically require oven curing at high temperature.
 Since available ovens were $\leq 20$ m$^3$, 
a new kind of fluorocarbon paint that cures at normal temperatures was adopted for the $\sim$100 m$^3$ SSV.   
Different paint formulas were applied to 304L stainless steel samples. 
The chosen  black-matte fluorocarbon paint, a mixture of diluting agent, solidifying agent and fluorocarbon resin,  
 adhered well to the stainless steel, had good uniformity, and was scratch and wear resistant.
The diffuse reflectivity of the black matt fluorocarbon paint was measured over wavelengths ranging from 250 to 800 nm
as shown in Fig.~\ref{fig-ssv-reflectivity}.
Reflectivity was roughly a constant (4.5\%)  above 300 nm. 
The radioactivity of the paint was measured by a very sensitive $\gamma$-ray detector  consisting of a 
high-purity Ge detector and anti-Compton system to be
$\leq 0.31$ Bq/Kg  for $^{214}$Pb, $\leq 0.21$ Bq/Kg  for $^{212}$Pb, 
$\leq 0.29$ Bq/Kg  for $^{238}$Ac, 
and $\leq 0.30$ Bq/Kg  for $^{40}$K.

\subsubsection{Radioactivity Control}
Stainless steel plates and welds typically have natural radioactive contaminants which could generate backgrounds in the inner detectors.
A Monte Carlo study established requirements for allowable radioactivity levels from the major contaminants.
Quality assurance procedures were implemented  to ensure that construction materials met these goals.
A total of 260 tons low radioactivity 304L stainless steel was specially made by the Shanxi Taigang Stainless Steel Co., Ltd. in one batch. 
The steel sample was radioassayed before the steel ingots were rolled to sheets of more than 10 different thickness as required by the SSV design. 
The contamination goals and measured radioactivity of materials used in  SSV construction are shown in Table~\ref{table-ssv}.  
Welding materials were selected from many vendors after assaying radioactivity in samples. 
The radioactivity of the chosen welding materials was measured to be $0.14\pm 0.08$ Bq/kg for $^{238}$U, $<0.30$ Bq/kg 
for $^{232}$Th, and $<0.48$ Bq/kg for $^{40}$K.  The contribution to the overall background rate was negligible due to the 
small mass of the welds compared to the total SSV mass. 
A stainless steel disk sample with a welding line was also measured to verify that no additional radioactivity was introduced during the welding process.
The radioactivity of all SSV materials was well below the requirements set by Monte Carlo studies.

\begin{table}
\centering
\caption{Measured radioactivity levels in the stainless steel used in  SSV construction are compared to the maximum allowable
contamination levels determined from Monte Carlo Studies. The expected contribution to the AD trigger rate from each contaminant 
at the allowed level is also shown.}
\begin{tabular}{|c|c|c|c|}
\hline
Isotope  &  Requirement & Measured  &  AD Trigger\\
          &   &  Value &       Rate (Hz\\
 &    (Bq/Kg)  &      ( Bq/Kg)   &  $>1$ MeV)\\
\hline
$^{238}$U    &  $<0.03 $  &  $<0.00124 $ &  8.5\\
$^{232}$Th  &  $<0.02$  & $<0.006$ &  8.5 \\
$^{40}$K   & $ <0.13 $  & $<0.013$ &    5\\
$^{60}$Co   &  $<0.03 $  & $<0.002$ &  11\\ 
\hline
\end{tabular}
\label{table-ssv}
\end{table}

\subsection{Acrylic vessels }
All of the inner components of the AD which come into contact with the liquid scintillator (LS or GdLS) 
are constructed from acrylic or teflon
to  ensure chemical compatibility with the LS. Short  and long-term compatibility tests of  all materials 
were performed  before selection for use in the AD.
Ultraviolet transmitting (UVT) acrylic was chosen both for its increased transmission of the light produced 
by the liquid scintillator at short wavelengths
and for the absence of chemical additives that protect against yellowing from UV light. 
Special care was taken to avoid prolonged exposure 
of the acrylic components to sunlight during construction and assembly~\cite{AV2}.  

Both acrylic vessels were built with  conical lids to allow bubbles from the filling process
to escape easily into the overflow tanks. 
Tanks were visually inspected during construction and assembly for any areas of cracks or crazing.
Any flaws discovered were repaired by cutting out the affected area and replacing with fresh material. 
This visual inspection  also served as the primary leak test since any imperfections were
clearly visible in the optically clear materials. 
More detailed descriptions of the acrylic vessel design, construction and assembly are found in references~\cite{AV,Bryce}. 

\subsubsection{ Inner Acrylic Vessels }
\begin{figure}
\centering
\includegraphics[clip=true, trim=40mm 30mm 30mm 20mm,width=3.1in]{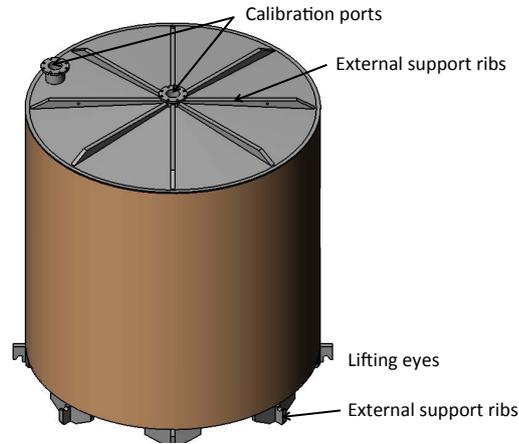}\hfil%
\caption{Inner acrylic vessel (IAV).}
\label{fig-IAV}
\end{figure}

The walls of the inner acrylic vessels (IAV) 
were constructed at Nakano International Limited, in Taipei, Taiwan ~\cite{Nakano} from 10 mm thick 
UVT sheets manufactured by PoSiang.    The bottom and top lid were made from 
15-mm-thick UVT sheets. Cut sheets were first bonded into large panels and then heat formed 
around cylindrical or conical forms to achieve the desired shape. Ribs made from 55-mm-thick material were bonded onto
the bottom and lid to stiffen the shape and to provide lifting points. 
All components were polished until optically clear. 
The lid and cylinder were bonded together as a unit. All surfaces were then cleaned inside a class 10,000 clean 
room before the final bond between the base of the cylinder and bottom were made. 
A drawing of the IAV showing the external ribs and calibration ports is shown in Fig.~\ref{fig-IAV}.

The physical size and position of the IAV within the AD was tracked to provide accurate data for simulation studies
and to ensure that the calibration ports were properly aligned.
The diameter of the IAVs were calculated from the measured circumference at several heights. The height was
 measured from the outside of the bottom to the top of the lid rim excluding ribs. Wall thickness was measured by an ultrasonic 
 thickness gauge. These data are shown in Table~\ref{table_IAV}
 and compared to the design dimensions. Although the specifications were mostly met, 
 IAV5 was nearly 20 mm taller than the other ADs due to
 a manufacturing error. The inner volume while dry is calculated from the derived internal dimensions. 
 Since the thin  acrylic walls are not perfectly rigid,  the GdLS filled volume
 is expected to differ from the dry volume.
 The overall IAV acrylic mass was determined from a spring scale during assembly.

\begin{table*}[htp]
\caption{Outer IAV dimensions  from the as-built surveys. Errors are the r.m.s. of the measured points.  
The  acrylic mass resolution was 5 kg.
The inner volume (dry) was estimated from the outer dimensions, wall thickness, and the design  angle of the conical lid.}
\begin{center}
\begin{tabular}{|c|c|c|c|c|c|c}
\hline

 AD & height (mm)  & diameter (mm) & wall (mm) & mass (kg)&volume (m$^3$)\\
 \hline
design & 3100   & 3120   & 10.0 &   &    \\
\hline
EH1-AD1  & 3101  $\pm$ 2   & 3123  $\pm$ 2 & 10.7 $\pm$ 0.8 & 907   &  23.39  \\
\hline
EH1-AD2  & 3106  $\pm$ 1   & 3123  $\pm$ 1 & 10.6 $\pm$ 1.0 & 916   &  23.43  \\
\hline
EH2-AD1  & 3102  $\pm$ 2   & 3125  $\pm$ 2 & 10.9 $\pm$ 0.9 & 915   &  23.42  \\
\hline
EH2-AD2  & 3109  $\pm$ 2   & 3120  $\pm$ 1 & 10.7 $\pm$ 0.9 & 950   &  23.40  \\
\hline
EH3-AD1  & 3101  $\pm$ 2   & 3122  $\pm$ 2 & 10.9 $\pm$ 0.8 & 945  &   23.37  \\
\hline
EH3-AD2  & 3123  $\pm$ 2   & 3117  $\pm$ 3 &    no data                      & 965   &  23.79  \\
\hline
EH3-AD3  & 3102  $\pm$ 2   & 3115  $\pm$ 3 &     no data                     & 920   &  23.57  \\
\hline
EH3-AD4  & 3102  $\pm$ 2   & 3118  $\pm$ 2 & 10.8 $\pm$ 0.6 & 945   &  23.31  \\

\hline
\end{tabular}

\end{center}
\label{table_IAV}
\end{table*}

Except for the IAV prototype, IAVs were constructed in pairs using the same techniques and procedures. 
As previously described,   the IAV which was used in AD7 was a replacement for an IAV damaged during
an annealing procedure. The replacement IAV was built one year after the construction of the other IAVs.  The IAV in AD8 was  damaged
during tests of the manual calibration system. The damaged section of IAV8 was cut out of the bottom floor and new acrylic was bonded
in place. Neither of the affected AD's performance differs significantly from other  ADs in studies to date.

\subsubsection{Outer Acrylic Vessels  }

\begin{figure}
\centering
\includegraphics[clip=true, trim=40mm 30mm 30mm 20mm,width=3.1in]{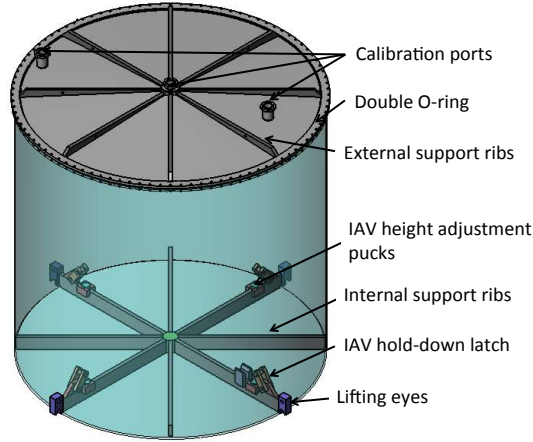}\hfil%
\caption{Outer acrylic vessel (OAV).}
\label{fig-OAV}
\end{figure}

The outer acrylic vessels (OAV) were manufactured by Reynolds Polymer Technology, Inc.~\cite{Reynolds}
in Grand Junction, Colorado, USA.  The cylindrical walls were formed from eight 18-mm-thick Polycast UVT acrylic sheets bonded together
to form  half height cylinders.  Reynolds cast  large,  thick UVT sheets for the top and bottom.
Two such sheets were bonded together to make blanks for the 3D milling machine. The conical shape was cut into 
the underside of the lid. The top of the lid was thinned down to 18 mm thickness except for the six ribs and central hub.
Holes were cut for the calibration ports. Additional material was bonded to the ribs and central hub to strengthen the OAV lid.
The cylinder flange and  a half cylinder were bonded together as were the bottom and the other half cylinder.
Bonding the two half cylinders together completed the process. A drawing of  a completed OAV is shown in Fig.~\ref{fig-OAV}.

All of the acrylic surfaces were sanded smooth and polished with 1-3 micron grit aluminum oxide polishing powder
until the surfaces were optically clear.  The vessels were surveyed in a manner similar  to the IAVs. Leak checks of the O-ring seals
on the lid and calibration ports were made before the vessels were mounted in steel shipping frames for transport.
The OAVs were shipped by truck to  the port of Long Beach, CA and placed on a container vessel for shipment to China. 
The long term testing of acrylic properties  is described in~\cite{AV2, IAV_2}. 

As with the IAVs, the physical size and position of the OAVs were measured to provide accurate data 
for simulation studies and to ensure  proper alignment of the ports. 
The outer height and diameter, wall thickness, and mass of the OAVs are shown in Table~\ref{table_OAV}.

\begin{table*}[htp]
\caption{OAV dimensions from the as-built surveys. Errors are the r.m.s. of the measured points. 
The acrylic mass resolution was 5 kg.
The inner volume (dry) was estimated from the outer dimensions, wall thickness, and the design  angle of the conical lid.
}
\begin{center}
\begin{tabular}{|c|c|c|c|c|c|c}
\hline
 AD & height (mm)  & diameter (mm) & wall (mm) & mass (kg)&volume (m$^3$)\\
 \hline
design  &  3982. & 4000.  &  18.0  &     &   \\
\hline
EH1-AD1  &  3981.5  $\pm$ 1.0  & 3997.4  $\pm$ 0.4   &  16.80  $\pm$ 1.06   & 1851    &  49.33  \\
\hline
EH1-AD2  &  3997.4  $\pm$ 0.4  & 4002.3  $\pm$ 0.7   &  16.42  $\pm$ 0.95   & 1876    & 49.46  \\
\hline
EH2-AD1  &  3986.3  $\pm$ 1.3  & 3998.9  $\pm$ 0.6   &  16.96  $\pm$ 0.70   & 1810    &   49.42\\
\hline
EH2-AD2  &  3978.8  $\pm$ 1.8  & 3998.0  $\pm$ 0.6   &  17.57  $\pm$ 0.56   & 1860    &   49.27 \\
\hline
EH3-AD1  &  3980.8  $\pm$ 0.9  & 3999.9  $\pm$ 0.3   &  17.25  $\pm$ 0.70   & 1860    &    49.37 \\
\hline
EH3-AD2  &  3983.0  $\pm$ 0.5    & 3999.8  $\pm$ 0.9   &  17.05  $\pm$ 0.97   & 1960    &  49.40   \\
\hline
EH3-AD3  &  3983.3  $\pm$ 1.6  & 3998.4  $\pm$ 0.8   &  17.61  $\pm$ 0.76   & 1985    & 49.34  \\
\hline
EH3-AD4  &  3985.0  $\pm$ 1.0  & 3996.3  $\pm$ 1.5   &  17.24  $\pm$ 0.80   & 1910    & 49.33 \\

\hline
\end{tabular}
\end{center}
\label{table_OAV}
\end{table*}

\subsection{Overflow tanks and calibration tubes }

\begin{figure}
\centering
\includegraphics[clip=true, trim=20  58mm 5mm 45mm,width=3.1in]{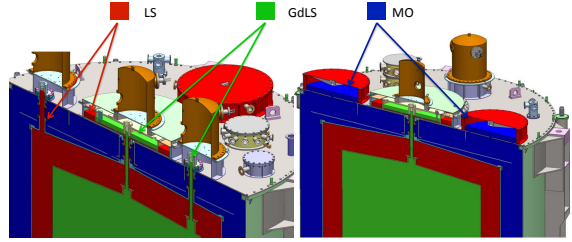}
\caption{Orthogonal slices through the top of an AD showing the GdLS (green), LS (red), and MO (blue) overflow tanks and calibration tubes.}
\label{fig-covergas}
\end{figure}

 All three AD liquids are connected to overflow tanks on the lid of the SSV to allow for thermal expansion 
 or contraction of the AD liquids while keeping the internal volumes full. Figure~\ref{fig-covergas} shows the 
 concentric LS and GdLS overflow tanks connected to the central calibration tube and the two MO
 overflow tanks. The MO overflow tanks are part of the SSV lid structure and are connected to the MO volume below
 by several 10 cm holes in the SSV lid.

Two additional calibration tubes near the edge of the IAV and near the edge of the OAV provide
paths for calibration sources at different radii to study the positional dependence  of the detector energy measurement.
The OAV tube was constructed from a 10 cm convoluted teflon tube, 16.5 cm long, connected to the OAV and SSV lid by
acrylic transition pieces. A double O-ring sliding seal at the top transition adjusts for any variation in detector height
and decouples the OAV from movement of the SSV lid when the water pool is filled. Similar double O-ring
seals on the top of the 6.4 cm diameter, 65 cm long IAV calibration tube connect the tube to the lid and overflow tanks.

\begin{figure}
\centering
\includegraphics[width=3.1in]{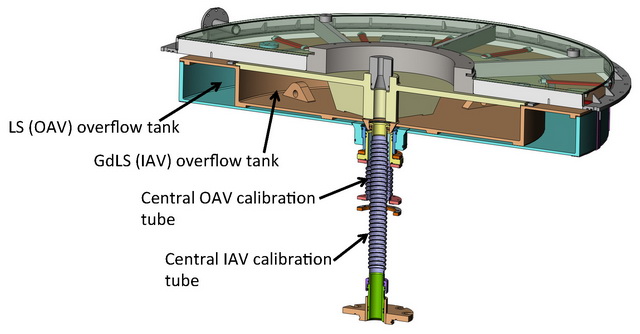}
\caption{Central overflow tanks and calibration tubes are shown. The central IAV tube connects the bottom of the IAV overflow tank (brown) 
to the top of the IAV and is open at the top to allow the insertion of sources from the central ACU. The central OAV tube 
connects the bottom of the LS overflow tank (aqua) to the top of the OAV. The overflow tanks are covered by a steel lid (grey) with
a mating flange for the central ACU. }
\label{fig-overflow}
\end{figure}

As seen in Fig.~\ref{fig-overflow}  of the central overflow region, the IAV tube goes through
the OAV tube before connecting to the IAV overflow tank. Likewise, the OAV tube connects to the OAV overflow tank. 
The tanks are constructed from 10-mm-thick acrylic at the Reynolds factory  to ensure chemical compatibility with the GdLS. 
The inner GdLS (IAV) tank has an inner diameter of 1.28 m and is 134 mm tall. The outer LS (OAV) tank has an inner diameter of  1.82 m 
and is 139 mm tall. 
The total volume of the GdLS (LS) tank is $\approx$200~l  (165~l). 
Both overflow tanks are covered with acrylic lids which have mounts for height, temperature, and tilt  sensors. 
Gas  volumes above and  below the acrylic lids  are flushed by the AD cover gas system.

\subsection{Reflective panels  }

Daya Bay ADs were designed to use optical reflective panels at the top and bottom of the inner detector cylinder to increase the number of detected photons. 
In simulation studies, end reflectors   increased  the total light collection  by 40 to 50\% and provided a more uniform response 
to light produced at different heights within the GdLS volume. 
Both specular and diffuse reflectors were considered at the design stage. 
Monte Carlo studies show that specular and diffuse reflectors have the same total photoelectron yield if the total reflectivity is the same. 
Specular reflectors were chosen since position reconstruction is simpler with specular reflectors. 
A maximum likelihood vertex reconstruction method~\cite{ML_recon} was developed to handle the difficulties of position 
reconstruction with multiple  light reflections.

\begin{figure}[!htb]
\centering
\includegraphics[clip=true, trim=5mm  10mm 5mm 5mm,width=3.1in]{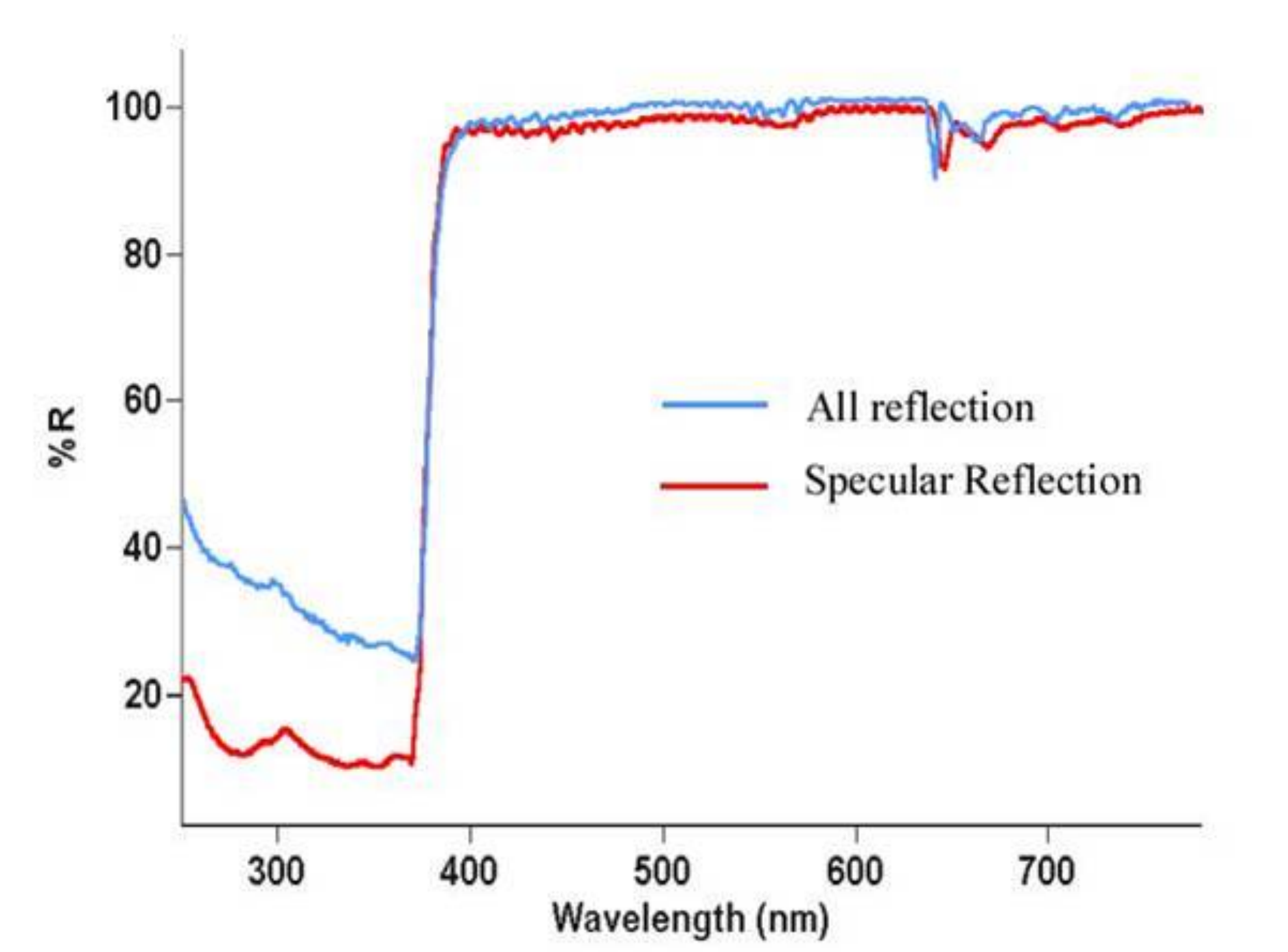}
\caption{Measured specular and total reflectivity of ESR.}
\label{fig-esr}
\end{figure}

Highly specular reflective materials, ESR (Enhanced Specular Reflector, 3M$^{\textregistered}$) and MIRO-Silver (silver film coated on aluminium base, MIRO$^{\textregistered}$), were found to be the best candidates for the AD reflectors. 
ESR film is a highly reflective, mirror-like optical enhancement film with a typical thickness of 65~$\mu$m. It is a non-metallic film made using multi-layer polymer technology and is almost free from radioactivity. The specular reflectivity of ESR was measured to be $>$98.5\% in the visible region using a Varian-5E spectrophotometer, as shown in Fig.~\ref{fig-esr}. 
MIRO-Silver film can also reach 98\% total reflectivity, with a diffuse reflectivity less than 6\%. 
It was observed that the reflectivity of ESR in air is almost unchanged at large incident angles, while the reflectivity of MIRO-Silver degraded at incident angles $>70\textordmasculine$. 
Furthermore, large-size ESR sheets were commercially available,  unlike MIRO-Silver. 
ESR was therefore selected for use in the AD reflectors.

The reflector design was first tested in a $\approx$1/3 size prototype~\cite{ad_prototype}. 
Good agreement of the energy response between data from this prototype and Monte Carlo simulations validated our understanding of the optical properties. 
In this prototype, the ESR was immersed in mineral oil for more than a year, demonstrating good long-term compatibility.

In a Daya Bay AD, the PMT surface is 20~cm from the OAV, thus the ideal reflector diameter is 4.5m, to maximize the effective PMT coverage while maintaining enough clearance for installation. To avoid scratches, the ESR film is sandwiched between two acrylic panels. The thickness of each acrylic panel was determined to be 1cm by the FEA analysis to ensure enough mechanical strength during lifting and installation. The transmittance of each acrylic panel was measured by UV-Vis spectrophotometer, and the results showed very good transparency in both visible and UV region. The final assembly of each reflector was in a 10,000 class clean room.  
Special care was taken to clean the acrylic panels, the ESR and the lifting fixtures before the assembly. 
A bulk polymerization technique using methyl methacrylate (MMA) was used to seal the sandwich structure. 
Several small-size and full-size prototypes of the AD reflector were made to resolve various technical difficulties in making the ESR film lie flat 
and attach tightly  to the acrylic panels. 
Excellent optical surfaces were achieved by continuous vacuum-pumping of  the air gap between the acrylic panels during the  polymerization  
process and the gap vacuum sealing stage.

\begin{figure}[!htb]
\centering
\includegraphics[width=3.0in]{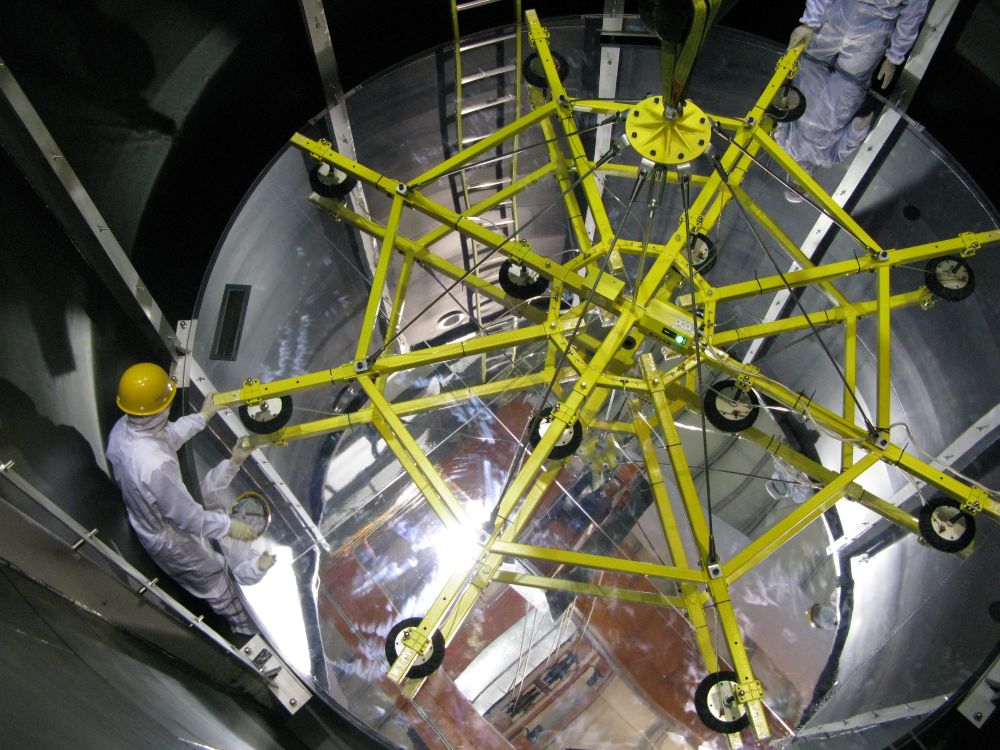}
\caption{Installation of the bottom reflector}
\label{fig-reflector}
\end{figure}

A custom lifting fixture equipped with suction cups was designed to move the reflectors as shown in Fig.~\ref{fig-reflector}.
The lower reflector rests  on top of the SSV bottom ribs and supports the OAV. 
The upper reflector rests on the top of the OAV lid ribs but is additionally cable-tied to the ribs to prevent movement
during transportation or liquid filling.

\subsection{PMT system}

Light from the liquid scintillators is measured by 192 PMTs
mounted on ladders positioned around the SSV perimeter.
Effective  photocathode coverage with the top/bottom reflectors was  about 12\%.

The performance of the PMTs in an AD plays a critical role for the
success of the experiment since all antineutrino events are reconstructed from
the signals provided by the PMT arrays. Based on physics requirements,
specifications were developed~\cite{tdr} to guide our selection of the PMTs.
We chose the Hamamatsu~\cite{Hamamatsu}  R5912 ten-stage 20-cm photomultiplier tubes 
with low-radioactivity borosilicate glass and a tapered base. The R5912 can 
withstand pressure up to 7 atmospheres. 
The photocathodes of the PMTs are biased with positive voltage, allowing 
the use of a single coaxial cable to provide the HV as well as the transmission
of the PMT signal to the front-end electronics. 

The base of each PMT is sealed inside an acrylic shell with epoxy so that it
will not contaminate the mineral oil. All PMTs were thoroughly tested before being installed in the ADs.

\subsubsection{PMTs  }

After passing visual inspection, PMTs were burned-in for three days while  operating at a gain of about  $1 \times 10^7$ to reduce infant  mortality. 
Only  0.5\% of the $\approx$ 2500 PMTs failed the burn-in tests.
A test stand capable of  evaluating 16 PMTs simultaneously was setup to ensure that each PMT met the requirements listed in Table~\ref{tab-pmt}. 
An additional 1\% of the PMTs failed these requirements.

All the measured quantities were saved in the Quality Control data base. PMTs passing all the requirements were randomly distributed among the eight 
ADs to ensure statistically identical performance. Each PMT was wrapped with
a truncated conical magnetic shield made of FINEMET$^{\textregistered}$
to reduce
variation in charge collection due to the Earth's magnetic field~\cite{PMT-Mshield}.

Based on five samples of glass used  by Hamamatsu in fabricating the Daya Bay R5912 PMTs,
the average radioactivity of the PMT 
glass was measured as $1.9 \pm 0.3$ Bq/kg for $^{238}$U, $1.4 \pm 0.4$ Bq/kg for $^{232}$Th, and 
$5.1 \pm 1.4$ Bq/kg for $^{40}$K.

\begin{table}[htp]
\caption{PMT requirements}
\begin{center}
\begin{tabular}{|l|l|}
\hline
 Quantum Efficiency & $>25\% $ @420 nm \\
\hline
Gain &  $1 \times 10^7$ \\
\hline
Single photo-electron (SPE)  & $ \geq 2.5$\\
  peak to valley ratio       &          \\
\hline 
Pulse Linearity &  $2\%$ @40 mA \\$1 \times 10^7$
 & with tapered base\\
\hline
Dark Rate & $<10$ kHz \\
\hline
After-Pulse Ratio &  $<10\%$  \\
      &         for SPE main pulse \\
\hline
Rise and Fall Time & $ t_{rise} \leq 6.5$ ns\\
      &  $t_{fall} \le$ 10 ns \\
\hline
Transit Time Spread for SPE & $< 3$ns  (FWHM) \\

\hline
\end{tabular}
\end{center}
\label{tab-pmt}
\end{table}

During commissioning and early operation of the ADs about 5\% of 
AD triggers were found to be caused by ``flasher" events. These events seemed to be caused
by electrical discharges within the PMT bases, generating visible light which was then 
detected by other PMTs. The distribution of observed PMT signals was distinctive and
analysis cuts were developed to reject this background source~\cite{nim_sidebyside}. 
Although some PMTs were more likely than others to cause  "flasher" type events,
the inefficiency introduced by the analysis cut  ($\leq$0.02\%) and the residual
contamination of IBD events  ($\leq 10^{-3}$) were low enough that it was not necessary 
to turn off any PMTs.

\subsubsection{PMT support ladders }
\begin{figure*}
\centering
\subfigure[]{
\begin{minipage}[b]{0.35\textwidth}
\includegraphics[width=1\textwidth]{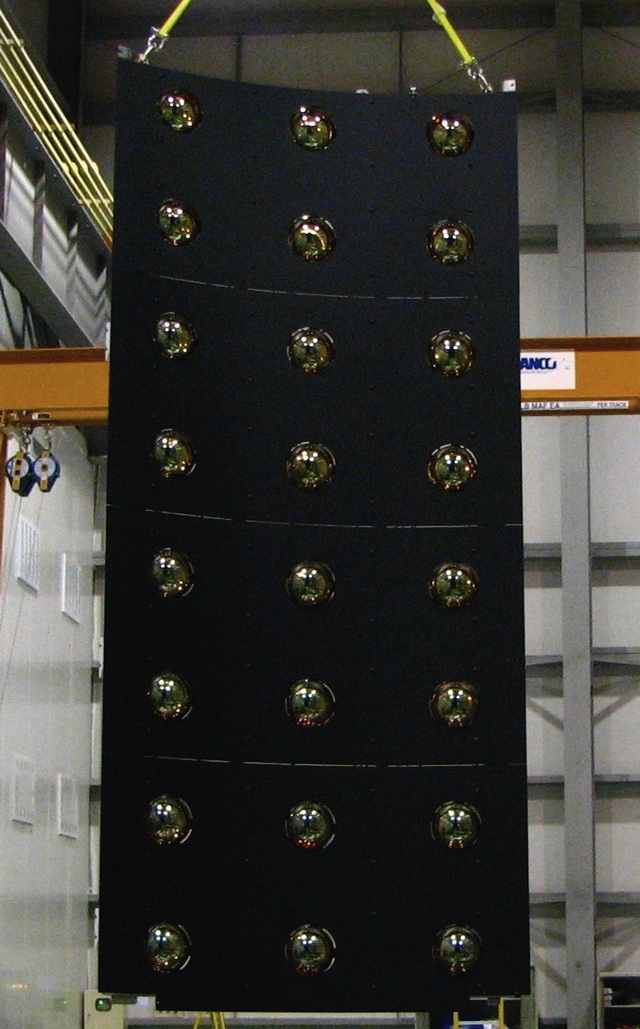} 
\end{minipage}
}
\subfigure[]{
\begin{minipage}[b]{0.35\textwidth}
\includegraphics[width=1\textwidth]{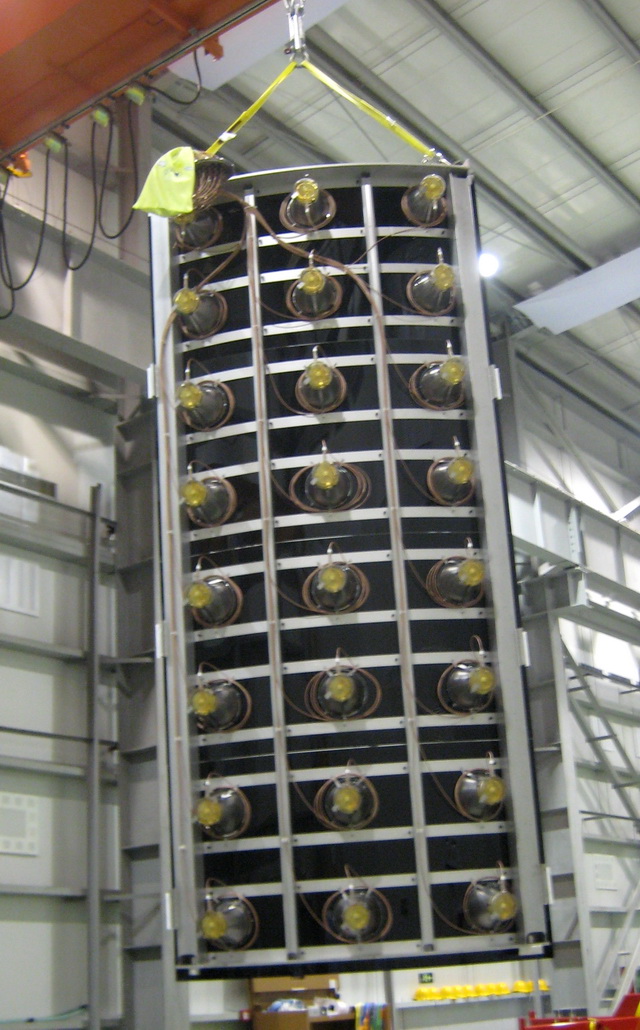} 
\end{minipage}
}

\caption{Front (a) and back  (b) of an assembled PMT ladder. Eight ladders hold
the PMTs in a cylindrical shell centered on the IAV and OAV.}
\label{fig-PMTladder}
\end{figure*}

The PMTs are mounted on 304-stainless steel tripod brackets which are bolted to a
304-stainless steel curved-frame ladder.
Each ladder holds 24 PMTs  arranged in three columns and eight  rows
To reduce the complexity in event reconstruction due to reflected photons, 
a 3.2-mm-thick matte-black acrylic radial shield (reflectivity 4-5\%) is attached to the inner curved
surface of the ladder frame. The inactive sections of the PMTs are hidden behind
the radial shield.  Figure~\ref{fig-PMTladder}  shows the front and back of an assembled PMT ladder.  
 A total of eight ladders holding 192 PMTs were installed in each SSV.   
The 24 7-m-long RG305 50-$\Omega$ coaxial cables  from the 
PMTs of each ladder were routed to 
one of eight cable feedthroughs on the wall of the SSV. 
Plugs and O-rings previously installed on each cable were mated to holes in the feedthrough flange.
These cables were later connected to 35-m-long JUDD C07947 coaxial cables just outside the SSV in 
dry pipe elbows which are later flushed with nitrogen gas~\cite{Gas}. 
Vacuum grade stainless steel bellows between the elbow and the cable tray running 
 around the water pool perimeter keep each group of  cables dry.

\subsubsection{HV system  }
\begin{figure}
\centering
\includegraphics[clip=true, trim=15mm 50mm 130mm 0mm,width=3.1in]{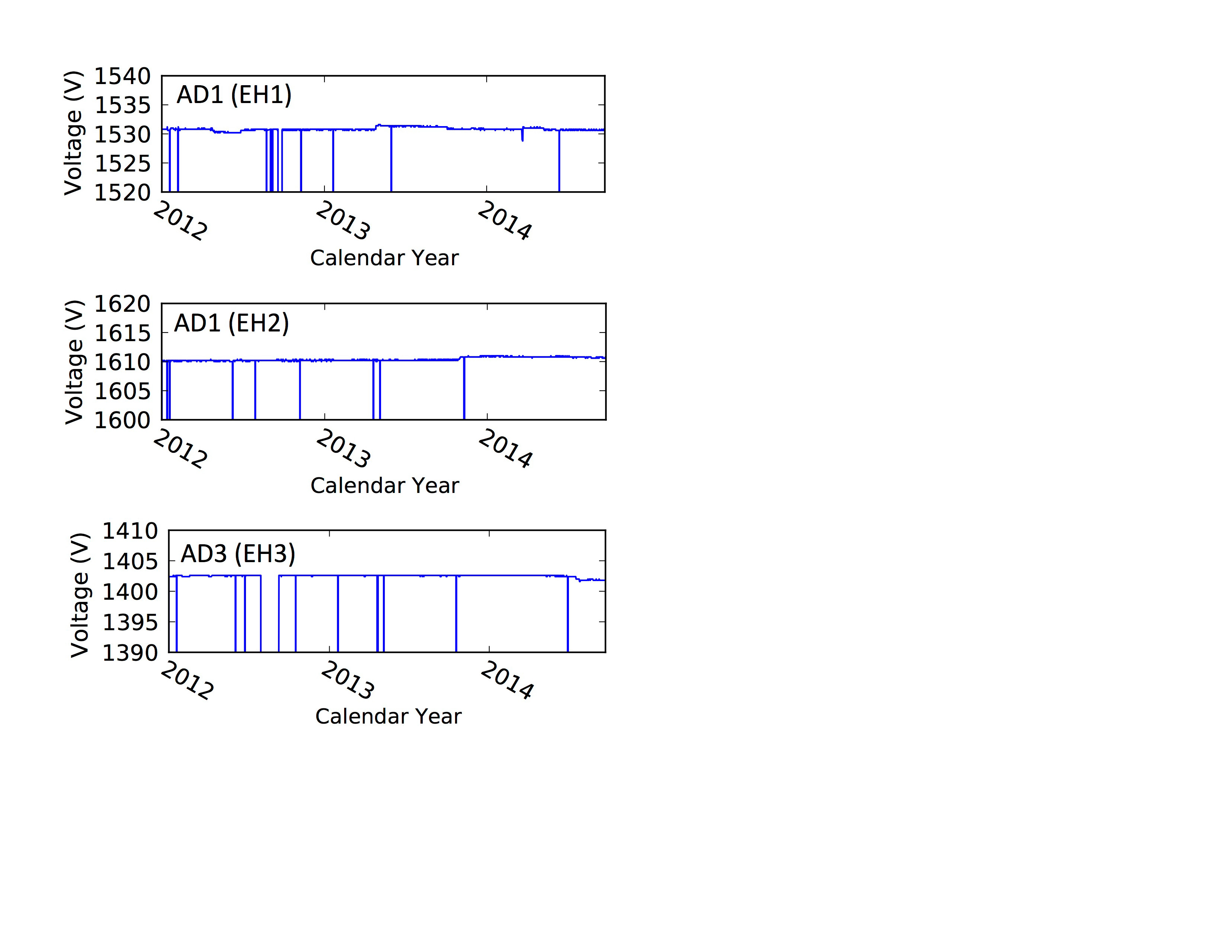}\hfil%
\caption{History of the operating PMT HV of a few  typical PMTs. 
High voltages are set to zero during extended accesses or when recovering from power outages.}
\label{fig-PMThv}
\end{figure}

The positive high voltages of the PMTs are provided by a HV system based 
on a CAEN SY1527 LC main frame housing up to eight 1934A 48-channel 
high-voltage distribution modules. Each HV channel is controlled with a 
LabView$^{TM}$ program. 
Individual voltages and  currents  are monitored and archived in the slow-control database. 
Hardware limits are set on the output voltage and current to protect the PMTs in case of any 
error or failure.
Custom-built splitter boxes holding 48 channels of decoupling circuits were 
installed in standard racks containing  the front-end and trigger electronics. 
The high voltage is filtered out, leaving only the analog signals of the PMTs 
to be sent to the front-end electronics.
PMT gains were set to be $1\times10^7$ for normal data taking. 
The operating HVs for the PMTs in the three experimental halls 
were quite stable as shown in Fig.~\ref{fig-PMThv}.

\subsection{AD liquids  }

Liquids for the ADs  were produced and stored in the  dedicated underground Liquid Scintillator hall (LS hall).   
The GdLS for all eight detectors was produced in Jan., 2011 and stored in  five 40-ton acrylic storage tanks. 
All of the LS was produced in Mar., 2011.
In total, 185 tons of gadolinium-loaded (0.1\% by mass) liquid scintillator and 200 tons of unloaded scintillator were produced.  
Details of the production  and characterization of the liquid scintillator  are given in~\cite{LS_dyb}.

\subsubsection{Gadolinium doped liquid scintillator  }

With a high thermal neutron capture cross section of about 49000 barn, gadolinium has been  added to 
liquid scintillators in previous neutrino experiments~\cite{chooz,paloverde} to 
reduce backgrounds by shortening the average capture time 
and to provide a delayed $\approx$8  MeV energy signal from  the de-excitation of the Gd nucleus after the neutron is captured.

Dissolving inorganic salts of Gd with organic liquid scintillator is chemically challenging. Previous 
studies showed that gadolinium could be either extracted or dissolved into a scintillator 
through chelating ligands, such as phosphors, diketones, or carboxylates~\cite{GdLS_Yeh, 
GdLS_Ding}.  
The Daya Bay liquid scintillator uses linear alklylbenzene (LAB), a straight alkyl chain of 10-13 carbons attached to a benzene ring~\cite{GdLS_Chen}, 
as the solvent, 3g/L of 2,5-di-phenyloxazole (PPO) as the fluor, and 15 mg/L of p-bis-(-o-methylstyryl)-benzene (bis-MSB) as the wavelength shifter. 
TMHA (3,5,5-tri-methylhexanoic acid)  was chosen as the ligand to complex with the gadolinium because of its high solubility in LAB. 
The solid Gd-complex was dissolved into LAB to form a Gd-LAB solution with 0.5\% concentration of Gd by mass. 
This Gd-LAB solution was then mixed with a pre-prepared concentrated liquid scintillator and 
then diluted with more LAB to form the final GdLS.

\begin{figure}
\centering

\includegraphics[width=3.1in]{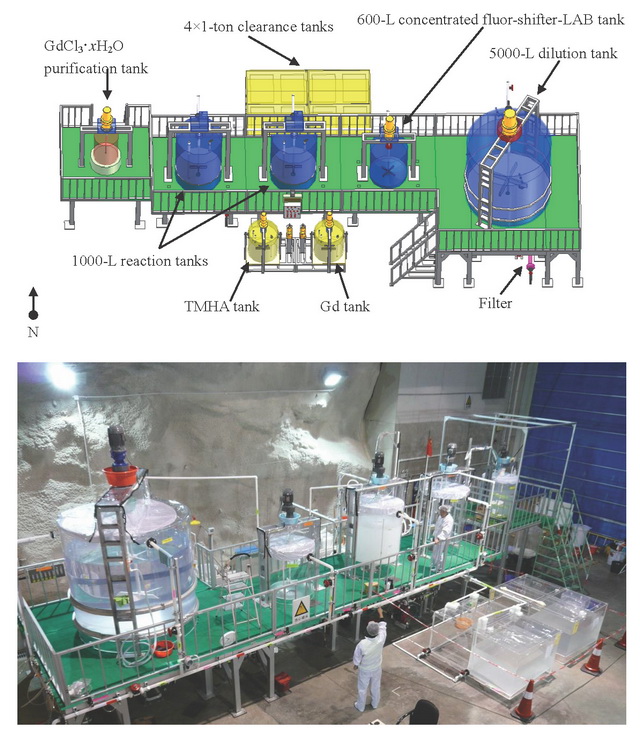}
\caption{Prototype production system for the Daya Bay liquid scintillators. The upper  diagram is of the prototype
production system in IHEP. The lower photograph shows operation of the production system in the Daya Bay LS hall.}
\label{fig-GdLSproduction}
\end{figure}

Production methods were tested and developed during a prototype stage~\cite{GdLS_Yeh, 
 GdLS_Ding}. GdLS test samples of 800 and 600 liters were separately prepared  and 
 deployed at IHEP~\cite{ad_prototype} and at the Aberdeen Tunnel Underground Laboratory~
 \cite{GdLS_Aberdeen} for prototype testing.   Performance was monitored~\cite{GdLS_Ding} 
 for over a year before  production  at Daya Bay to ensure  that  
the chemical stability, optical transparency, and light-yield of the GdLS met the required criteria.   
The prototype production system, seen in Fig.~\ref{fig-GdLSproduction}, was  reassembled in the LS 
 hall for Daya Bay  liquid production.  Production procedures can be found in reference~\cite{LS_dyb}.

The quality of liquid scintillators (GdLS and LS)  was controlled by purification of raw materials 
and by an optimized LAB production procedure at the manufacturer. 
Colored contaminants such as  iron (Fe) and cobalt (Co),   were removed since they  degrade 
optical transparency and affect the chemical stability of the scintillators. 
Other impurites of concern  were  radioactive daughters from naturally occurring uranium (U) and 
thorium (Th) decay chains. These contaminants were required to be $\leq$ 1~ppb.
 
A  self-scavenging purification~\cite{GdLS_Yeh_purity} using pH adjustment and fine-filtration of 
GdCl$_3\cdot$xH$_2$O  dissolved in water removed U, Th, and color impurities effectively. Based 
on an assessment of nearly 1 kg  purified GdCl$_3\cdot$xH$_2$O , a U/Th level of $\leq1$ ppb was 
achieved with  no trace of Fe  detected.   Long-lived radium from the U/Th decay chain could not be 
removed but was unlikely to complex with TMHA.  A total of 
1.2 tons of THMA was purified by thin-film vacuum distillation operating at 20 kg per day. The UV 
spectra of TMHA before and after purification showed a factor of two improvement in the optical 
transparency over the  wavelength range  in which the Hamamatsu R5912 PMTs are sensitive.

As the solvent, the LAB  determines the optical properties of the GdLS. 
The LAB manufacturer,  Jinling Petro-chemical Corporation in Nanjing, China, provided
388 tons of  high-quality LAB using an optimized production process. 
Heavier alklylbenzene components were removed by narrowing the allowed range of distillation temperatures and altering the timing
of catalyst use in the production cycle.
Transfer pipes, containers and transportation were also carefully planned to avoid contamination. 
PPO, provided by the Joint Institute for Nuclear Research (JINR), Russia, was further purified by an 
analytical laboratory (Huashuo Technology Co., Ltd. in Wuhan, China) via filtration after melting, 
distillation and recrystallization.  A total of 1.4 tons of PPO was purified and passed the 
Quality Assurance (QA) 
requirement.  Water for  production processes was produced by  a resin-bed water purification system  outside 
the LS hall. The  resistivity  of the purified water was about 15~M$\Omega$-cm with no detectable metallic impurities.

\subsubsection{Liquid Scintillator   }

LS production used the same production system as the GdLS production.    
All parts which had contacted  GdLS were cleaned thoroughly before LS production by
acid extraction and rinsing with deionized water. 
Rinsing  with a strong acid such as a diluted HCl solution extracts the gadolinium carboxylate 
into an aqueous phase. The equipment was then washed with deionized water 
several times to remove the gadolinium. 
After washing, the water was measured by X-ray fluorescence (XRF) to verify the absence  of residual gadolinium. 
The LS was produced by adding LAB to a concentrated PPO-bis-MSB LAB solution. 
Fifty batches 
of LS were pumped into a  200-ton-capacity storage bag made of nylon and 
polyethylene composite membrane (PA/PE). LS is compatible with nylon based on our 
compatibility studies. The bag was placed in a concrete storage pool which is coated with 
thermoplastic Olefin as a secondary containment.

\subsubsection{Mineral Oil  }

Optically  transparent, low radioactivity mineral oil fills the space  between the SSV and the outer acrylic 
vessel to attenuate radiation from the PMT glass, SSV materials and other sources outside the detector 
module.  
With a minimum of $\sim$20 cm of space between the photocathode and the OAV, the detector response is relatively uniform for 
events occurring in the LS.

\subsubsection{Liquid procurement, storage and handling}

\begin{figure}
\centering
\includegraphics[clip=true, trim=30mm 40mm 30mm 35mm,width=3.1in]{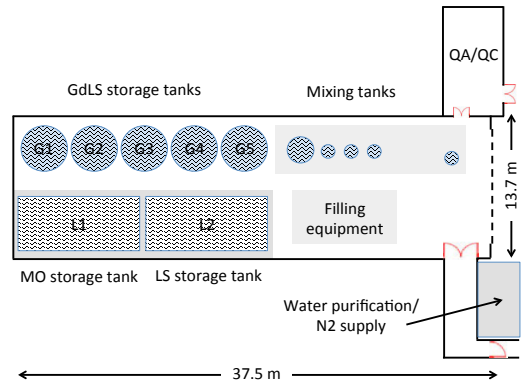}
\caption{Footprint of the LS hall. G1-G5 label the five 40-ton storage tanks for the GdLS. L1 and L2 are two 200-ton storage pools. Mixing equipment and filling equipment are installed close to the door. There is a Quality Assurance (QA)/ Quality Control (QC) laboratory and an area for nitrogen and water.}
\label{fig-hall5footprint}
\end{figure}

As shown in Fig.~\ref{fig-hall5footprint}, there are two  concrete storage tanks (L1 and L2) of 200 ton 
capacity, five 40-ton storage tanks (G1-G5)  for GdLS, and related piping and pumping systems in the LS hall. 
Each concrete pool is about 250 m$^3$ (10.5~m long, 4.75~m wide, 5~m high) and is lined with a rectangular nylon bag. 
Bags are made from a 200~$\mu$m thick co-extruded 
multi-layer of polyamide (PA) and polyethylene (PE) composite film, with the PA surface contacting the liquid.
Since the PA  cannot be thermo-melt bonded, the outside 
of the bag is  PE, which allows thermo-bonding of the seams. 
Limiting the intrusion of air  into the bags helps keep the liquids  below the required radioactivity upper limits.
The bags were tested for leak-tightness, with liquid leak rates less than 
5$\times$10$^{-2}$ cm$^3$/s, corresponding to an air leak rate of $\sim$ 1 cm$^3$/s.

The GdLS storage tanks are 4~m in diameter, 4~m high, 15~mm thick acrylic vessels, reinforced with 5 mm thick fiber reinforced plastic and aluminum belts. The GdLS tanks were sealed and purged with nitrogen gas before filling  and continuously purged with nitrogen during liquid storage.

The 388 tons of  LAB were synthesized by Jinling Petro-chemical Corporation, Nanjing, China,  in a single batch over three days. 
The synthesis parameters and procedures were adjusted (based on more than 2 years
of collaborative research with the company) to satisfy  Daya Bay requirements. 
The synthesized LAB was directly pumped into shipping tanks via temporary PVDF piping. 
The ISO shipping tank was lined with a 20 ton PE/PA bag.  A total of 21 tanks were shipped via sea and truck to Daya Bay. 
The LAB was unloaded and pumped into the two 200-ton storage pools, L1 and L2,  also using  PVDF piping (Fig.~\ref{fig-hall5footprint}).
All wetted parts of the diaphragm pump were made of fluoridated plastics. 
The LAB produced in this manner has an attenuation length of 15 meter, significantly more transparent than  normally available LAB.

The LAB stored in L2 was used to make the 185 tons of  GdLS in 50 batches. 
The synthesized GdLS was pumped into the storage tanks G1-G5 in sequence. 
After GdLS production, the L2  tank was emptied and the piping system was switched to  send LAB from the L1 tank to the mixing equipment
for  synthesizing  LS which was then stored  in L2.

The Hangzhou Sinopec Co. Ltd produced 305 tons of  mineral oil  in one production run.
The MO was custom made for Daya Bay  
to be between  99\% and 100\% of the LS density. 
The density (at $20^{o}$C) of the Daya Bay GdLS, LS, and MO liquids are 0.860 g/cm$^3$, 0.859 g/cm$^3$, and 0.851 g/cm$^3$, respectively. 
The production MO  was split into two batches. The first half 
was shipped to Daya Bay using  liquid bags similar to the LAB bags, and emptied into a new PE/PA bag in the L1 storage tank.
The remaining oil was stored at Hangzhou and later shipped to Daya Bay  after the first 4 ADs were filled.

\subsubsection{Liquid scintillator performance and stability }

\begin{figure}
\centering
\includegraphics[width=3.1in]{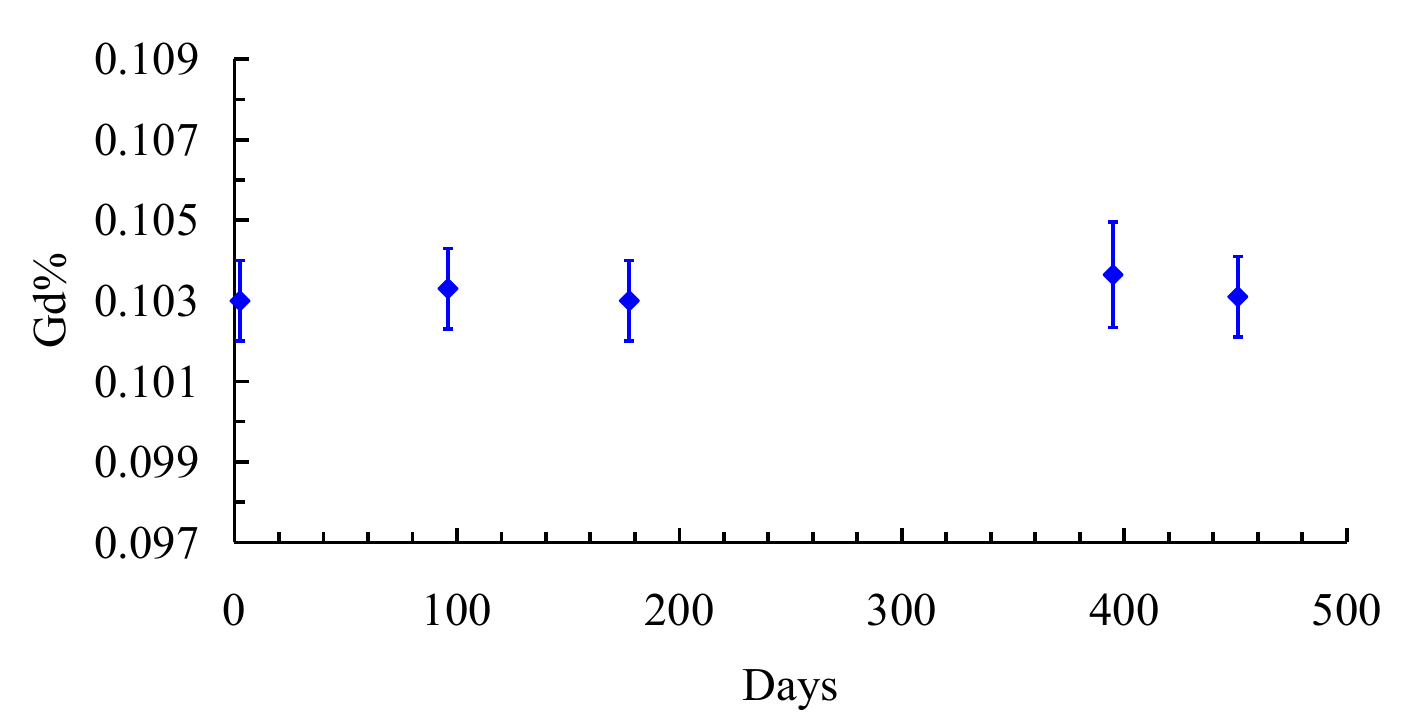}
\caption{Gd content of a GdLS sample taken during AD filling as a function of time}
\label{fig-Gdcontent}
\end{figure}

Many liquid properties, such as  the gadolinium concentration in the  GdLS, optical
transmission, carbon to hydrogen ratios,  liquid density, light yield, and radioactivity, were measured 
when the ADs were filled and monitored over time.
 Liquid samples [GdLS (two samples, before and after filling), LS,  and MO] were taken from each AD for these measurements. 
The Gd concentration was measured using X-ray fluorescence spectrometers  with an  uncertainty of about 2\%.  
Two years of  monitoring the Gd concentration in GdLS samples at the BNL and IHEP labs  show that the Gd concentration 
remains stable with time as shown in Fig.~\ref{fig-Gdcontent}.  
A second, indirect measurement of the GD concentration relies on the correlation between the Gd concentration and the neutron capture time.
Since neutron capture times in all  ADs are stable~\cite{LS_dyb}, the Gd concentration must also be stable.

\begin{figure}
\centering
\includegraphics[width=3.1in]{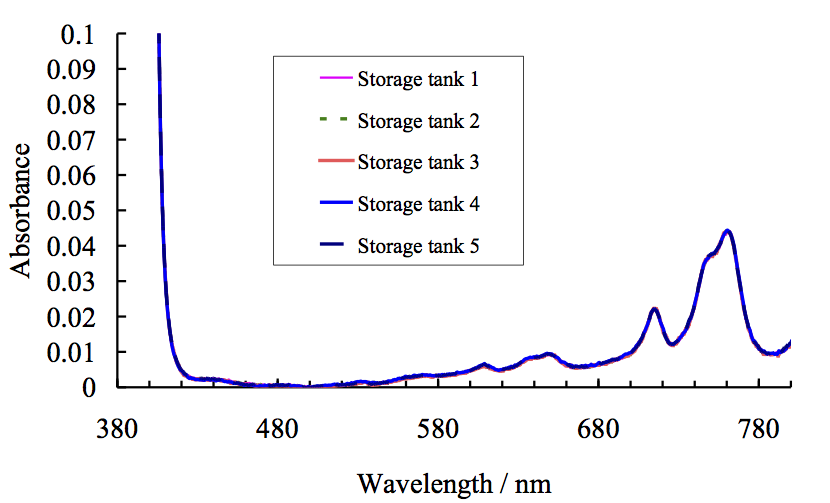}
\caption{Comparison of absorbance spectra of GdLS samples from the five 40-ton storage tanks at Daya Bay.
The spectra are nearly identical with the last curve obscuring the other curves. }
\label{fig-GdLSstorage}
\end{figure}

\begin{figure}
\centering
\includegraphics[width=3.1in]{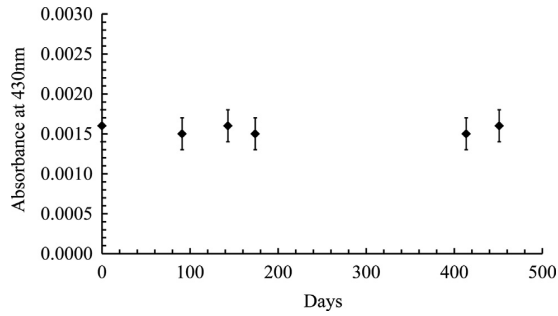}
\caption{Optical absorbance  at 430 nm of GdLS as a function of time}
\label{fig-Gdabsorption}
\end{figure}

Figure~\ref{fig-GdLSstorage}  shows the absorbance spectra measured by a UV-vis spectrometer of GdLS samples taken
 from storage tanks G1 to G5 which had been filled from different production batches of GdLS. 
The curves are nearly identical,  indicating  the uniformity of the produced GdLS optical-transmission properties.  
The optical transmission of the AD samples were also studied over time. 
A representative curve of  GdLS  absorption at 430 nm is shown in Fig.~\ref{fig-Gdabsorption} and has been stable for more than 500 days.

The emission spectrum from the GdLS excited at 260 nm were consistent with the bis-MSB emission spectrum~\cite{LS_dyb}.
LAB-based liquid scintillator usually has a high light yield. 
According to our preliminary research, the addition of the Gd-TMHA complex to the LS does not significantly degrade the light yield.
The light yield of both GdLS and LS are about half of that of anthracene, 
as measured at BNL and IHEP.

C/H/N weight and atomic ratios were obtained directly from combustion analysis. 
Gd concentration together with C/H weight ratios of the GdLS samples from 
the  eight  ADs are summarized in  Table~\ref{table-Gd}.  Table~\ref{table-C_HinLS}  lists the percentage weights of C and H in LS samples from the eight ADs.

\begin{table*}
\caption{Percentage weights of Gd, C, and H in  GdLS samples from the eight ADs.  
The relative Gd concentration error is less than 1\%.  The relative error in the computed C/H ratio is less than 0.7\%. }
\centering
\begin{tabular}{|c|c|c|c|}
\hline
 Sample   & Gd concentr.(\%)& Carbon(\%) &  Hydrogen(\%)   \\
\hline
EH1-AD1  & $0.1031\pm0.0003$  & $87.85\pm 0.94 $ & $12.01\pm0.31$\\
EH1-AD2  & $0.1029\pm0.0006$  & $87.91\pm 0.91 $ & $11.97\pm0.38$\\
EH2-AD1  & $0.1032\pm0.0009$  & $88.02\pm 0.92 $ & $11.95\pm0.14$ \\
EH2-AD2  & $0.1034\pm0.0005$  & $87.90\pm0.92  $ & $11.95\pm0.13$\\
EH3-AD1  & $0.1034\pm0.0009$  & $87.97\pm 0.92 $ & $12.00\pm0.13$ \\
EH3-AD2  & $0.1031\pm0.0009$  & $88.05\pm 0.93 $ & $11.97\pm0.13$\\
EH3-AD3  & $0.1029\pm0.0009$  & $87.86\pm 0.95 $ & $12.04\pm0.43$ \\
EH3-AD4  & $0.1020\pm0.0010$  & $87.81\pm0.95  $ & $11.98\pm0.16$ \\       
\hline
\end{tabular}
\label{table-Gd}
\end{table*}

\begin{table}
\caption{Percentage weights of C and H in  LS samples from the eight ADs. }
\centering
\begin{tabular}{|c|c|c|c|c|}
\hline
Samples  & Carbon & Hydrogen    \\
\hline
EH1-AD1  &  $88.00\pm 0.93 $   &   $11.91\pm 0.13 $  \\
EH1-AD2  &  $88.09\pm 0.92 $   &   $11.98\pm 0.12 $  \\
EH2-AD1  &  $88.16\pm 0.92 $   &   $11.89\pm 0.13 $  \\
EH2-AD2  &  $88.08\pm 0.88 $   &   $11.93\pm 0.13 $  \\
EH3-AD1  &  $88.23\pm 0.92 $   &   $11.96\pm 0.13$  \\ 
EH3-AD2  &  $87.90\pm 0.92 $   &   $11.97\pm 0.13$ \\
EH3-AD3  &  $88.14\pm 0.92 $   &   $12.02\pm 0.13$ \\
EH3-AD4  &  $88.09\pm 0.93 $   &   $11.95\pm 0.13 $ \\
\hline
\end{tabular}
\label{table-C_HinLS}
\end{table}

\subsubsection{Filling system  and target mass measurement  }

Each AD was moved underground to the filling hall and filled with precisely measured amounts of GdLS, LS, and MO~\cite{Filling}.
This irreversible operation was the single highest risk element of an AD's construction as there was no feasible method
to remove liquid from the AD which would not compromise the mass measurements or the liquid purity. The ultimate
precision of the relative measurements at the near and far detectors is directly driven by the GdLS mass measurements.

The filling system was made of three metering pump circuits, each with a Coriolis flowmeter, Teflon plumbing, filling lines and  level probes
which were inserted into the AD. The LS and MO circuits pumped directly from the storage tanks. 
The GdLS circuit utilized a
PFA Teflon-lined ISO tank to collect equal amounts of GdLS from the five acrylic GdLS storage tanks. 
The ISO tank rested on precision weigh-bridge load cells. 
Weighing the tank before and after detector filling determined the GdLS mass pumped into the detector. 
A peristaltic pump was used to top-up the GdLS overflow tank before the final weight measurement.

All components of the filling system were tested for compatibility with the AD liquids. 
Piping was constructed  from PVDF (polyvinylidene fluoride) or Teflon materials.
Particular care was taken with the GdLS pump circuit which used a non-metallic PVDF pump head.
Remaining stainless steel parts in the GdLS pumps and valving were passivated with nitric acid.

The filling process took several days to complete. First, MO was pumped into the SSV until the MO 
reached the bottom of the OAV. Then MO and LS were pumped at speeds adjusted to keep
the liquid levels within $\pm 5$~cm of each other. When the LS reached the bottom of the IAV
the GdLS pump was started. Filling-level probes and special cameras inside the AD were used to
monitor the filling levels. Transitions between the filling stages were identified by changes of slope in the
liquid height vs pumped mass plots. 
Filling was stopped periodically to add and remove four one-ton calibration masses to
the ISO tank to track any calibration drifts.  

The GdLS pump was stopped when the top of the IAV
was reached. Likewise the LS pump was stopped when the OAV lid was reached. MO pumping 
continued until the MO partially filled the overflow tanks on the SSV lid. The added pressure of the MO
on the non-rigid OAV and IAV lids and walls squeezed the inner volumes. As a result both the GdLS 
and LS liquid levels rose into their respective overflow tanks without further pumping. However,
the resulting liquid levels were not well matched. Additional liquids were added so that all
three overflow tanks were at the same level and were filled to about 1/3 capacity.
It was necessary to add several liters of  MO and LS  after the ADs were moved into the 
experimental halls since trapped gas bubbles at the top of the SSV and OAV were dislodged during transport. 
It was not necessary to add GdLS.

The dominant uncertainty in the GdLS mass measurement was due to drifts in the load cell readings.
Studies over several days with empty and full ISO tanks saw a maximum variation of $\pm~2$~kg 
(0.01\% of 20 ton).  The calibration mass measurements determined a 0.18\% correction to the absolute
mass scale, probably due to different values of gravitational acceleration ($g$) at Daya Bay and the manufacturer's location. 
An additional 0.13\% correction was made to correct for the weight of the nitrogen gas which filled the
empty ISO tank volume. The AD target mass was defined as the mass of the GdLS in the IAV vessel.
Thus the estimated mass of the GdLS in the overflow tank and calibration tubes is  subtracted from
the total GdLS mass determined from the before and after load cell measurements. 
The masses of each AD are shown in Table~\ref{tab:masses}. 
The error in the GdLS mass, $\pm~3$~kg (0.015\%), is well below the design goal of $\pm~0.2$\%.

\begin{table*}[htp]
\caption{AD liquid masses. The GdLS mass is calculated from the change in weight of the ISO storage 
container before and after filling with corrections for the amount of liquid in the overflow tank and connection tubes
and for the weight of the gas in the ISO tank. The LS and MO mass are determined from  Coriolis flow  meters. }
\begin{center}
\begin{tabular}{|c|c|c|c|c|c}
\hline
AD  &   GdLS   & LS &MO \\
 &    mass (kg) & mass (kg) &  mass (kg) \\
\hline
EH1-AD1    &  19941   $\pm$ 3    & 21623  $\pm$ 28    &  36444   $\pm$ 76  \\
\hline
EH1-AD2    & 19967    $\pm$ 3    & 21570  $\pm$ 28    &  36472   $\pm$ 76   \\
\hline
EH2-AD1    &  19891   $\pm$ 4    & 21637  $\pm$ 24    &  36240  $\pm$ 76   \\
\hline
EH2-AD2    & 19944    $\pm$ 5    & 21500  $\pm$ 24    &  36348   $\pm$ 77    \\
\hline
EH3-AD1    & 19917    $\pm$ 4    & 21616  $\pm$ 24    &  36292   $\pm$ 76   \\
\hline
EH3-AD2    & 19989    $\pm$ 3    & 21459  $\pm$ 28    &  36248   $\pm$ 76   \\
\hline
EH3-AD3    & 19892    $\pm$ 3    & 21702  $\pm$ 28    &  36414   $\pm$ 76    \\
\hline
EH3-AD4    &  19931   $\pm$ 3    & 21524  $\pm$ 24    &  36520   $\pm$ 133   \\

\hline
\end{tabular}
\end{center}

\label{tab:masses}
\end{table*}

\subsection{AD monitoring}

An AD lid monitoring system~\cite{Mass} measures the heights of the liquids in the overflow tanks, 
temperatures of the AD liquids and possible tilts of the AD lid. 
Temperature sensors mounted at various depths inside the AD monitor the MO temperature.
In addition to these monitors which provide continuous monitoring during normal data taking,  there are 
several specialized systems which take data as needed.
Cameras and lights~\cite{Mass} view the liquid levels in the off-center calibration ports during calibration periods,
providing a cross-check of the LS and GdLS heights from the level sensors.
A mineral oil clarity system mounted on the SSV lid shines light at different wavelengths
through the mineral oil to monitor mineral oil clarity.
Two other sets of cameras and 
lights~\cite{Camera} were  used to monitor the AD interior liquid levels  during filling.

\subsubsection{Liquid level sensors }

\begin{figure}
\centering
\includegraphics[clip=true, trim=0mm 10mm 70mm 0mm,width=3.1in]{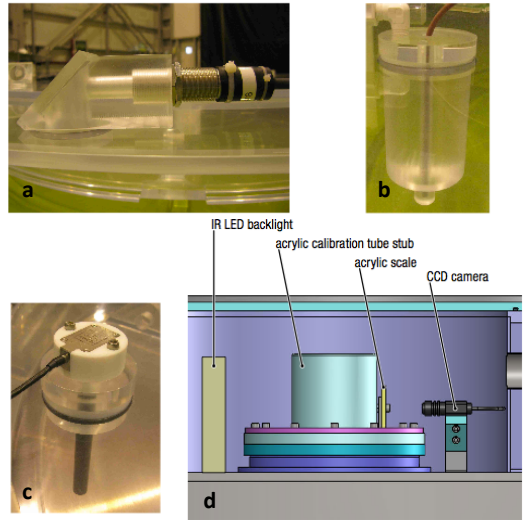}\hfil%
\caption{AD level monitoring sensors. (a) Ultrasonic, (b) temperature, (c) capacitance, (d) calibration port cameras}
\label{fig-AD_sensors}
\end{figure}

Two sets of liquid level sensors are incorporated into the central overflow tanks as shown in Fig.~\ref{fig-AD_sensors}. 
Each tank contains an
ultrasonic level sensor as well as a capacitance level sensor to provide redundant measurements of the GdLS or LS heights. 
The ultrasonic sensors (Senix Corporation model TSPC-30S1)  are mounted parallel to the liquid surface 
on the overflow tank lid and view the liquid level via a flat acrylic mirror built into the mount. 
Sensors were individually calibrated in air. 
Reflections off the bottom of the unfilled acrylic tanks were used to define the bottom of the measurable range. 
Calibration errors were less than 1 mm over the 400 mm range. 
Monitor readings during data taking are corrected  for the difference in the speed of sound 
between air and the dry nitrogen AD cover gas. 

As a  cross check of the height versus volume calculation,  one of the overflow tanks was filled with water in precisely measured increments while monitoring the 
reported height. The largest deviation from the expected volume versus height curve was 1.5 liters which is taken as the systematic error
in the overflow tank liquid volume measurement.

The calibration port cameras shown in Fig.~\ref{fig-AD_sensors}(d) are used once a month to verify the GdLS and LS liquid levels
in the off-axis calibration ports.  
The camera views the backlit liquid level in the clear acrylic port next to a fixed scale. 
Long-term drifts in the level sensors are measured by comparing the pictures and sensor data. 
No significant drift has been observed in any of the AD ultrasonic sensors.

\begin{figure}
\centering
\includegraphics[clip=true, trim=0mm 0mm 0mm 0mm,width=3.1in]{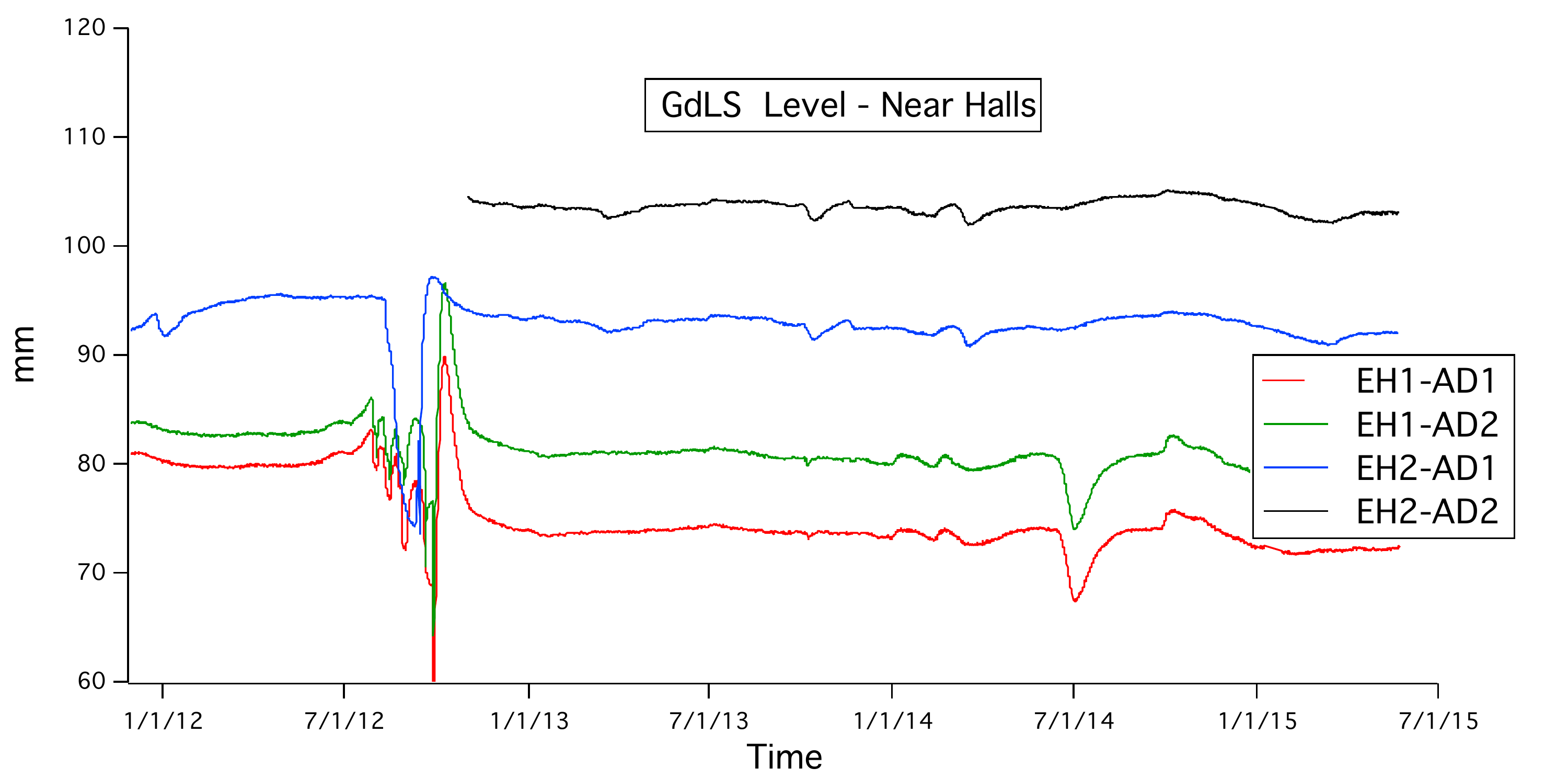}\hfil%
\caption{Time history of the GdLS levels in the near detector AD overflow tanks as measured by the ultrasonic sensor.
The rapid variations in Aug.-Sep., 2012 data correspond to multiple EH1 pool draining and filling cycles  during manual calibration tests. EH2 was drained
and filled once in the same time frame to install EH2-AD2.}
\label{fig-level}
\end{figure}

A time history of the GdLS levels in the near hall detectors is shown in Fig.~\ref{fig-level}. The  large observed fluctuations 
in the liquid level are generated when the SSV is compressed by the weight of the water during pool filling. 
Smaller fluctuations are correlated with changes in the water pool temperature.

Custom capacitance sensors were constructed by Gill~\cite{Gill} from  Polytetrafluoroethylene (PTFE) for compatibility with the LS. 
Calibration was provided by the factory for the LAB base of the LS and GdLS. 
A standard stainless steel capacitance sensor
(Gill Type R) was installed in one of the two MO overflow tanks. 
The capacitance sensors lack the precision of the ultrasonic sensors,
and have proved to be more susceptible to long term drifts. Therefore  they are used as backups to the ultrasonic sensors.

\subsubsection{Temperature monitoring }

\begin{figure}
\centering
\includegraphics[clip=true, trim=0mm 0mm 0mm 0mm,width=3.1in]{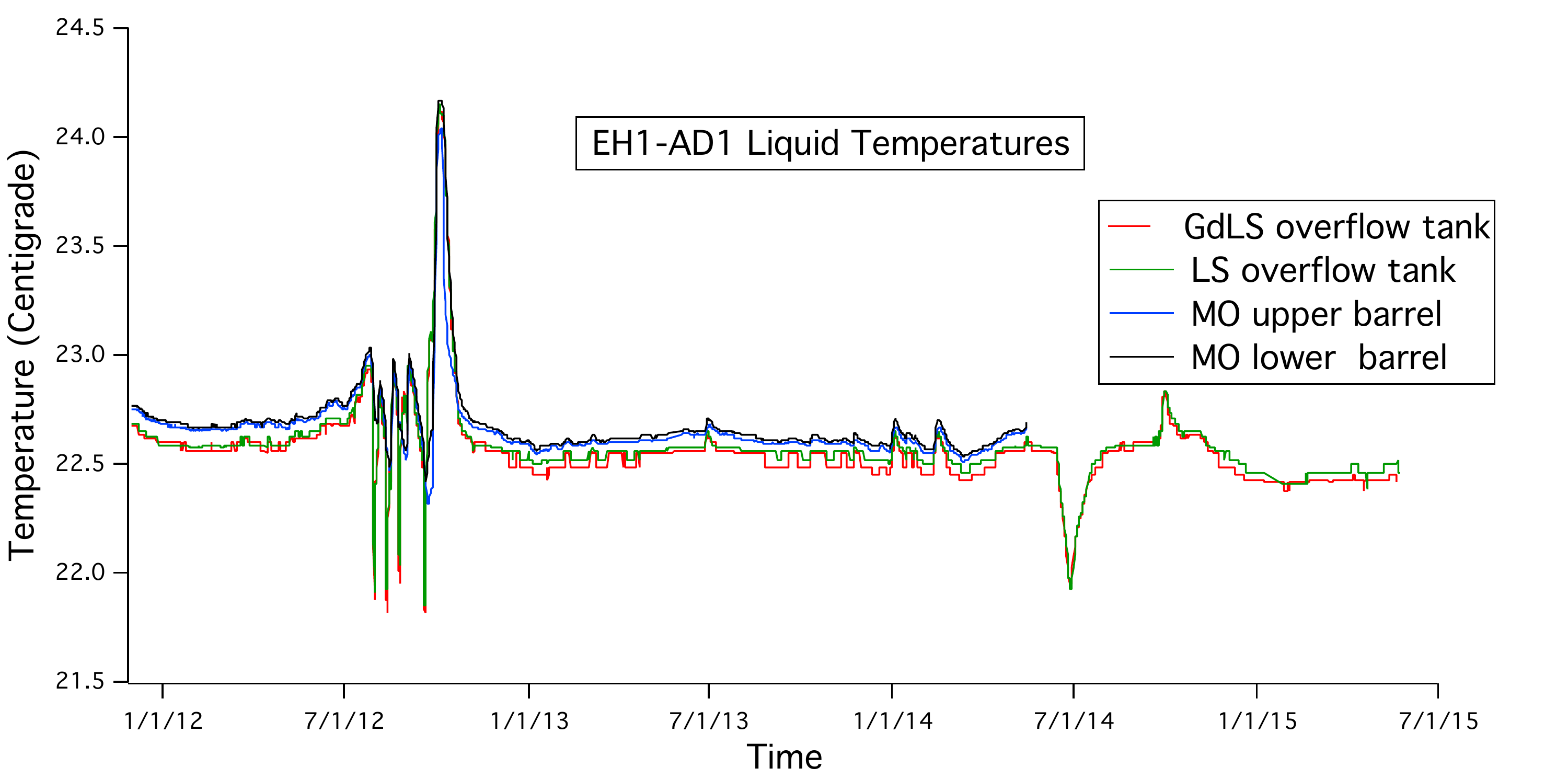}\hfil%
\caption{Time history of  liquid temperatures in EH1-AD1.  Temperatures measured in the GdLS overflow tank (red), LS overflow tank (green), 
and at two heights in the MO volume(blue or black) track each other very well.}
\label{fig-temp}
\end{figure}

AD liquid temperatures are measured by Pt100 platinum resistance thermometers in the GdLS and LS overflow tanks and at four
depths in the MO volume. The sensors have a nominal accuracy of $\pm 0.2^\circ$C at $25^\circ$C. A time history of recorded temperatures
in  AD1 can be seen in Fig.~\ref{fig-temp}. The temperatures clearly track each other and are
correlated to the small changes of liquid level observed in Fig.~\ref{fig-level}.

\subsubsection{Mineral Oil clarity sensor }

An automated MO clarity system was designed and installed on the AD lid 
to monitor the optical transmission properties of the MO.
The system utilizes a high power LED to send light pulses to the AD through a 50 m optical fiber. 
A stepping motor driven monochromator and collimator produce beams at different wavelengths. 
The light is directed vertically down into the MO parallel to the PMT faces.
A corner cube retro-reflector mounted above the bottom AD reflector
reflects the light back to the top of the AD, 
thus doubling the light path in the mineral oil to around 8 m.
Reflected light is received by a 2" PMT and digitized in a flash ADC. To monitor the light intensity,
a small mirror in the light path generates a secondary beam of light directed to the 2" PMT without entering the MO.
The ratio of reflected and reference signals at each wavelength measures the light attenuation in the MO as a function of wavelength. 

A second system within the MO clarity enclosure uses a different method to monitor the attenuation in the MO.
A diffuser ball mounted beneath the AD lid is driven by a broad band LED pulse. The ball
illuminates the AD PMTs directly below. Since the PMTs are at different MO path lengths from the diffuser ball
the ratio of PMT responses is sensitive to the MO attenuation length. 
Systematic differences in the PMT gain and acceptance limit the precision of an absolute  attenuation length measurement. 
However, changes with time of  the ratios are a clear indication of changes in the  MO attenuation length.

\begin{figure}
\centering
\includegraphics[width=3.1in]{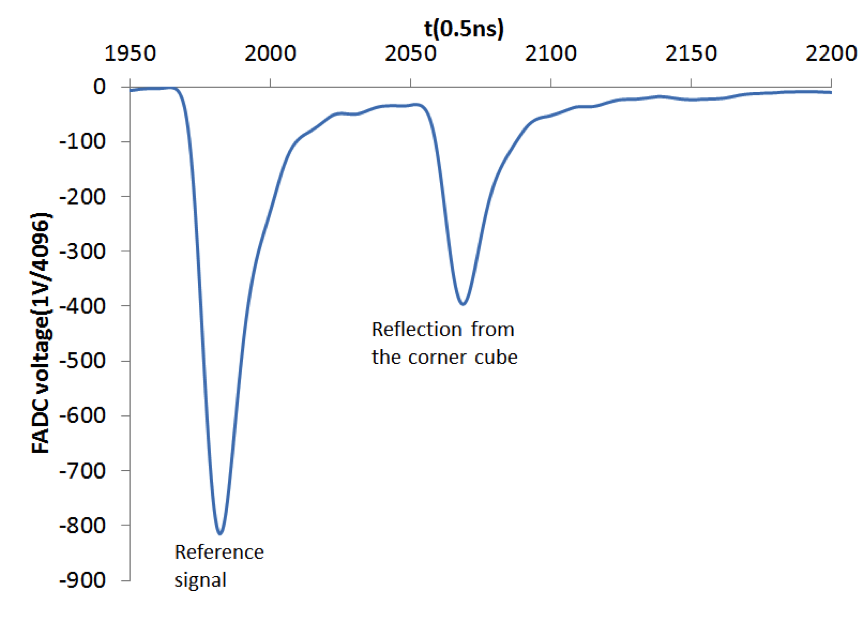}
\caption{: Averaged waveform of 2000 samples of a typical MO monitoring run in AD2, at 410 nm wavelength. The unit of the vertical axis is the voltage resolution and the unit of the horizontal axis is the time resolution of the FADC}
\label{fig-MOMonit}
\end{figure}

One diffuser ball run and nine monochromator/collimator runs (from 390 nm to 430 nm, at 5 nm intervals) are performed weekly. 
LEDs for the diffuser ball or collimator are triggered by a pulser board operating  at 20 Hz  for 2000 triggers in each run.
In collimator runs, the total sampling time window of the FADC is 1280 ns, of which 100 ns is 
occupied by two clear peaks. The reference signal and the corner cube reflection are shown in Fig.~\ref{fig-MOMonit}.
The time separation  between the two signals is 40 ns, consistent with the speed of light in the MO over the $\approx$8~m path length.  
The area of the two signals is obtained by fitting the peaks with a double Crystal Ball function~\cite{CB} or by numerical integration. 
The ratio of the first peak to the second peak is monitored for the stability of the MO attenuation.  
Tests demonstrate that this ratio is stable even if the PMT gain changes by a factor of two 
from the nominal operating value of  $1\times 10^7$.

\subsection{AD  gas system}

Liquids in the overflow tanks and calibration tubes are covered by  inert nitrogen gas supplied by an
AD  gas system.  The cover gas system constantly flushes the gas volumes above the liquid
scintillator with dry nitrogen to minimize oxidation of the scintillator over the five year lifetime
of the experiment. This constant flush also prevents the infiltration of radon or other contaminants
into these detecting liquids, keeping the internal backgrounds low. Since the Daya Bay antineutrino
detectors are immersed in the large water pools of the muon veto system, other gas volumes are
needed to protect vital detector cables or gas lines. These volumes are also purged with dry gas.
Return gas is monitored for oxygen content and humidity to provide early warning of potentially
damaging leaks. 

The cover gas circuit  flushes the gas volumes above the overflow tanks and calibration tubes
 (as shown in Fig.~\ref{fig-covergas}), the ACUs, Mineral Oil clarity Monitor, and the LED cabling port,  all of 
 which  are located on the SSV lid.  
 The cover gas volume of approximately 1500~l is refreshed at the rate of two volume exchanges per day.
 The single 1/2 inch supply line is split ten ways by a manifold on the lid. 
 Separate 3/8 inch supply and return lines go to each detector volume. 
 Another  tenfold manifold combines the return flows. This manifold arrangement and the low flow rate ensure that  no
 significant pressure differential exists between the gas volumes. As a further safety measure 
 the cover gas is kept at $\approx1$  cm of water equivalent pressure
 above ambient atmosphere by an oil bubbler in the return line.

 Since the  ADs are beneath about 2.5 m of water,  considerable design effort went into preventing a single seal failure from 
 causing a water leak into the AD liquids.  
 Almost all of the lid and ACU seals are of a double O-ring design. 
 The cover gas volumes are isolated from other gas volumes on the lid by electrical and gas feed-through flanges. 
 All of the cover gas lines 
 are protected by vacuum grade stainless steel bellows which are flushed with an additional purge gas circuit.
 Similarly, electrical lines for the ACUs, lid sensors, and LEDs are run through bellows flushed by  a third gas circuit.  
 A fourth gas circuit flushes the eight bellows and dry box elbows which protect the PMT high voltage lines from the AD to above the
 water level.

\section{AD Calibration }

\label{sec-adcalib}

Detailed characterization of the AD detector properties and frequent monitoring of the detector performance are 
required to understand  detector differences at the $0.1\%$ level and to identify $0.1\%$ changes in a particular detector 
module over time.  This is achieved via full and frequent detector calibration. 

\subsection{Automated Calibration Units }

\begin{figure}
\centering
\includegraphics[clip=true, trim=5mm 0mm 0mm 0mm,width=3.1in]{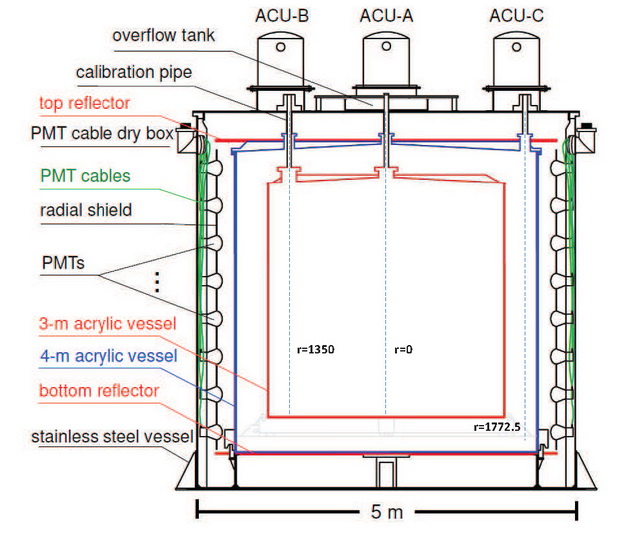}
\caption{An illustration of the Daya Bay detector showing the three ACUs.}
\label{fig-acuAD}
\end{figure}

Three fully automated calibration units (ACUs) were mounted and sealed on the top of each detector. 
Deployment positions are along  the AD central axis, near the  edge of the GdLS cylindrical volume (r = 1350 mm),
and in the LS between the inner and outer acrylic vessels  (r = 1773 mm), as shown in Fig.~\ref{fig-acuAD}.    
All eight ADs (using one ACU at a time) can be calibrated simultaneously.
ACUs deploy one of three possible calibration sources within the ACU into the detector at different z positions. 
Multiple calibration runs for each  antineutrino 
detector were taken every  week for $\approx 3$ hours during 2011-2012.

\begin{figure*}
\includegraphics[width=0.99\textwidth]{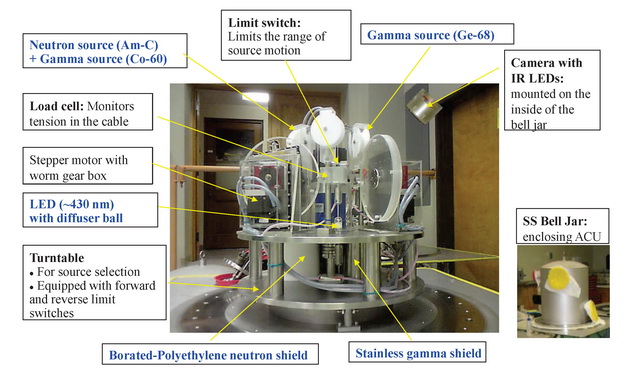}
\caption{An overview picture of the ACU}
\label{fig-acu}
\end{figure*}

Calibration sources are deployed with a precision of 7~mm along the vertical axis (z-axis)~\cite{acu}. 
Figure~\ref{fig-acu} shows the major ACU components.   
The three calibration sources are a LED, a $^{68}$Ge source   and a  $^{241}$Am-$^{13}$C and  $^{60}$Co combined source.
$^{68}$Ge decays into $^{68}$Ga via electron capture with a half life of 271 days. The  $^{68}$Ga decays via positron emission 
with a half life of 68 minutes. The positron typically annihilates within the source enclosure, yielding two 511 keV gammas. The 
$^{68}$Ge  source is thus effectively a 10 Hz gamma source.
The $^{60}$Co sources emit two gammas  at 1.173 and 1.332 MeV at about 100 Hz.   
Alphas from the 28 $\mu$Ci $^{241}$Am source interact with $^{13}$C within the source, emitting  neutrons  at a rate  of 0.7 Hz~\cite{AmCsource}.

\begin{figure}
\centering
\subfigure[LED source]{
\begin{minipage}[b]{0.45\textwidth}
\includegraphics[clip=true, trim=0mm 0mm 0mm 0mm,width=0.7\textwidth]{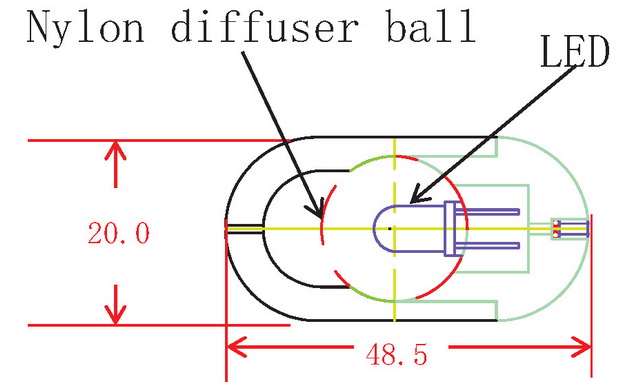} 
\end{minipage}
}
\subfigure[Radioactive sources]{
\begin{minipage}[b]{0.45\textwidth}
\includegraphics[clip=true, trim=0mm 0mm 0mm 0mm,width=1\textwidth]{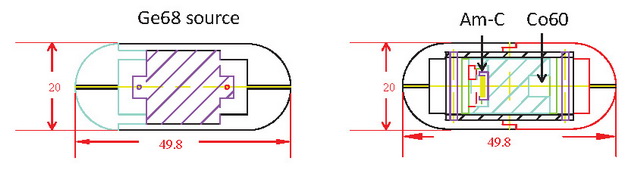} \\
\end{minipage}
}

\caption{Calibration sources}
\label{fig-acu-sources}
\end{figure}

As shown in Fig.~\ref{fig-acu-sources}, each source is enclosed in an acrylic shell and connected to the turntable by a stainless steel (radioactive sources) or coaxial cable (LED)  in Teflon sleeves. 
All materials having contact with liquids were thoroughly studied  for compatibility.  Radioactivity 
of the components meets the required specification  of $\leq0.6$~Bq/kg of 
$^{238}$U, 0.4~Bq/kg of $^{233}$Th  and 2.6~Bq/kg of $^{40}$K.  

During the summer shutdown of 2012, the calibration sources used in each ACU were modified based on experience with the  6 AD data set.  
Neutrons from the off-axis $^{241}$Am-$^{13}$C sources could produce  IBD-like  signals in the AD, even with the source in the parked position
above the SSV lid.  
Source neutrons could produce gammas in the LS or GdLS  either by scattering  or capture in the Fe-Cr-Mn-Ni SSV walls followed by the subsequent capture of the neutron in the GdLS.
The  $^{241}$Am-$^{13}$C sources in the EH3 off-center ACUs were removed  to reduce the background level. 
$^{68}$Ge and $ ^{60}$Co sources were swapped in some ACUs to improve the   neutron source calibration data for   H/Gd ratio studies. 
The final sources inside all 8 AD ACUs  are listed in Table~\ref{table-calibsources}.

\begin{table}[htp]
\caption{Calibrations sources in the ACUs after  the start of  the 8-AD data-taking in Dec. 2012.   The name of an ACU stands for the installed position (A/B/C)   in an AD.  See Fig~\ref{fig-acuAD} for details.}
\begin{center}
\begin{tabular}{|c| l|l|l|}
\hline
       AD-ACU &   1   &  2   & 3 \\
\hline
EH1-AD1-1A   & LED   &  $^{241}$Am-$^{13}$C/$^{68}$Ge   &  $^{60}$Co \\
EH1-AD1-1B   & LED &   $^{241}$Am-$^{13}$C/$^{60}$Co   & $^{68}$Ge  \\
EH1-AD1-1C   & LED & $^{241}$Am-$^{13}$C/$^{68}$Ge    &   $^{60}$Co \\
        \hline
EH1-AD2-2A   & LED   &  $^{241}$Am-$^{13}$C/$^{68}$Ge   &  $^{60}$Co \\
             EH1-AD2-2B   & LED &   $^{241}$Am-$^{13}$C/ $^{60}$Co   & $^{68}$Ge  \\
            EH1-AD2-3C   & LED & $^{241}$Am-$^{13}$C/$^{68}$Ge    &   $^{60}$Co \\
\hline
EH2-AD1-3A   & LED   &  $^{241}$Am-$^{13}$C/ $^{60}$Co   & $^{68}$Ge \\
         EH2-AD1-3B   & LED &   $^{241}$Am-$^{13}$C/ $^{60}$Co   & $^{68}$Ge  \\
         EH2-AD1-2C   & LED & $^{241}$Am-$^{13}$C/$^{68}$Ge    &   $^{60}$Co \\
\hline
   EH2-AD2-8A   & LED   &  $^{241}$Am-$^{13}$C/$^{68}$Ge   &  $^{60}$Co \\
      EH2-AD2-4B   & LED &   $^{241}$Am-$^{13}$C/$^{68}$Ge   &  $^{60}$Co \\
          EH2-AD2-4C   & LED & $^{241}$Am-$^{13}$C/$^{68}$Ge    &   $^{60}$Co \\
\hline
EH3-AD1-4A   & LED   &  $^{241}$Am-$^{13}$C/ $^{60}$Co   & $^{68}$Ge \\
      EH3-AD1-8B   & LED &    $^{60}$Co   & $^{68}$Ge  \\
       EH3-AD1-8C   & $^{40}$K &   $^{60}$Co     &  $^{68}$Ge \\
\hline
     EH3-AD2-5A   & LED   &  $^{241}$Am-$^{13}$C/ $^{60}$Co   & $^{68}$Ge \\
        EH3-AD2-5B   & LED &       $^{60}$Co   & $^{68}$Ge  \\
         EH3-AD2-5C   & LED &  $^{60}$Co    &  $^{68}$Ge \\
\hline
    EH3-AD3-6A   & LED   &  $^{241}$Am-$^{13}$C/ $^{60}$Co   & $^{68}$Ge \\
        EH3-AD3-6B   & LED &       $^{60}$Co   & $^{68}$Ge  \\
      EH3-AD3-6C   & LED &   $^{60}$Co    &  $^{68}$Ge \\
\hline
       EH3-AD4-7A   & LED   &  $^{241}$Am-$^{13}$C/$^{68}$Ge   &  $^{60}$Co \\
       EH3-AD4-7B   & LED &       $^{60}$Co   & $^{68}$Ge  \\
       EH3-AD4-7C   & LED &  $^{60}$ Co    &  $^{68}$Ge \\
\hline
\end{tabular}
\end{center}
\label{table-calibsources}
\end{table}

 A special source made from $^{137}$Cs deposited at the center of a spherical scintillator was used during dry detector commissioning in the SAB before the ADs were filled.
 A detailed description of the automated calibration system can be found in ~\cite{acu}.

\subsection{Manual Calibration System}
\label{section-mcs}
\begin{figure}
\centering
\includegraphics[width=3.1in]{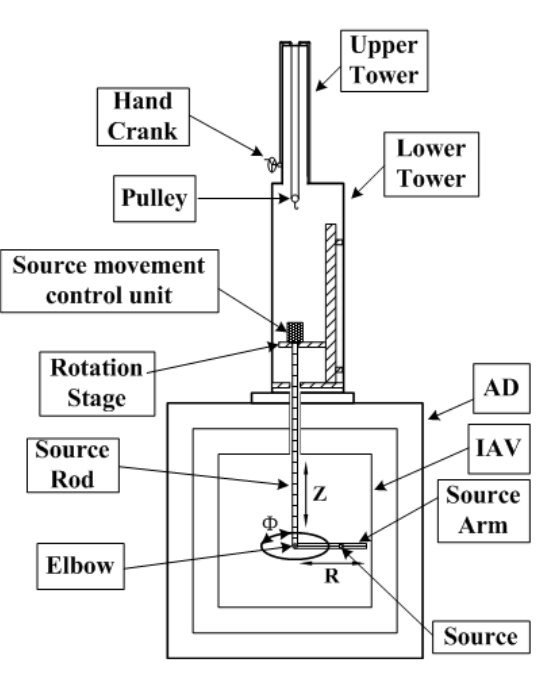}
\caption{Schematics of the MCS deployment}
\label{fig-mcs}
\end{figure}

 A manual calibration  system (MCS) was built to extend source calibrations to the full detector volume. 
The MCS consisted of
three main components: a calibration tower, a source rod assembly and  control system. 
Figure~\ref{fig-mcs} shows a schematic of the MCS installed on an AD. 
The calibration tower protected all of the inner components from the water pool since manual
calibrations were performed with normal water levels.
The rotatable source rod could be moved up or down to control the height of the calibration
source and rotated to any desired azimuthal ($\Phi$) position.
The source was moved inside the hollow source arm (an acrylic tube) 
by a teflon coated stainless steel 
wire and pulley system to position the source at the desired radius.
The source rod, articulation joint, and elbow were made of stainless steel coated 
with Teflon for compatibility with AD liquids. 
The source could be deployed to almost any position within the IAV except  very near the IAV wall. Calibration sources were placed in the full volume of the AD with a positional accuracy of 25 mm radially, 12 mm vertically  and $0.5^\circ$ in azimuth.

 A combined  $^{60}$Co  and $ ^{238}$Pu-$^{13}$C  source was used in the MCS. 
The source strength was selected to minimize data acquisition time while  avoiding 
saturation of the PMT or readout systems. 
 The  $^{60}$Co source strength was  150 Bq.  
 The $ ^{238}$Pu-$^{13}$C source produced 2000 Bq of neutrons and 40 Bq of 6.13 MeV gamma rays.  
The overall trigger rate at a trigger threshold of 0.4 MeV was about 3 kHz.  
With these sources, calibrations at  energies from 1 MeV to 8 MeV were possible, using  1.02 MeV of gammas
from positron annihilation, 2.2 MeV gamma rays from neutron capture on hydrogen,
4.43 MeV from $^{12}$C(n,n)$^{12}$C$^*$, 4.95 MeV from gamma ray emission of the $^{12}$C(n,$\gamma$) reaction, 6.13 MeV of the de-excitation energy release of $^{16}$O$^*$ from $^{13}$C($\alpha$,n)$^{16}$O$^*$ reaction and 8 MeV from neutron capture on gadolinium.

A manual calibration was performed on AD1 in EH1  during the summer shutdown of 2012. The performance is detailed in section~\ref{sec-performance}. 
A detailed description of the MCS can  be found in~\cite{mcs}. 

\subsection{2-inch PMT calibration system }

Three pairs of Hamamatsu 2-inch R7724 PMTs with standard Borosilicate glass 
were mounted in each AD to monitor the attenuation length of the liquids. 
The radioactivity of the PMTs was measured to be  $\leq15$~Bq/PMT from $^{40}$K, 
$\leq1.5$~Bq/PMT from $^{238}$U, and $\leq$ 0.25 Bq/PMT from $^{232}$Th, contributing negligibly  to the AD singles rate.  
One PMT of each pair was installed on the top lid facing downwards and the other was installed  
on the bottom of the steel vessel facing upwards. 
10-cm circular optical apertures were cut into top and bottom reflectors to allow 
the PMTs to see into the liquids. 
The glass windows of the PMTs have a nominal clearance of 28 (90) mm to the top (bottom) reflector. 
The radial position of the three PMT pairs was chosen to be 370.5, 1790.0, and 2150.0 mm,  to look into  
the GdLS target,  LS gamma catcher, and MO liquid zones. 
The attenuation length of the liquids can be monitored  by the ratio of the charge collected by 
the top and bottom PMTs with a mono-energetic calibration source deployed in the AD.

\section{Water pool and Muon Detectors }
 \label{sec-mudet}

\begin{figure}
\centering
\includegraphics[width=3.1in]{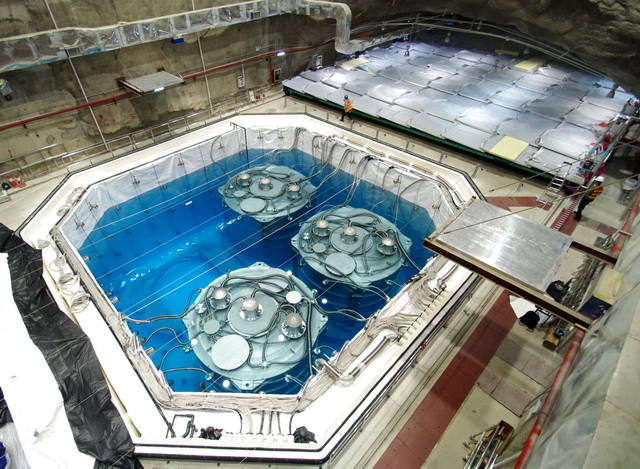}
\caption{The partially filled far hall water pool containing three ADs is shown.  
Inner Water Shield (IWS) PMTs are visible above and below the water line. 
Outer Water Shield (OWS) PMTs are hidden beneath the white Tyvek$^{TM}$ sheets
separating the IWS and OWS.  Previously installed RPC arrays are 
rolled into the RPC garage area on the top-right to provide access to the water pool.  
Two sets of telescope RPCs are mounted on the hall walls as shown.  }
\label{fig-farHallInstation}
\end{figure}

The  photograph in Fig.~\ref{fig-farHallInstation} shows a partially filled water pool in the far experimental hall 
and the major components of the water pool and muon systems. 
The highly purified water attenuates environmental background radiation while allowing the detection of 
 cosmic muons via Cerenkov radiation. 
Two groups of PMTs monitor the water pool and detect muons. 
White Tyvek$^{TM}$  sheets mounted on stainless steel Unistrut frames separate the pool into two regions.  
The central region, containing the ADs,  has inward facing PMTs  and is 
designated as the Inner Water Shield (IWS).
A 1m-thick Outer Water  Shield (OWS) forms an open shell around the  IWS (except for the top).
Rolled off to the top right side of Fig.~\ref{fig-farHallInstation} 
is an array of modules, each containing four layers of
resistive plate chamber (RPC),  which provide additional muon detection. 
Two modules of RPCs are mounted on the hall walls
to provide a subset of muons with well measured angular parameters. 
Not seen is the  black rubberized cover supported by cables which is pulled over the water pool
after filling.
Nitrogen gas flows beneath the cover to reduce the intrusion of  radon which is
commonly found in underground granite environments.
Detailed descriptions of the muon and  water purifying systems 
can be found in~\cite{muon} and ~\cite{Water}.

\subsection{Water system}

\begin{figure}
\centering
\includegraphics[clip=true, trim=5mm 8mm 5mm 0mm,width=3.1in]{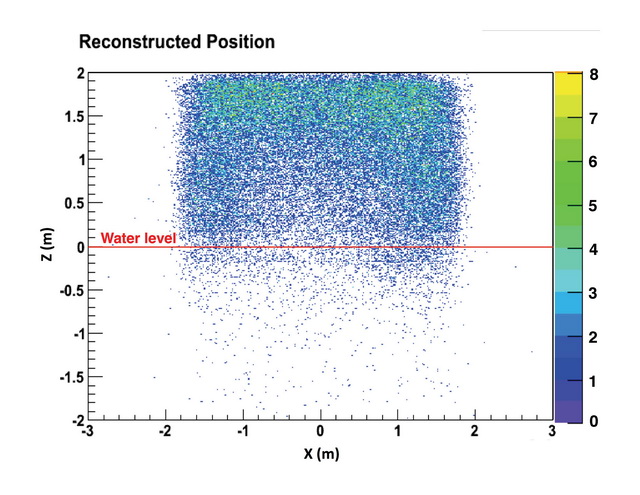}
\caption{Reconstructed positions of AD background triggers with the water pool partially 
filled showing the strong suppression of background events by the water. 
  }
\label{fig-filling}
\end{figure}

The antineutrino detectors are surrounded by more than  2.5-meter of water shielding in all directions
reducing backgrounds from  rock radioactivity by nearly a factor of $10^6$.
The effectiveness of the water shield can be clearly seen in 
Fig.~\ref{fig-filling} which shows the reconstructed position of background hits within an AD
when  the water pool  was about 1/2 full. Backgrounds below the water level are strongly suppressed.

The near (far) experimental halls contain nearly 1250 tons (2000 tons) of highly purified water 
with an optical attenuation length of  about 40~m at 420 nm (based on indirect studies~\cite{muon}).   
Producing and maintaining this water are several water processing systems.
A central treatment system in the Water Hall (WH) produces ultra-pure water  from 
the city supply for filling and topping off the muon pools. 
Polishing systems, located in utility rooms of each experimental hall, 
recirculate 5-8~m$^3$ of water per hour. 
Pool water passes through 1 and 0.1 micron filters,  a de-gasser, heat exchanger, and is exposed to  UV at 185 and 254~nm  
to  sterilize any biological contaminants.
The water is returned to the pool at $22.7^{\rm{o}}$C  with  
a resistivity of 18.2 M$\Omega$-cm  and $\leq10$~ppb dissolved oxygen. 
Radon levels measured after polishing in the polishing station are between $30-50$~ Bq/m$^3$~\cite{muon}.

\subsection{Muon PMTs}

The water pools EH1 and EH2 are instrumented with $8^"$  PMTs, 
with 121 PMTs in the IWS and 167 PMTs in the OWS. 
The larger, EH3 contains 160 PMTs in the IWS and 224 PMTs in the OWS. 
The  PMTs are  619 Hamamatsu R5912 assemblies with water-proof bases  and  341
EMI 9350KA and D642KB assemblies~\cite{Chow} recycled from the MACRO experiment.
Positive high voltage and PMT signals are transmitted via a 52-m-long 
coaxial cable. 
FINEMET shields,  identical to those used with AD PMTs, protect muon PMTs against ambient magnetic fields.
The EMI PMTs were used primarily in the upper region of the water pool
because they where shown to fail at pressures corresponding to depths of about 5~m.
All muon PMTs were tested to the same standards as the AD PMTs. 
Characterization of each PMT determined the HV setting for the  nominal data-taking gain 
of $1.0\times 10^7$, which yields about 20 ADC counts per photoelectron. 

\begin{figure}
\centering
\includegraphics[clip=true, trim=10mm 10mm 5mm 0mm,width=3.1in]{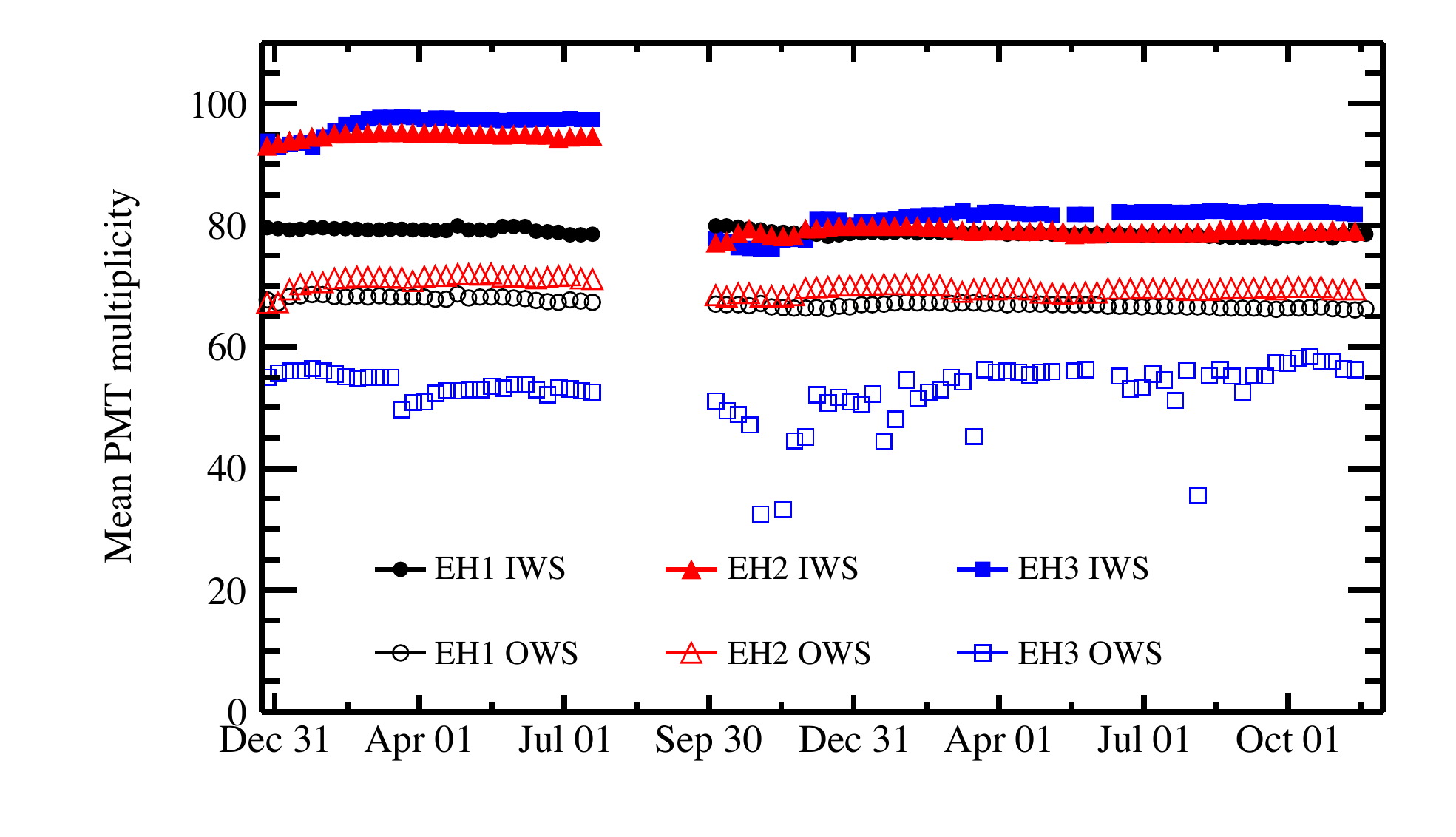}
\caption{Mean PMT multiplicity per muon vs time. The drop of mean PMT multiplicity in EH2-IWS and EH3-IWS duirng  the  8 AD period is  due to  the reduced mean muon track lengths since one more AD was put into the water pool. The instablility of EH3-OWS is induced by  electronics noise from unknown sources, which create triggers that have much lower PMT multiplicity.  This instablity has negligible impact on muon detection efficiency, as  shown in Fig.~\ref{fig-WPtrigger}.}
\label{fig-wp-nPMTvsTime}
\end{figure}

During normal data taking the PMT gain is constantly calibrated with dark noise based on data collected 
via periodic triggers. 
The average PMT gains are relatively stable  with a slight upward trend over time.   
As an indicator of the muon tagging efficiency, the mean PMT multiplicity per muon is 
monitored over  time as shown in Fig.~\ref{fig-wp-nPMTvsTime}.

\subsection{Resistive Plate Chamber Detectors }

Muon detection modules utilizing resistive plate chambers (RPCs) are
mounted on a rolling frame which is positioned over the water 
pool~\cite{muon} for additional muon detection. 
The modules~\cite{RPCmodule}, 2.17~m $\times~2.20$~m $\times~8$~cm,  contain 
four layers of RPCs 
constructed from non-oiled Bakelite sheets~\cite{RPC}, each instrumented with 
eight 26~cm~$\times$~2.10~m long readout strips oriented 
in either the x or y direction to provide a position resolution of about 10 cm. 

Modules are purged  with dry nitrogen to protect the enclosed RPCs against the moist underground environment.
The RPC gases are mixed, distributed and monitored locally by a gas system 
located in a dedicated utility room near  each hall.  
Argon, freon(R134a), isobutane and SF$_6$ are mixed in the ratio 65.5:30:4:0.5~\cite{RPC_Ma}. The 
gas mixture is periodically verified with a gas chromatograph to ensure stable performance. 
RPC layers of the same module are  connected to  different HV channels to ensure 
that only one layer is lost during the failure of a HV channel. 

\section{Electronics and data acquisition}
\label{sec-electronics-daq}

The primary data for the Daya Bay experiment originates in seventeen independent readout crates (eight ADs, three inner and 
three outer water shields, and three RPC detectors) in the three experimental halls.  
Collecting, compiling, and coordinating these data  requires a configurable, synchronized, and flexible data acquisition (DAQ) system, 
one that can record data from a single readout crate, or combine data from any number of readout crates.  
Data from readout crates within a single detector hall are time ordered and merged to create a data stream.  
The streams from each hall are saved independently for offline analysis.  
Each detector system (AD, IWS, OWS, RPC) can self-trigger  as well as receive triggers 
from other detector units via  a master trigger board (one per hall)  that coordinates hall wide trigger or calibration requests. 

\begin{figure}
\centering
\includegraphics[width=3.1in]{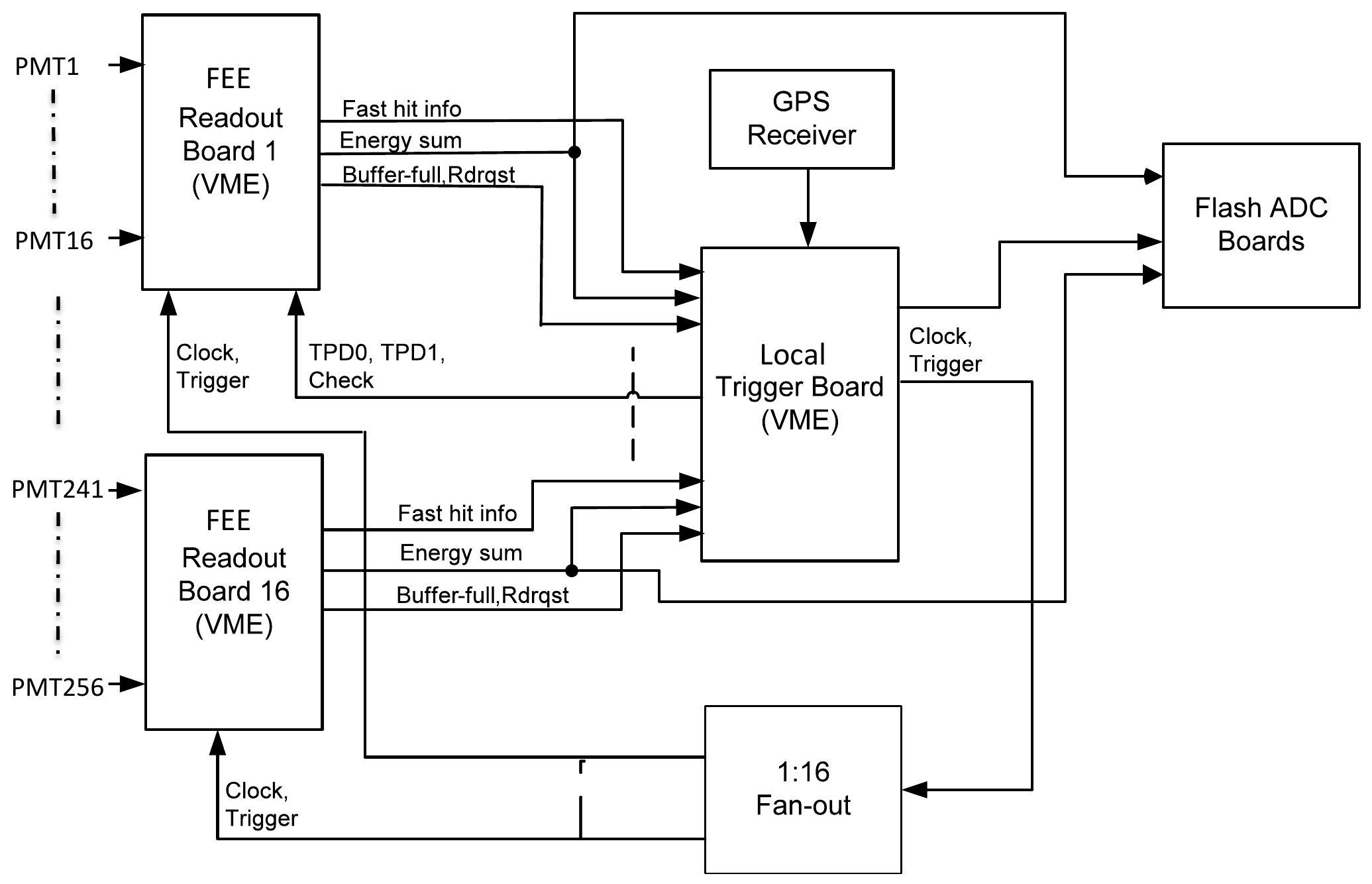}
\caption{ The PMT readout system block diagram is shown.
PMT signals are sent to the FEE channels via coaxial cables after passing through an AC decoupling unit.   
}
\label{fig-PMTreadout}
\end{figure}

The primary function of the PMT electronics is to amplify, discriminate, shape, digitize and record the PMT waveform, 
and to provide precise timing information between PMTs.
To simplify  design and maintenance, all PMT-based detector systems use identical electronics.  
The PMT readout system consists of one MVME Power PC controller, one Local Trigger Board (LTB), 
one trigger and clock fan-out board (FAN), one flash ADC board and up to fourteen Front End Electronics (FEE) 
boards. The functionality of the readout system is shown in Fig.~\ref{fig-PMTreadout}.  
RPC detectors have their own electronics system to record muon hit, time, and location information.  

\begin{figure}
\centering
\includegraphics[width=3.1in]{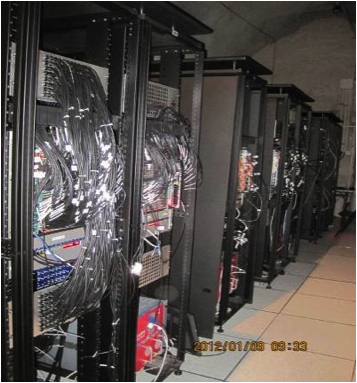}
\caption{Photograph of the readout electronics in EH2.}
\label{fig-electronicsLA}
\end{figure}

All readout electronics systems follow the VME 64xp standard.  
There is one VME crate per AD, two VME crates per experimental hall to read out the IWS and OWS, and one VME crate per experimental
hall for the RPC system.
Figure~\ref{fig-electronicsLA} is a photograph taken within the electronics room of EH2. 

\subsection{PMT-based FEE}
Raw signals from PMTs are transmitted via a coaxial cable to
an AC decoupler unit that filters out the HV and 
sends the  fast PMT signals to the front end electronic (FEE) cards. 
Essential  FEE functions are:
\begin{itemize}

\item{To calculate the  number of over-threshold PMT signals within a given time window (Nhit) and send the information to the trigger system.}
\item{To create the  linear sum of  signals (Esum) within the card and send it to the trigger system.}

\item{To shape and digitize  the height of each PMT signal, values that correspond to the energy deposition in the detector.}

\item{To digitize and record the precise timing data for each over-threshold PMT signal for position reconstruction and background rejection.}
\end{itemize}

Detailed requirements are documented in~\cite{tdr}. 
Design and testing information have also been documented elsewhere~\cite{fee}. 
In summary, each channel in the FEE board accepts signals from up to sixteen PMTs. 
Signals are amplified for use in three distinct circuits: a discrimination circuit, a charge summing circuit, and a pulse shaping circuit.  
Discriminator outputs are used by the local trigger board to form a multiplicity trigger and to start onboard time-to-digital converters (TDC).  
Outputs of the FEE charge summing circuits are  used by the trigger board as inputs to an energy summing circuit.  
The pulse-shaping circuit shapes and then samples the signal using a flash analog-to-digital converter (ADC).  
A trigger signal serves as the common stop for all the TDCs throughout the VME crate, and initiates read out of the ADC and TDC data.

\subsubsection{Charge Measurement}

\begin{figure}
\centering
\includegraphics[width=3.1in]{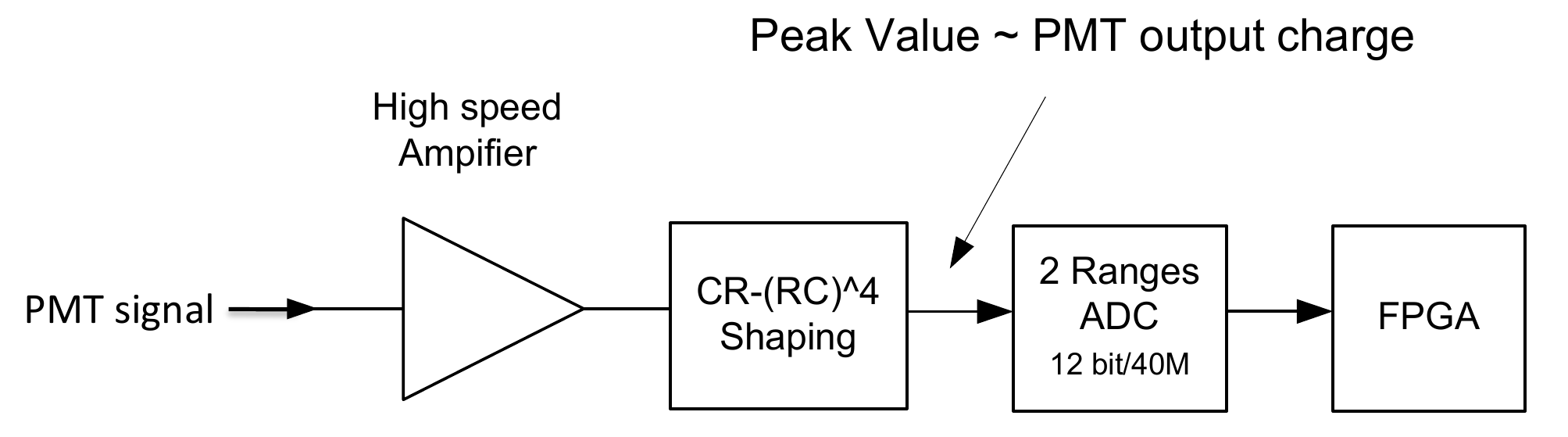}
\caption{ Charge Measurement Scheme of the FEE}
\label{fig-FEEQ}
\end{figure}

Typical output signals from  the front-end linear amplifiers in each FEE channel  have widths between 10 and 20 ns and 
heights proportional to the number of photo electrons (P.E.) produced at the photocathode of the PMT.  
Amplified PMT signals are shaped via 4-resistor-capacitor network after a capacitor-resistor (CR - (RC)$^4$) circuit before being  digitized by  two
12-bit  ADCs with a sampling rate of 40 MHz.  
One ADC is set to have a fine-grain resolution for measuring small amplitude signals. 
The second ADC range is set to sample large signals.  
The ADC samples are fed into an FPGA which selects and stores the largest value (peak value) and the baseline value, as shown in Fig.~\ref{fig-FEEQ}.  
Further details are available in~\cite{fee}.

\subsubsection{Time Measurement}

The arrival time of the PMT signals is measured relative to the trigger signal. 
Signals within 1300 ns prior to the trigger are recorded. This time window
comfortably contains  the range of times expected  between the first and the last PMTs to see 
either direct or reflected light from an IBD event, and also allows for  trigger latency. 
Precision of the timing measurements are limited by jitter in the rising edge of the amplified PMT signal (2-3 ns) 
and  by time walk of the signal ($\sim$ 1 ns). 
The digitization process begins when the amplified PMT signal is fed into a high-speed discriminator circuit.  
The  discriminator output  starts a TDC. 
To adjust for possible differences in
PMT gains the threshold for each channel can be set via a VME controller.
TDC functionality is programmed into the FEE's FPGA.  
Major components of the TDC are two high speed gray-code counters,
the first changes at the rising edge of a 320 MHz clock, while the other changes at the falling edge. 
A discriminator output starts the TDC, while the trigger signal is used to stop the TDC.  
The time difference between the start signal and stop signal is measured via these two counters with a time resolution of less than 0.8 ns.  
Further details can be found in~\cite{fee}.

\subsection{Flash ADC Board }

An independent eight channel 1~GHz Flash ADC board is used to crosscheck the triggers based on Esum signals.   
Each channel employs an 8-bit ADC, although the effective number of bits is seven. 
Each channel samples the output of an FEE's charge summing circuit, which is an integration of all sixteen amplified PMT signals on an FEE board.  
The output is recorded so that offline cross-checks can be carried out.  Additional information on the Flash ADC board can be found in~\cite{fadc}.

\subsection{Trigger system} 
\label{sec-trigger}
The trigger system initiates electronic recording of detector information.  As shown in
 Fig~\ref{fig-trigger}, Triggers are formed primarily within an individual readout crate, but may also be initiated by a different crate (called a cross-trigger), or from other sources, such as the calibration system (called a calibration trigger).  The primary physics requirement is that the system achieve a greater than 99\% efficiency for detecting events with an energy deposition of 0.7 MeV or greater in an AD.  Detailed information can be found elsewhere~\cite{ltb}.  In brief, each PMT readout crate contains a local trigger board (LTB) that receives information as described in section 7.1.  The LTB can generate internal triggers or respond to external trigger requests.  Various trigger types can be enabled or disabled via VME interface. The enabled triggers are ORed to generate a final trigger signal (local trigger) that is distributed to the FEEs by front-panel cables.  This design is flexible, reproducible, and redundant.  The RPC sub-detector employs a self-triggering scheme.  Each hall also possesses one master trigger board (MTB) that coordinates external requests and cross-triggers.  For example, calibration trigger requests originating in the calibration system are initially sent to the MTB, which forwards the trigger to the appropriate LTB.  The trigger decisions implemented are:

\begin{figure}
\centering
\includegraphics[clip=true, trim=0mm 0mm 0mm 0mm,width=3.1in]{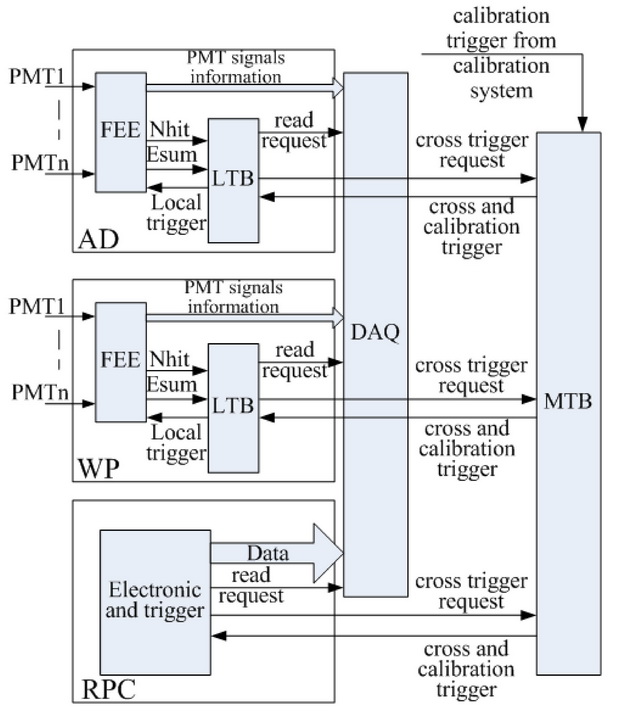}
\caption{Layout of the trigger system. In each PMT-based detector system (AD, IWS, or OWS),  the number of PMTs over the 
threshold (Nhit) and the analog sum of PMT signals (Esum) are sent to the Local Trigger Board (LTB), which will generate a local trigger 
once the prescribed trigger conditions are met.  
DAQ software reads PMT signal information from the FEEs and trigger information from the LTB upon receiving a trigger.  
The LTB in one subsystem  can also initiate a trigger in a different subsystem by sending a trigger request to the Master Trigger Board (MTB), 
which can send trigger requests (cross triggers and calibration triggers) to all of the subsystems in the same experimental hall.  
The RPC trigger board differs in detail, but can also send and receive triggers to the MTB.  }
\label{fig-trigger}
\end{figure}

\begin{itemize}
\item{Physics trigger:    Nhit $\ge45$ or Esum $\ge~65$ P.E. in  an AD, which corresponds to about 0.4 MeV in deposited energy.  }
\item{Periodic trigger:  A 10 Hz period trigger is turned on during normal physics data-taking to monitor detector stability and check random backgrounds.}
\item{Calibration trigger: Generated by the LED system in ACUs to routinely monitor PMT gains and timing.}
\item{Cross trigger: Used to  correlate activity  in different detector modules (such as water pool versus AD). 
An important cross trigger is the "look-back" trigger which is issued when  two physics triggers within a single AD
coincide within a time interval of 200 $\mu s$. The LTB for the AD generates a cross trigger (if enabled) which
is sent to water pool detectors to  initiate readout of all hit information.} 
\item{RPC trigger: Any three of the four RPC layers see a signal above the threshold.}
\end{itemize}

\subsection{Global timing system}

The timing system provides a global time reference for synchronizing triggers from individual detectors. 
It consists of three major parts: a GPS driver module, a central clock generator (CCG) and  local clock fanout boards (CFB). 
A commercial GPS driver is located in the control room, providing an IRIG-B coding signal with absolute 
UTC time information and a pulse per second (PPS) signal to the CCG.  To achieve  frequency accuracy and  long-term stability,  the 10 MHz signal generated by 
the Rubidium oscillator in the CCG  is synchronized by the PPS signal from the GPS  module.
The 10 MHz frequency is  quadrupled to 40 MHz via PLL (Phase-locked loops). 
The 40 MHz clock and the IRIG-B information are distributed separately to each detector hall via optical fibers. 
The CFB in each hall converts the optical signal and fanout timing information to all FEE boards in the hall. 
The CFB returns the IRIG-B signal immediately to the control room via another fiber link to measure the link delay. 
The GPS driver is equipped with an embedded system and acts as a Tier-1 network time protocol server for all experimental computers. 
A backup 10 MHz oscillator was built  in the CCG in case the Rubidium clock is not available.

\subsection{Data acquisition} 
\label{sec-daq}

\begin{figure}
\centering
\includegraphics[width=3.1in]{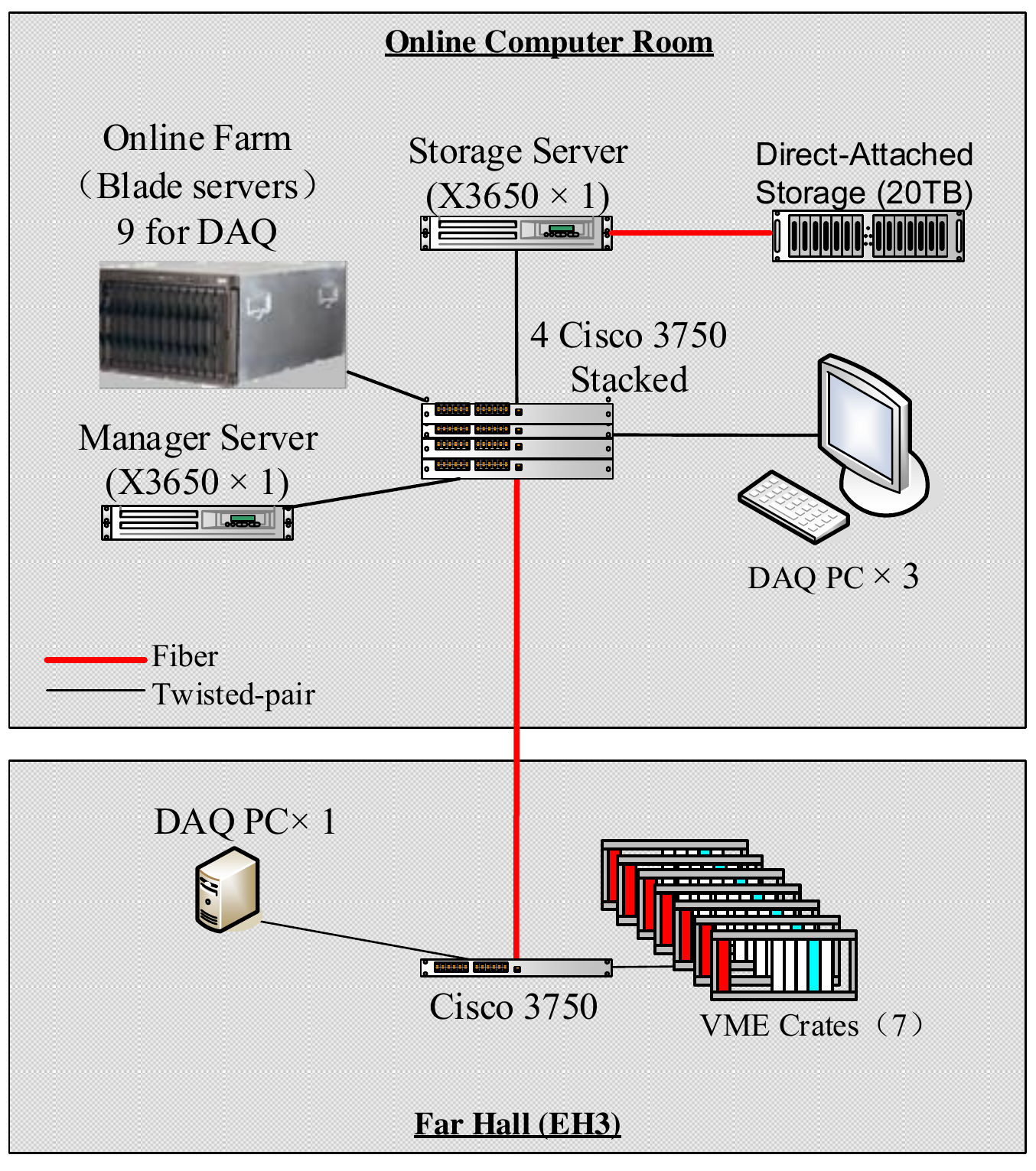}
\caption{Hardware deployment diagram of the DAQ system (Fiber connections from online computer room to EH1 and EH2 are removed from the figure for simplicity).}
\label{fig-daq}
\end{figure}

The DAQ architecture is a multi-level system using embedded Linux, advanced commercial computers and distributed network technology. Figure~\ref{fig-daq} shows the hardware deployed in the DAQ system.  Additional details can be found in reference~\cite{daq}.

The front-end of the DAQ system is a real-time embedded system based on the VME bus. 
Each VME crate holds a Motorola MVME 5500 system controller.  
The MVME 5500 is an embedded single-board computer based on a PowerPC MPC 7455 CPU  with a 1GHz clock and Universe II 
chip for interfacing to the VME bus.  The computer uses a TimeSys operating system (real-time Linux, kernel version 2.6.9). 
This front-end system can manage a 2.5 kHz event rate (2 kilobyte event size) and meets all experimental requirements~\cite{daq_ros}.    
The DAQ back-end is a blade server based computing farm~\cite{besIII_farm} utilizing a gigabit ethernet network.
Also included are  two x3650 file servers, and nine blade servers for data gathering and data quality monitoring. 
Daya Bay online software was customized and migrated to the front-end Linux and PPC platforms from the 
ATLAS online software~\cite{atlas_online}  with three high level components (packages) - Control, Databases, and Information Sharing.  
These packages work together to provide the functionality for the various configurations of the DAQ and detector instrumentation.

\begin{figure}
\centering
\includegraphics[width=3.1in]{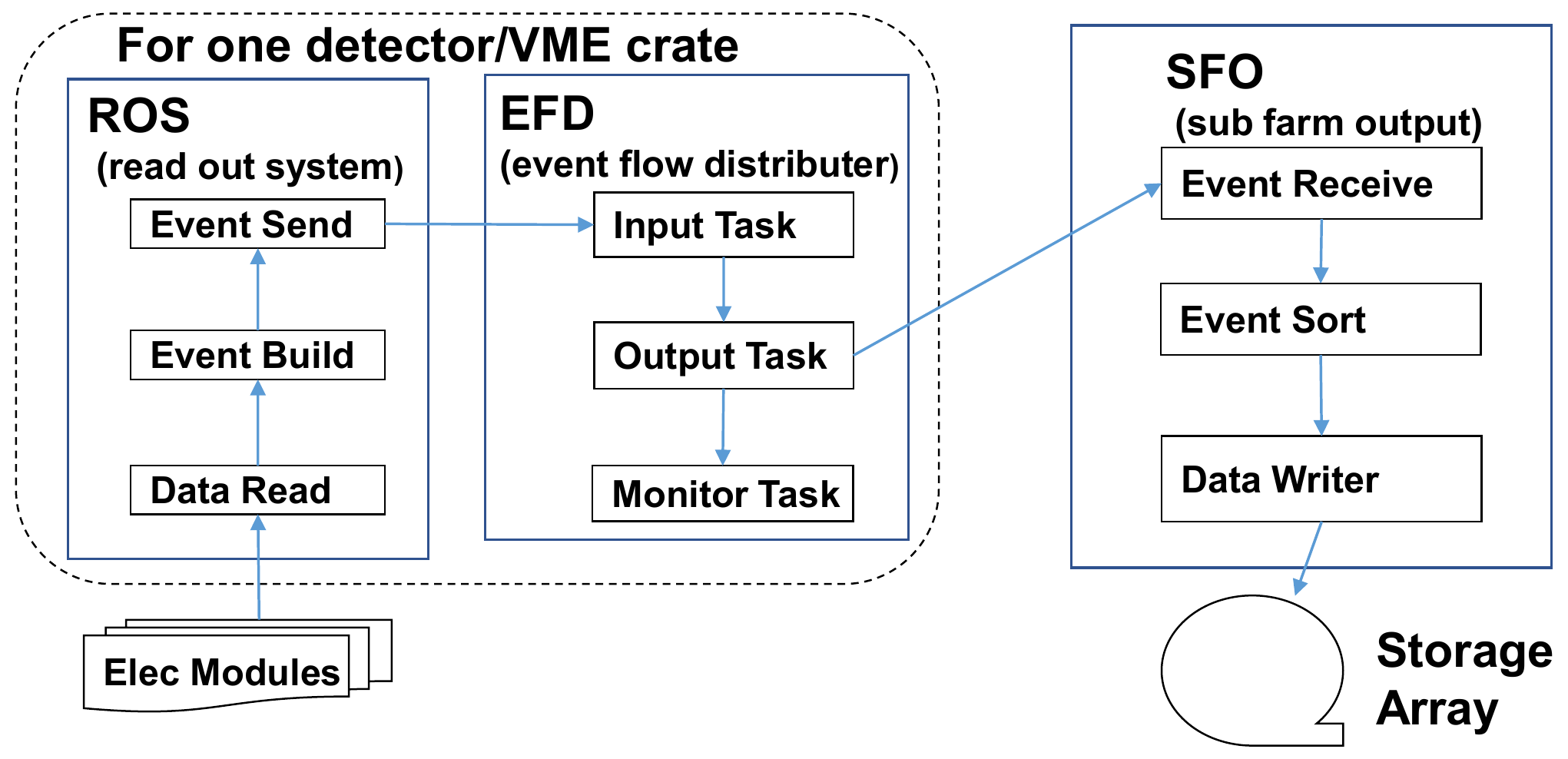}
\caption{ The data flow software. Details can be found in Sec~\ref{sec-daq}. }
\label{fig-daq-df}
\end{figure}

The data flow is shown schematically in Fig.~\ref{fig-daq-df}, and was developed using  BESIII front-end readout software 
and ATLAS back-end data flow software~\cite{besIII_ros}.  
Identical readout systems (ROS) run on each MVME5500 controller, reading and concatenating data to build individual events, packing and sending the data to the back-end~\cite{daq_ros2}.  
The Event Flow Distributor (EFD) runs on a blade server.  
It parses data, fills and publishes histograms for data quality monitoring, and sends data to the SubFarm Output (SFO).  
The SFO runs on file servers, merges data from EFDs, sort events by time, and saves files to a disk array. 

\begin{figure}
\centering
\includegraphics[width=3.1in]{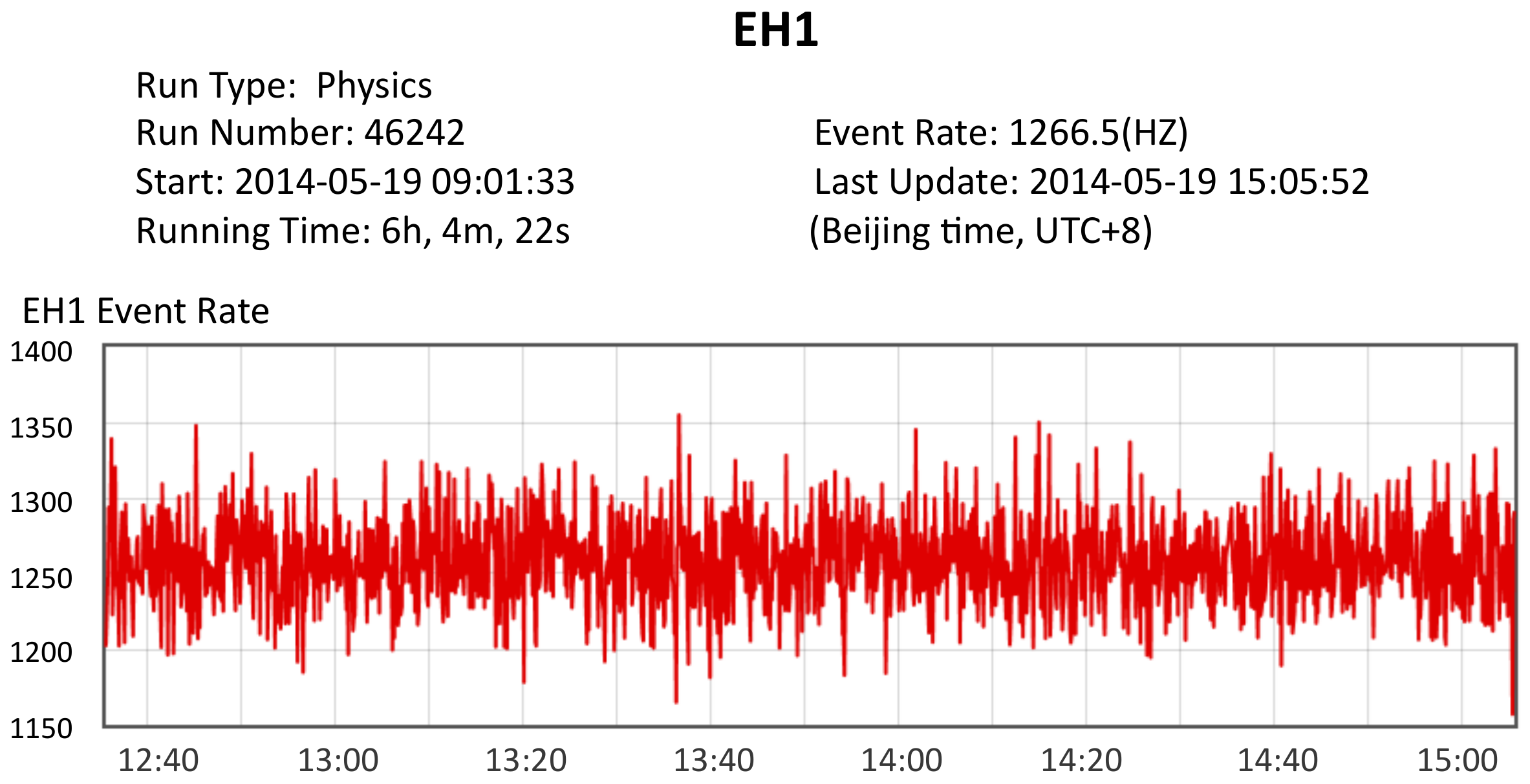}
\caption{ Snapshot of part of the overview page of  the DAQ web site. Event rate history as well as the related run information for EH1 are shown.  }
\label{fig-daqweb}
\end{figure}

The DAQ supports six run modes: physics, pedestal, electronics diagnosis, AD calibration, AD mineral oil monitoring, and water pool calibration.  
The DAQ exchanges information with the Distributed Information Management (DIM) system~\cite{DIM} and with 
external subsystems such as the calibration system.  
Online monitoring was implemented to aid shifters in spotting problems in real time with multiple graphic interfaces 
for monitoring and control.  
Web-based remote monitoring and control software was developed to enable remote shifting using CORBA Java API for 
DAQ access, and RESTful for web service~\cite{daq_web}.  Figure~\ref{fig-daqweb}  displays an example of web-based DAQ monitoring.
An online supernova trigger~\cite{Wei:2015qga} has been running in the DAQ system since August 2013. 

\subsection{Monitoring and control system}

The Daya Bay Detector Control System (DCS) monitors and controls the experimental hardware and environment. 
A commercial software package (Supervisory Control And Data Acquisition),  implementing the supervisory, control, and 
data acquisition  standard for the DCS, 
uses LabVIEW with Data logging and Supervisory control modules. 
The framework of the detector control system is divided into two parts, a  Local Control Layer (LCL) and a Global Control Layer (GCL). 
The LCL is composed of multiple local control IPCs (Industrial Personal Computer) 
which are responsible for the subsystem data acquisition and control. 
The GCL is composed of three parts: the Global Control Station (GCS),  a Database Server (DBS),
 and a Web Server (WS). 
The status and alarm information of all subsystems are collected into the GCS and stored into the DBS. 
Physics data recorded in the DCS can be retrieved by the Web Server.

The DCS monitors a wide array of detector parameters, including AD liquid levels, temperatures, 
humidity and air pressure, RPC high voltages, as well as photomultiplier high voltages and currents. It 
measures gas flow and pressures in the gas purification system and in the gas storage system. In  addition, the DCS monitors environmental parameters and auxiliary systems, such as the electronics crate, 
radon gas concentration inside and outside the experiment halls, extra environmental  temperatures  during the 
detector commission and installation process. The state of the water pool system (water pressure, water level, etc.) is also monitored.  
Data from environmental parameters 
and detector conditions (e.g. power supply voltages, temperature/humidity, gas mixtures, radiation, 
etc.) are collected from local IPCs and put into a data pool by DIM~\cite{DIM} and monitored by global 
control layer IPCs.
Details of the DCS system design and implementation can be found in~\cite{dcs}.

Some safety systems, such as rack protection and fast interlocks, are also included in the DCS. 
Safety interlocks can be activated to prevent undesired states in the  state machine, 
based on information from the  RPC HV, water  and cover gas systems.   
DCS can control the RPC HV status (on/standby/off). 
VME power is shutdown for over-current or over-temperature conditions. 
 Interlock signals can shutdown the gas system in the event of a HV trip. 
 Likewise RPC HV can be shutdown in the event of a gas system alarm.

A remote monitor and query system~\cite{dcs_web} was developed in the DCS to provide various functions 
including history data querying, real-time display, historical charts, threshold display and realtime alarm information. 

\section{Offline software overview  }
\label{sec-offline}

NuWa, the Daya Bay offline software adaptation of  the Gaudi framework~\cite{gaudi}, provides the full functionality 
required for simulation, reconstruction and physics analysis. NuWa employs Gaudi's event data service as the data manager. 
Raw data, as well as other offline data objects, can be accessed from the Transient Event Store (TES). 
The prompt-delayed coincidence analysis requires looking-back in time, which is fulfilled via a specific implemented Archive Event Store (AES). 
All the data objects in both TES and AES can be written into or read back from ROOT~\cite{root} files through various Gaudi converters. 

An alternative Lightweight Analysis Framework (LAF) was designed and implemented by Daya Bay to improve the analysis efficiency. 
LAF is compatible with NuWa data objects with higher I/O performance by the simpler data conversion, the implementation of lazy loading, and the flexible cycling mechanism. 
LAF allows  access to events both backwards and forwards through a data buffer, which also serves to exchange data among multiple analysis modules. 
The NuWa and LAF packages are available to collaborators using the Subversion (SVN) code management system~\cite{svn}.

Daya Bay simulation is based on GEANT4~\cite{geant4} with critical features validated against external data or other simulation packages. 
The Monte Carlo simulation is tuned to match the observed detector response. 
To simulate time correlations of unrelated events, different categories of simulated events are mixed in the output file. 
Reconstruction algorithms have been developed to construct the energy and vertex of the antineutrino event from the charge pattern of the PMTs. 
The detector-related parameters and calibration constants needed by the reconstruction are stored in an offline central database with a number of mirror sites located at different institutes. 
The algorithms can access the contents in the database via an interface software package called DBI.

\begin{figure}
\centering
\includegraphics[width=3.1in]{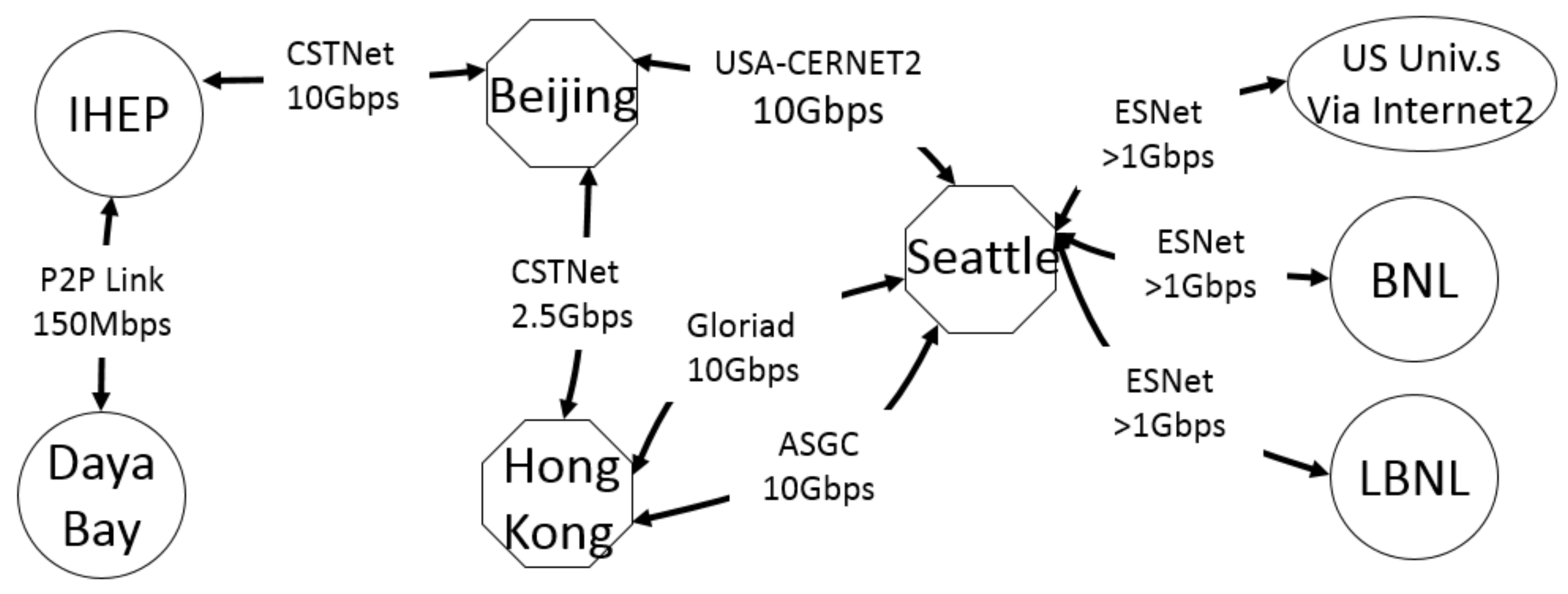}
\caption{Network topology}
\label{fig-data-move}
\end{figure}

\begin{figure}
\centering
\includegraphics[width=3.1in]{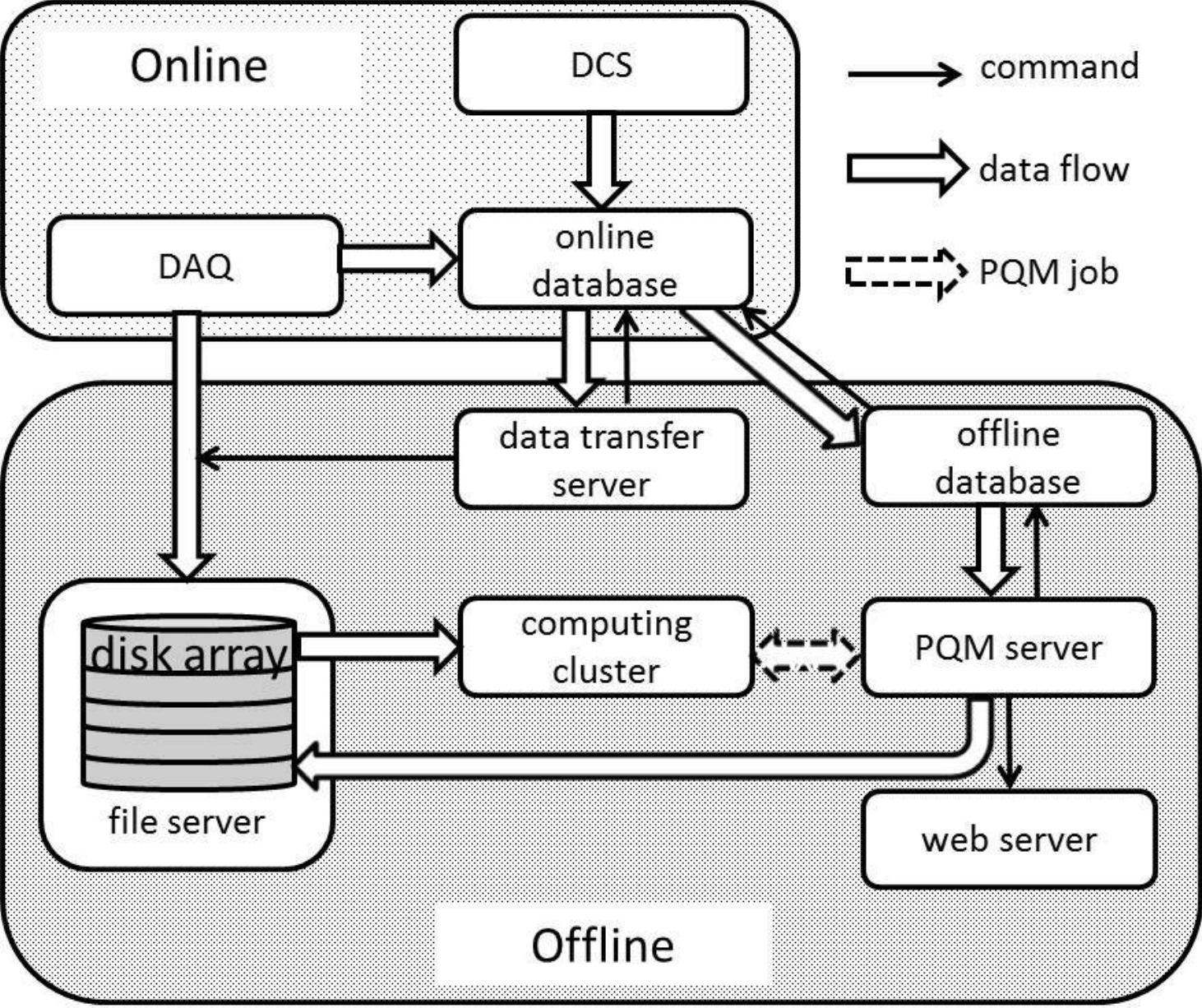}
\caption{Schematic diagram of the offline computing environment at the Daya Bay site  described in Sec~\ref{sec-offline}.}
\label{fig-farm}
\end{figure}

An automated system ensures real-time delivery of raw data and databases to the offline processes. 
Data are first  transferred to the onsite farm, then to  the Institute of High Energy Physics (IHEP) in Beijing and Lawrence Berkeley 
National Laboratory (LBNL) in California for  central storage and processing, and distributed to other 
institutions for validation and analysis. Figure~\ref{fig-data-move} shows the network topology. 
A dedicated small-size offline system is deployed at the Daya Bay site, consisting of a file server, a data transfer server, an offline database
server, a web server, two data monitoring servers and five servers forming user farms, as shown in Fig.~\ref{fig-farm}.  
Two large-size offline systems are deployed at IHEP and LBNL, separately. 
Each of them has about 1 PB of disk space capacity  and 800 CPU cores.
Additional computing resources have been planned to accommodate the increasing data volume.

A Performance Quality Monitoring system (PQM)~\cite{pqm}  runs onsite, using fast 
reconstruction algorithms and analysis modules in NuWa, to monitor the physics performance with a 
latency of around 40 minutes. 
A ``keep-up" data processing takes place as soon as the data reach IHEP or LBNL, using the full NuWa 
reconstruction and the latest calibration constants. 
The generated detector monitoring plots are distributed for collaboration-wide review and archived through an Offline Data 
Monitoring system  with a latency of around 3 hours. 
The extracted data quality information are filled in a dedicated database for  long-term monitoring.

\section{Operation and Performance}
\label{sec-performance}

Daya Bay has collected antineutrino data since Sep. 2011, first with 2 ADs, 
then with six ADs starting in Dec. 2011, and most recently with 8 ADs
since Oct.  2012.   Over one million (150 K) antineutrinos
have been detected at the near (far) site detectors by the end of 2013.

\begin{figure}
\centering
\includegraphics[width=3.1in]{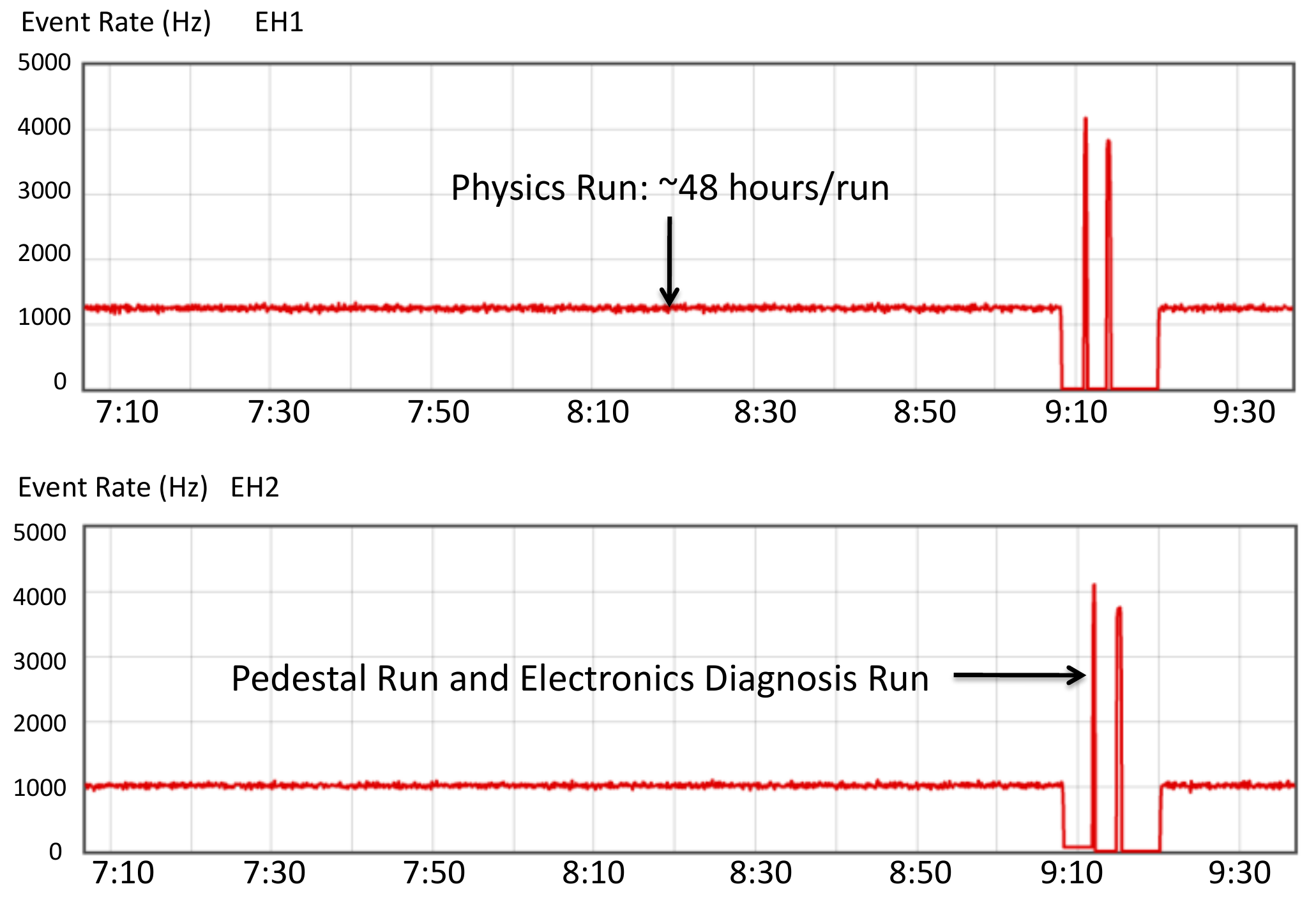}
\caption{Sample trigger rates in  the near experimental halls}
\label{fig-trigger-rate}
\end{figure}

Typical data taking  includes a $\approx 48$ hr physics run, 
followed by  $\approx$~1-minute pedestal  and  electronic diagnostic runs,  as shown in Fig.~\ref{fig-trigger-rate}. 
Data from each of  the three halls is recorded in separate runs using a universal clock to
record the begin/end run times. 
Triggers from any of the detectors within a hall can initiate readout of the electronics in that hall. 
The typical  trigger rate for all detectors at each  site in a physics run is  
1.3 kHz for EH1, 1.0 kHz for EH2, and 0.6 kHz for EH3.  Variation between halls is driven by the different overburden and  muon rate  at each hall. One example of the trigger rates for each subsystem 
is listed in Table~\ref{tab-triggerrate}.
Electronic diagnostic runs give additional information on the noise and gain of each channel  for nearby physics runs. 
About 320 raw data files, 1 GB per file, are generated per day and transferred to the computing farms.
Radioactive source and LED pulser calibrations are also routinely performed. 
Detector live time  for recording antineutrino events  was greater than 92\%  
with the majority of the down time dedicated to calibration.

\begin{table}
\caption{Trigger rates recorded for run 54261(EH1), run 54262(EH2) and run 54260(EH3)}
\centering
\begin{tabular}{ cc|cc}
\hline
         Site & Total rate & Subsystem      &  Trigger rate  \\
           &   (Hz)   &  & (Hz) \\
\hline
          &  &   AD1 & 273 \\
           &  & AD2  & 268 \\
 EH1   & 1301     & IWS &  220 \\
        & &OWS & 325 \\
        & &RPC & 215 \\
         \hline
&  &AD1 & 215\\
      & &AD2 & 211 \\
EH2     &966 &  IWS &  192 \\
         & & OWS &  245  \\
    & & RPC & 103 \\
    \hline
&   &AD1 &  131  \\
      &   &AD2 & 124 \\
      & &AD3 & 120 \\
EH3   & 635  & AD4 & 131 \\
    & & IWS &  39 \\
     & & OWS & 54  \\
   &   & RPC  & 36 \\
\hline
\end{tabular}

\label{tab-triggerrate}
\end{table}

\subsection{Muon System Performance}
\label{sec-muperform}
\begin{figure}
\centering
\subfigure[IWS]{
\begin{minipage}[b]{0.45\textwidth}
\includegraphics[clip=true, trim=15mm 8mm 10mm 0mm,width=1.0\textwidth]{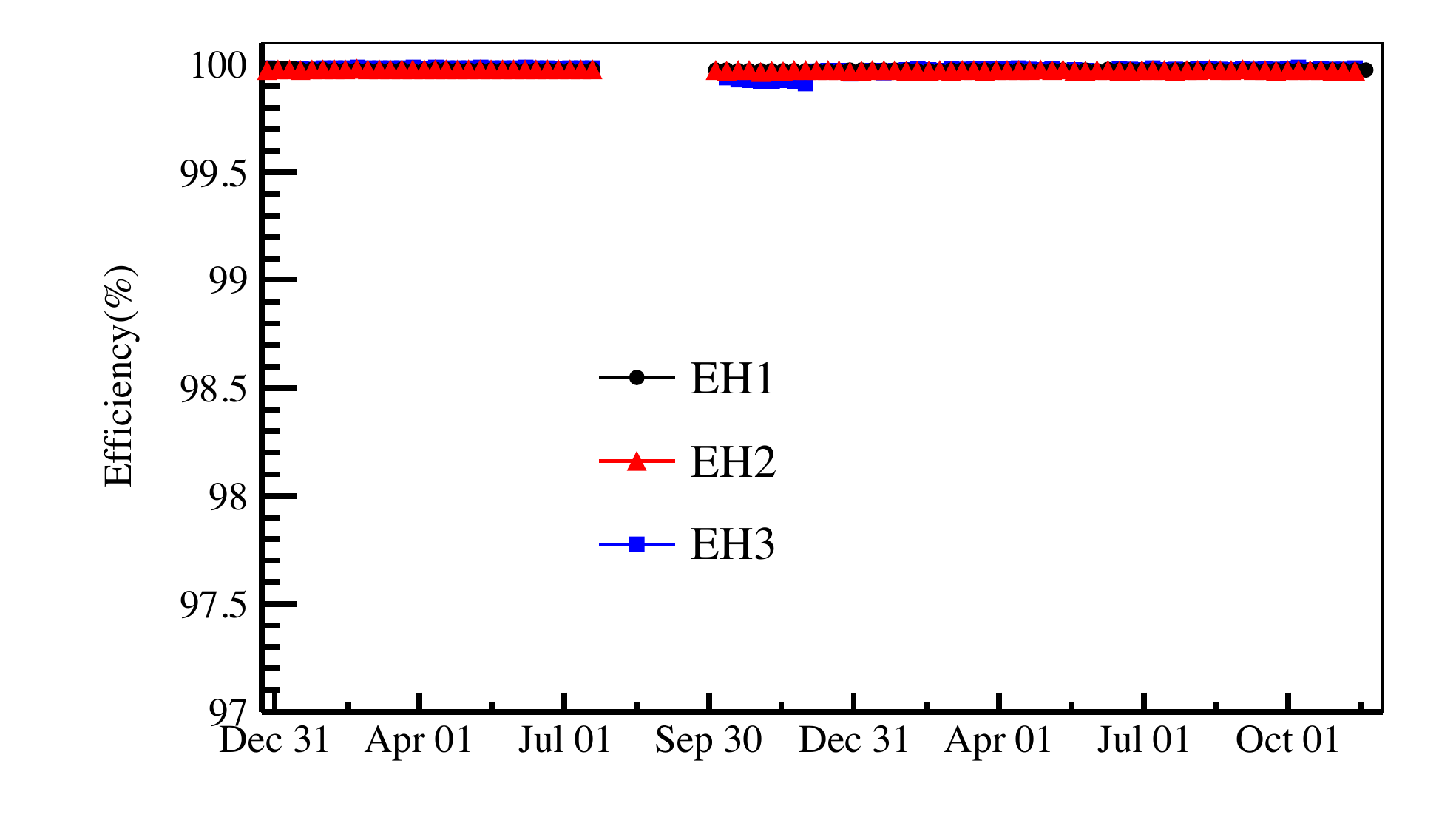}
\end{minipage}
}
\subfigure[OWS]{
\begin{minipage}[b]{0.45\textwidth}
\includegraphics[clip=true, trim=15mm 6mm 10mm 0mm,width=1\textwidth]{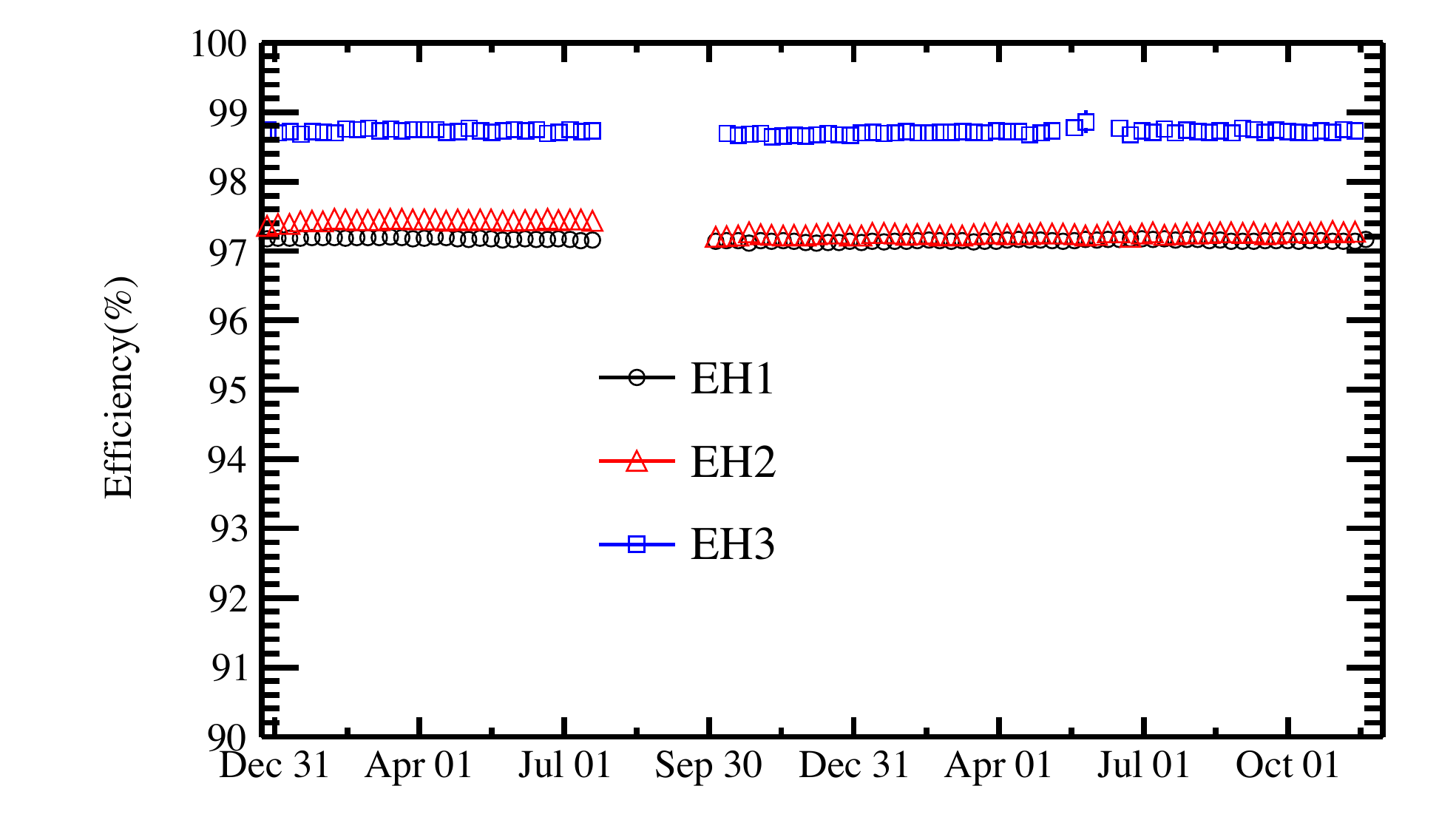}
\end{minipage}
}
\caption{WP muon  detection efficiencies from Dec. 2011 to  Dec. 2013 (See Sec.~\ref{sec-muperform} for details).
The gap corresponds to the summer shutdown in 2012.}
\label{fig-WPtrigger}
\end{figure}

The water pool systems provides essentially 100\% efficiency in detecting muons. 
This efficiency is studied using muons going through the ADs, which occur at  about 20 Hz in EH1, 
15 Hz in EH2 and 1 Hz in EH3. 
The  mean IWS efficiencies for all three halls is $99.98\pm0.01\%$ and is quite 
stable over time as shown in Fig.~\ref{fig-WPtrigger}(a).  
The OWS efficiencies  shown in Fig.~\ref{fig-WPtrigger}(b) are underestimated since there is no correction for
 stopping muons which deposit energy in the AD or IWS but not in the  OWS.

\subsection{AD performance - energy response}

To take full advantage of the unprecedented statistical precision of this data set, 
careful monitoring and control of systematic errors  are required.
ADs are calibrated periodically using the ACUs on the lid of each detector.  
Combined radioactive sources ( Am-$^{13}$C$/^{60}$Co or Am-$^{13}$C$/^{68}$Ge) are lowered to five vertical locations for all three ACUs separately, followed by a similar scan using the other  radioactive sources ($^{68}$Ge or $^{60}$Co).
Data are collected for four minutes at  each position, accumulating   
about 27$\times 10^3$ (4.8$\times 10^3$) $^{60}$Co ( $^{68}$G)  events in each AD  during early 2012. 
Low intensity LED runs collect  data which  can be used in the ADC calibration for each PMT readout channel. 
The demonstrated stability of the AD performance allowed a reduction in the number of calibrations 
starting Jan.~11, 2013.
The weekly calibration was reduced to calibrations with the LED and $^{60}$Co source at the detector center.
The full calibration previously described was performed every  fourth week. 

Single photoelectron (SPE) hit data from physics data runs were selected and fitted  
after baseline subtraction with  a convolution of a Poisson distribution with a Gaussian function  as shown 
below:
\begin{equation}
S(x) = \sum_n {\frac{\mu^n e^{-\mu}}{n!}}\frac{1}{\sigma_1\sqrt{2n\pi}} \exp \left(-\frac{( x-nQ_1)^2}{2n\sigma_1^2}\right)
\label{eqn-spe}
\end{equation}
where $\mu$ is the expected value of the Poisson distribution determined by the PMT hit occupancy, 
$\sigma_1$ and $Q_1$ are the resolution and mean value of the SPE distribution.  The summation over the 
number of photo-electrons is truncated at n=2. The SPE  typically has a mean value of about 19 ADC counts with $\sigma_1/Q_1$ of about 33\%.

\begin{figure}
\centering
\includegraphics[clip=true, trim=8mm 5mm 10mm 12mm,width=3.1in]{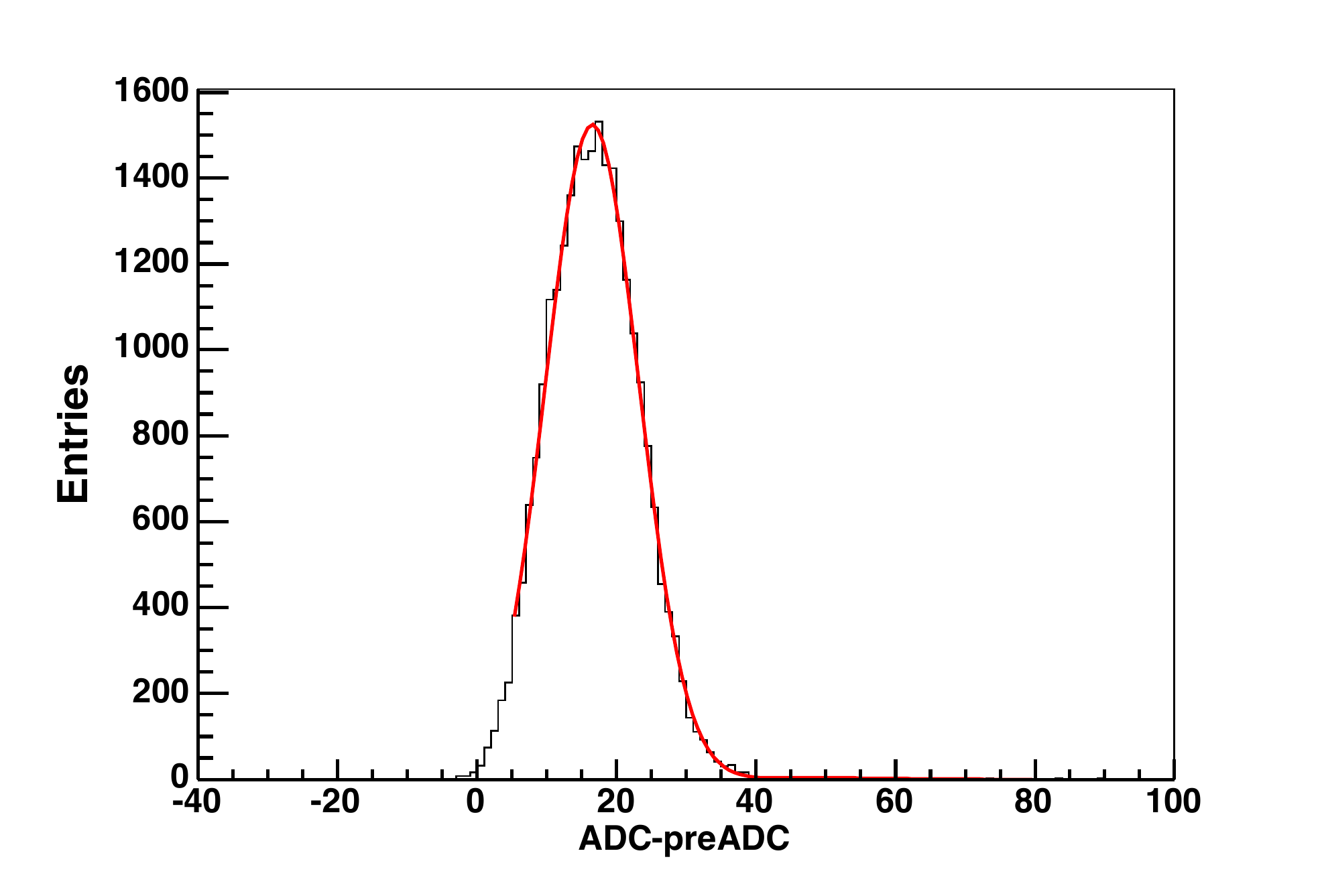}
\caption{Example baseline-subtracted SPE distribution with fit using the  function in Eq.~\ref{eqn-spe}. 
The x-axis is the peak ADC value minus  the ADC average baseline before the pulse.
}
\label{fig-spe}
\end{figure}

Figure~\ref{fig-spe} shows an example ADC distribution for SPE  hits with 
fitting results overlaid.   

\begin{figure}
\centering
\includegraphics[clip=true, trim=5mm 5mm 0mm 0mm,width=3.1in]{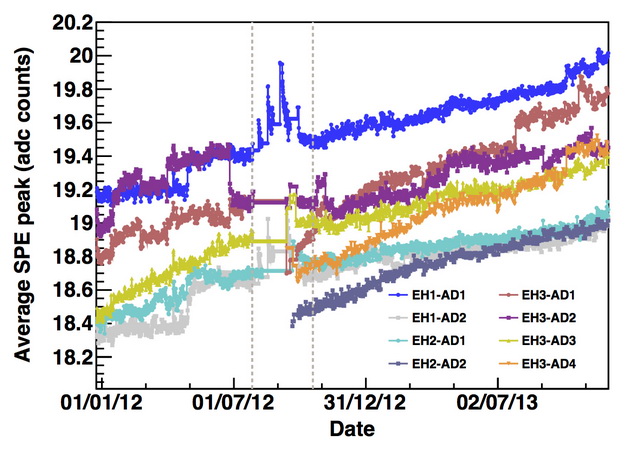}
\caption{Average number of ADC counts for single photoelectrons in each AD versus time. There were fewer 
measurements during the summer shutdown of 2012 (between the two grey vertical dash lines).  }
\label{fig-PMTgain}
\end{figure}

Combined with low intensity LED run data, the PMT gain and timing are monitored continuously.  
The stability of the ADC gains was measured every $\approx$6 hours  using SPE data from physics data runs  as 
shown Fig.~\ref{fig-PMTgain}. 
Slight gain drifts are seen in all ADs and  have been crosschecked with low-intensity LED data.  
Many parameters were investigated as possible causes  of the observed  gain drifts. 
Temperature drifts  were partially correlated to the observed drifts but were not  the only cause.    
Jumps  correlated among the ADs are suspected to be caused by power cycling of the HV mainframes.  
It is likely that the PMT SPE pulse height is slowly increasing with time since periodic calibrations of the FEE card ADCs
by onboard DACs are stable to better  than a percent per year.

\begin{figure}
\centering
\includegraphics[clip=true, trim=0mm 5mm 0mm 0mm,width=3.1in]{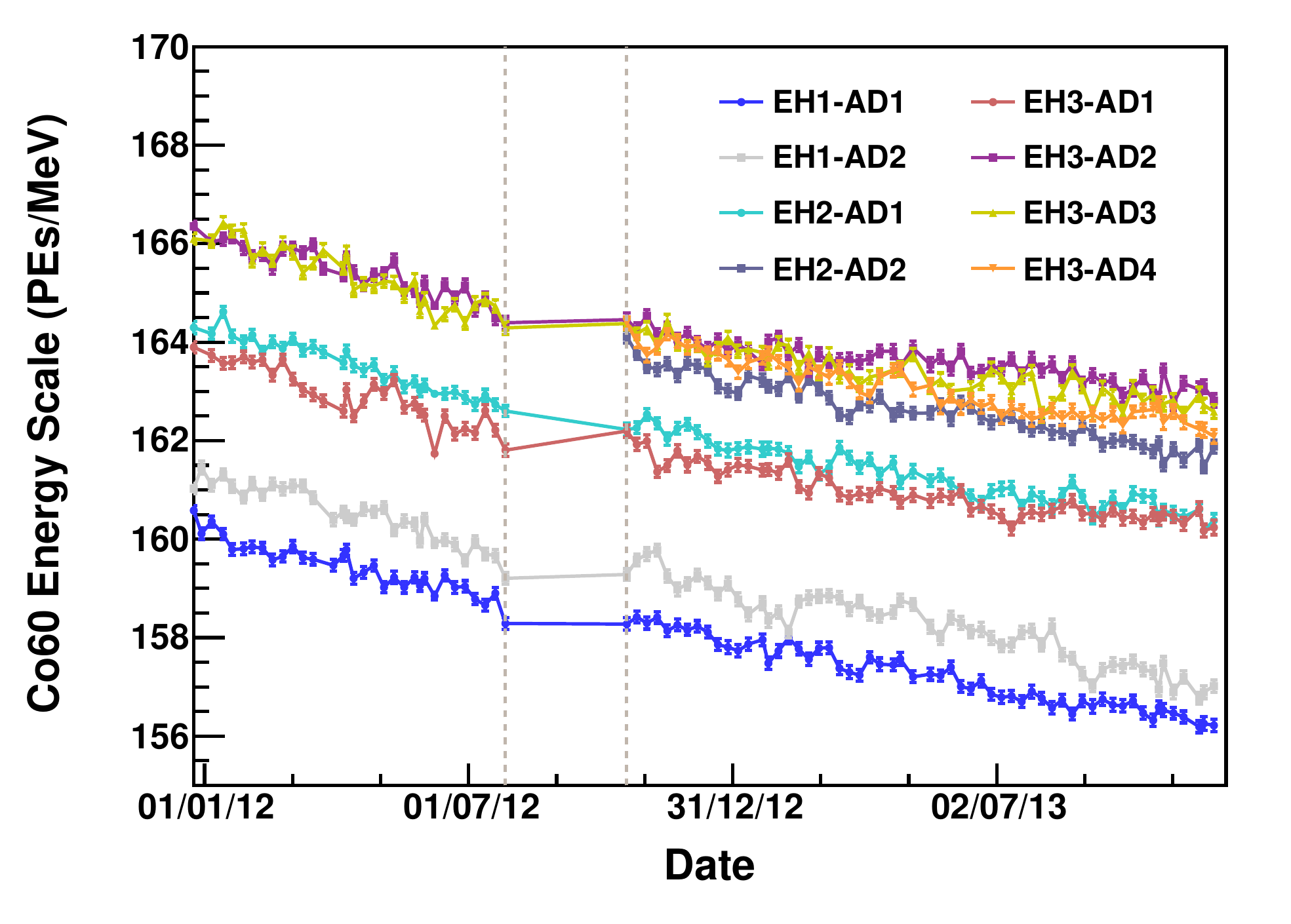}
\caption{The energy scale from $^{60}$Co calibrations at the detector center is monitored over time. 
The 2012 shutdown period occurs between the two grey dashed lines.}
\label{fig-EscaleVStime}
\end{figure}

The number of P.E. per MeV (energy scale)  based on $^{60}$Co calibrations at the detector center   is
 monitored over time as shown in Fig.~\ref{fig-EscaleVStime}.  
A slight downward slope ($\approx$1.5\% per year) is seen.  
Similar behavior is seen in energy scales derived from spallation neutrons. 
The reason for the downward slope of the energy scale is under study.
A downward slope could be consistent with a decrease in  LS and GdLS attenuation lengths,  
however, no degradation of light yield or attenuation length has been observed to date in the  LS and GdLS production samples 
preserved and tested in the lab.  Further study is clearly needed to fully understand all long-term trends. 
However, as will be shown in later plots, the AD energy scales after calibration are stable to 0.1\%.

Many special calibration runs were performed during the shutdown in  2012 with nonstandard sources.
A strong  60~Hz $^{241}$Am-$^{13}$C neutron source (same design as the weaker  ACU sources)  was deployed 
on the top lid of  AD5 in EH3 to 
better measure the backgrounds in physics data  due to the 
 $^{241}$Am- $^{13}$C neutron sources  while in the parked position inside the ACU.

Other dedicated calibrations were performed in EH1 and EH2 in parallel. 
 Gamma source ($^{68}$Ge, $^{60}$Co) and neutron source scans were performed with fine steps in z. 
 Two ACUs (9A and 8C)  were instrumented with special calibration sources and mounted on AD1 and AD2 in EH1.
 The sources in ACU 9A were AmC, AmBe, and a combined $^{60}$Co/ $^{137}$Cs source.  
 PuC, $^{60}$Co and $^{40}$K  were mounted in ACU 8C. 
 The $^{137}$Cs source in ACU 9A was replaced with a $^{54}$Mn source at a later time.

Additional calibration data included  full detector vertex scans with the gamma sources
in the z range from 1.8 m to -1.5 m in 10 cm steps and 7-hour-long neutron scans at four z positions
(1.775 m, 1.35 m, 0~m, -1.35 m). 
Data taking without calibration sources  was also performed.

These tests provided:
\begin{itemize}
\item{Multiple gamma source data for studying the non-linearity of the GdLS energy response along the center axes.}
\item{Neutron source data  at various positions for studying the H/Gd ratio, 
 the  prompt and delayed spectra in AD1/AD2,  and to study the spill-in/out correction uncertainties. }
\item{High statistics  $^{40}$K data for non-linearity studies at the detector center.}
\end{itemize}  
A manual calibration system was installed in AD1 in EH1 during the 2012 shutdown 
(Sec~\ref{section-mcs}).  
Calibrations  were performed over the full GdLS detector volume (r and z).

  \begin{figure}
\centering
\includegraphics[clip=true, trim=0mm 2mm 5mm 6mm,width=3.1in]{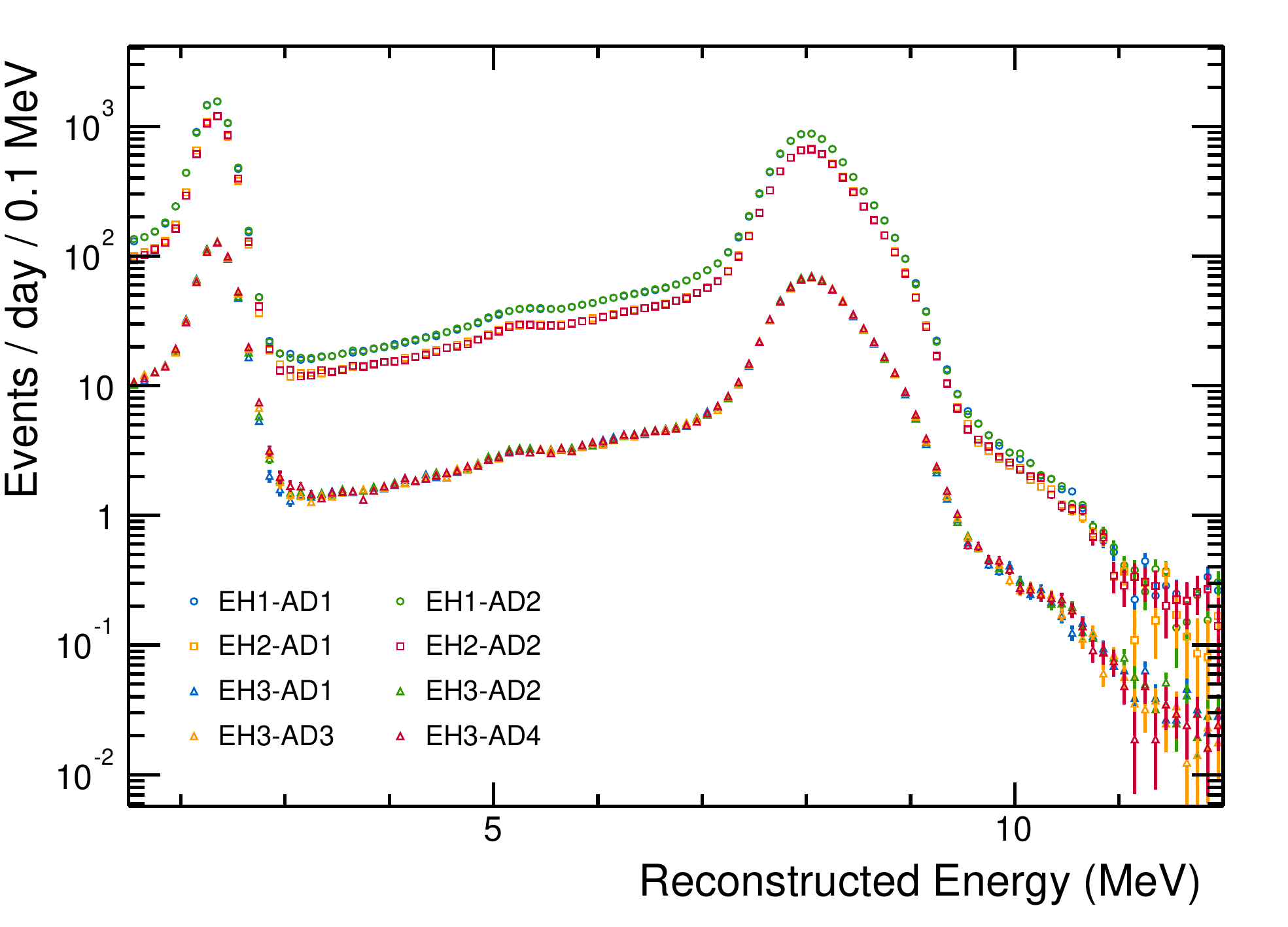}
\caption{Energy spectrum from muon induced spallation neutrons. Points  from detectors in the same hall are nearly identical and may be
plotted over  each other.}
\label{fig-spn-spectrum}
\end{figure}

\begin{figure}
\centering
\includegraphics[clip=true, trim=0mm 0mm 0mm 0mm,width=3.1in]{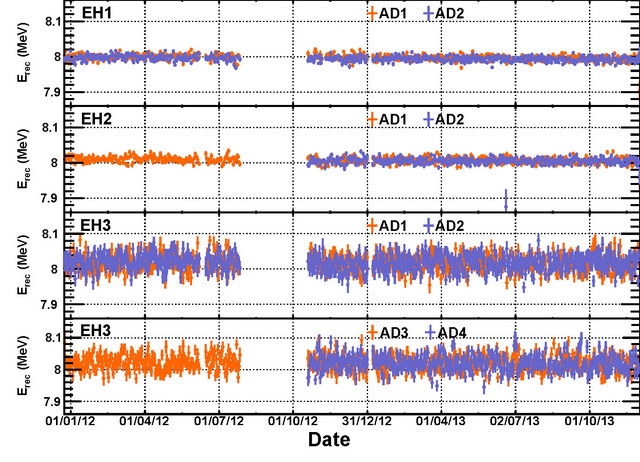}
\caption{ The time dependence of the reconstructed energy of spallation neutrons captured on gadolinium.  }
\label{fig-spn-Estability}
\end{figure}

Data from the calibration sources as well as data from  cosmogenic events such as spallation neutrons
(recorded with the physics data) 
are  used to study and track the energy response of the detectors.  
Spallation neutrons are cleanly selected by searching a time window from 20 to 200 $\mu$s after a muon passes through an AD.      
The energy spectra of  spallation neutrons in all 8 ADs are shown in Fig.~\ref{fig-spn-spectrum}.  
The 8 MeV peak from neutron capture on Gd and the 2.2 MeV peak from neutron capture on hydrogen are  clearly identified.  
 The neutron capture spectra of all  eight ADs are in excellent agreement. 
The calibrated energy is a convolution of  the ADC per PE and PE/MeV calibrations (see Fig.~\ref{fig-PMTgain} and 
Fig.~\ref{fig-EscaleVStime}  as examples). Although both calibrations change with time, the calibrated energy should be stable.
Figure~\ref{fig-spn-Estability} shows that the average energy  of spallation neutrons capturing on Gd for the eight ADs  is very stable.

\begin{figure}
\centering
\includegraphics[clip=true, trim=10mm 0mm 10mm 0mm,width=3.1in]{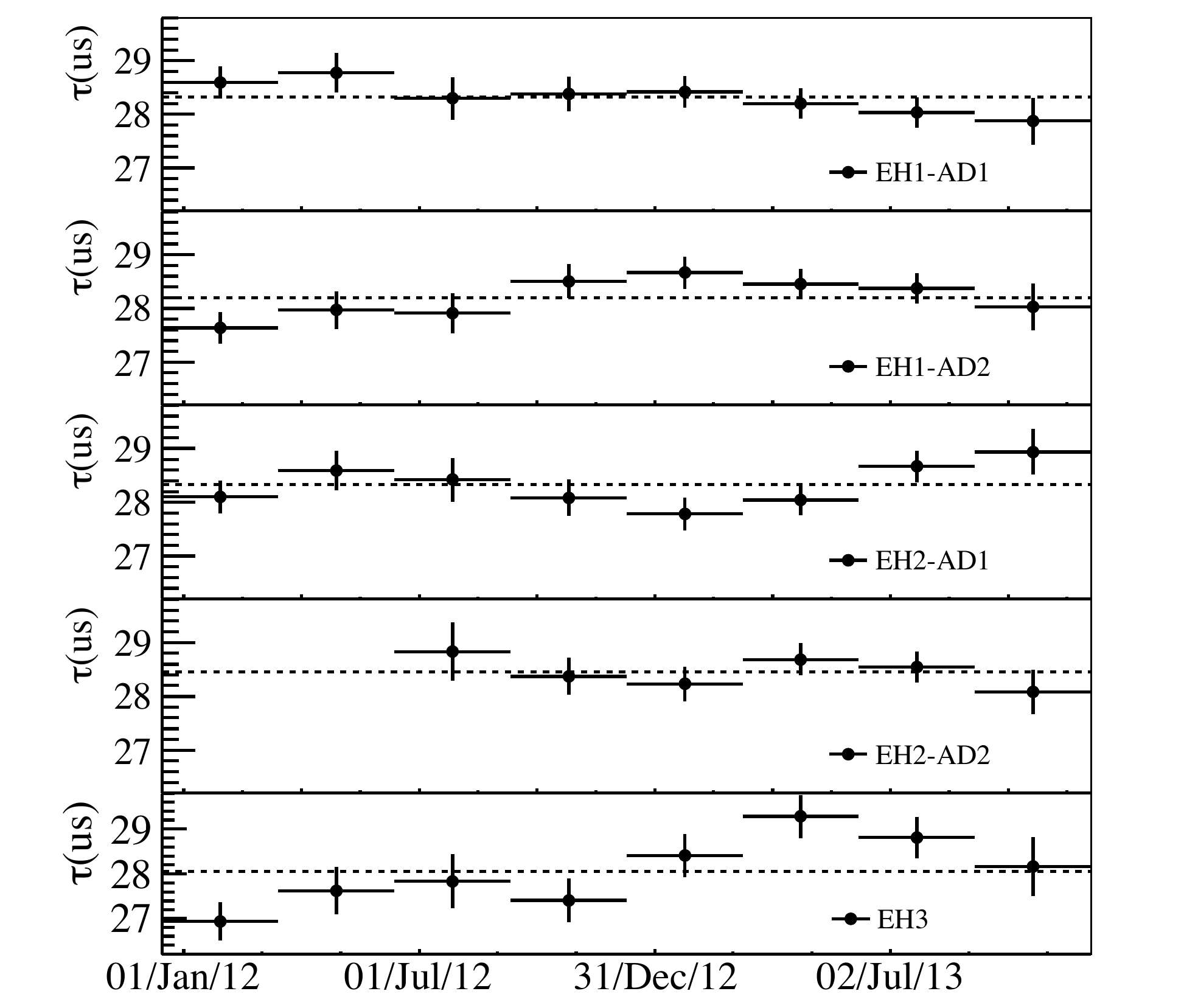}
\caption{Average capture time of IBD events  in the four near hall ADs   are shown as a function of time.
From the top the ADs are: EH1-AD1, EH1-AD2, EH2-AD1 and EH2-AD2.  
The bottom graph averages all four far hall ADs  together for better statistical precision.  
The dashed line shows the mean capture time of the AD(s).
}
\label{fig-CapT-Stability}
\end{figure}

 The neutron capture time is directly related to the Gd concentration in the detector and can be tracked to
 look for time variation in the Gd doping.
 Both neutrons from IBD interactions and muon spallation events are used to study the capture time spectra. 
 Figure~\ref{fig-CapT-Stability} shows the capture times of neutrons from IBD samples in the four near ADs.
 The four far ADs are added together in the bottommost plot to improve the statistical error.
 Capture times are stable, reaffirming that the Gd
 concentration in the GdLS does not change with time.

\begin{figure}
\centering
\includegraphics[width=3.1in]{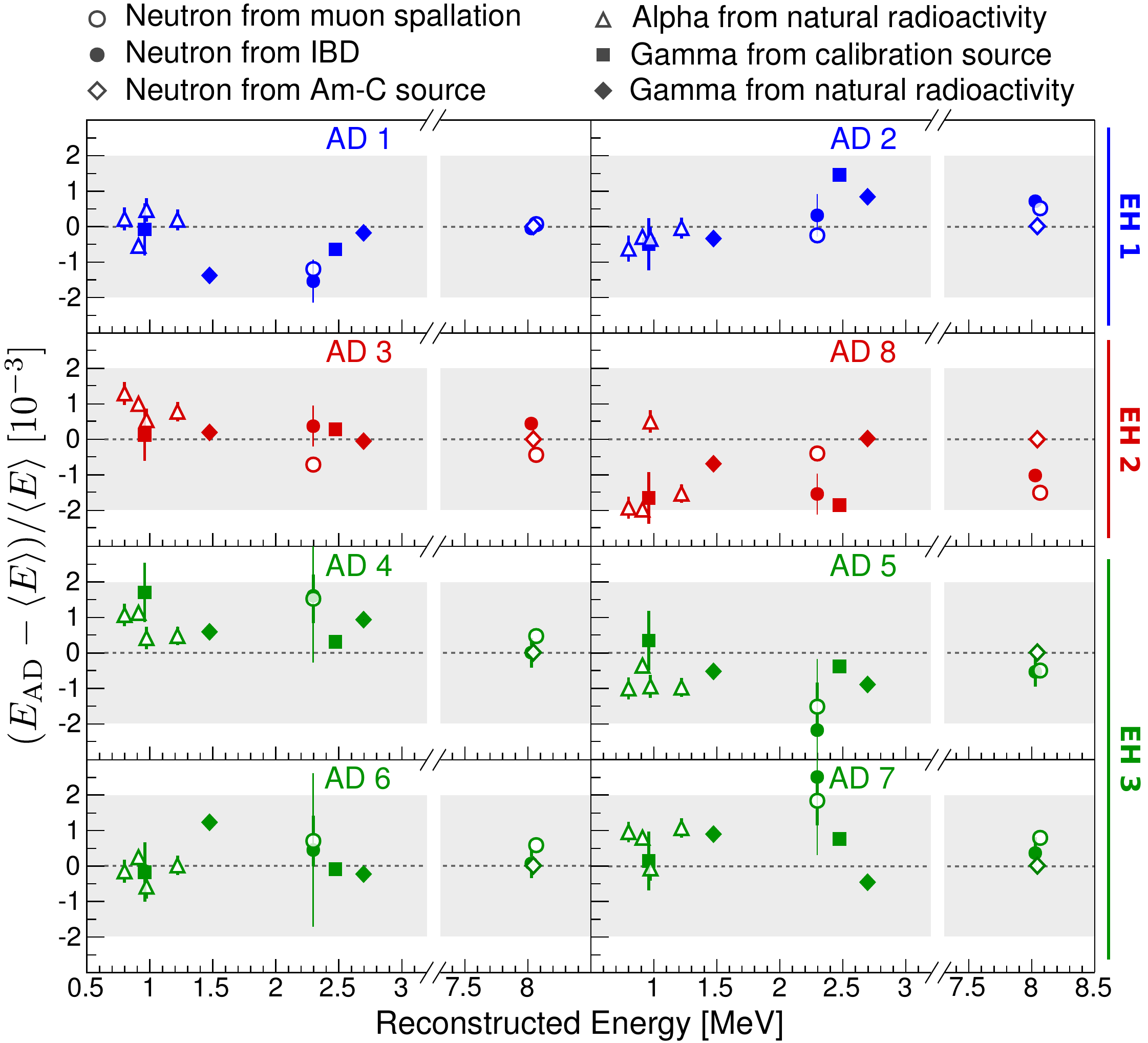}
\caption{Comparison of the reconstructed energy between antineutrino detectors for a variety of calibration references. $E_{AD}$ is the reconstructed energy measured in each AD, and $\langle$E$\rangle$ is the 8-detector average. Error bars are statistical only.  Systematic 
variations between detectors for all calibration references were $< 0.2$\%. The energy scale
of each detector was determined using the $\sim$8 MeV n-Gd capture gamma peaks from Am-C sources,  hence these points  show zero deviation.}
\label{fig-rel-Escale}
\end{figure}

Relative energy scale uncertainties between ADs are a major source of systematic errors.  
Particles at different energies are selected to study the relative energy scale in all ADs. 
Clean samples of alphas  are selected from decays of ~$^{214}$Po,~$^{212}$Po and~$^{215}$Po using 
the $\beta-\alpha$ and $\alpha-\alpha$ time and energy correlations. 
Calibration gamma and neutron sources at specific locations inside the AD, 
 as well as uniformly distributed IBD and spallation neutron sources
are studied in all eight ADs.   Figure~\ref{fig-rel-Escale}   shows that the relative energy scale differences 
among the eight ADs are all within a  $\pm 0.2\%$ band.

 A precise check of relative detector efficiencies can be made from the ratio of detected neutrino events 
 in the nearly identical detectors in each hall.
For the detectors in the same hall, background and reactor uncertainties cancel out, leaving only  small corrections for differing baselines and target masses.  
For IBD-like events with more than   0.7 MeV prompt energy (nearly 100\% of all prompt IBD positrons),
the expected  ratio in EH1 (AD1/AD2) is 0.982 and the measured value is $0.981\pm0.004$, 
agreeing within error.
The same is true in EH2, where the ratio of detected IBD events in AD3 and AD8 is measured to be $1.019\pm0.004$, consistent with the expected value of 1.012 within error.

\subsection{EH3-AD1 LS leak}

AD operating conditions were generally  stable during the first three years of operation.
Return cover gas humidities and oxygen levels  gradually declined in all ADs with no indications of water leaks. 
Liquid levels in the overflow tanks track small variations in the water pool temperature as the AD liquids expand or contract. 
The  largest change of GdLS levels from Dec. 2013 to May 2015 was  in EH3-AD1 with  a 11.4 mm decrease  corresponding
to a change in the AD target mass of 0.06\%. 

\begin{figure}
\centering
\includegraphics[clip=true, trim=0mm 0mm 0mm 5mm,width=3.1in]{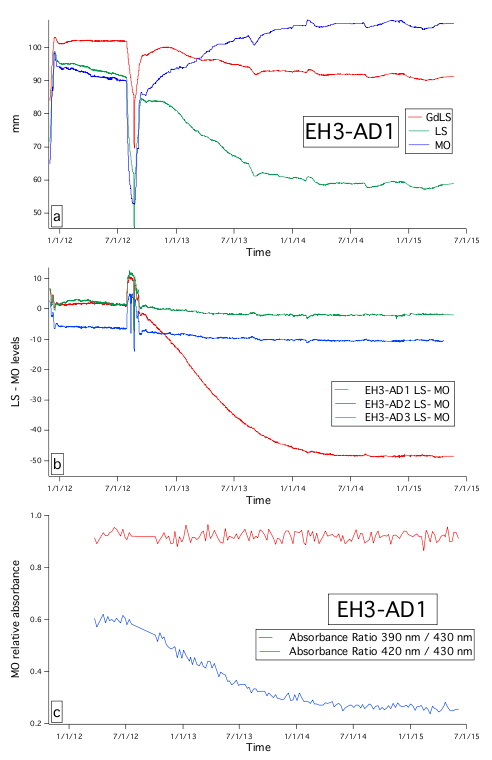}
\caption{(a) GdLS, LS, and MO levels in EH3-AD1 versus time. 
(b) Difference between the LS and MO levels versus time for the original 3 ADs in EH3.
(c) Ratio of the absorbance at 395 nm  (420) to that at 430 nm as measured by the EH3-AD1 MO clarity monitor.
The liquid levels and MO clarity data are both consistent with a small leak of  LS from the OAV volume to the MO volume
which started during the access in Aug. 2012.
 }
\label{fig-AD4}
\end{figure}

After the Aug. 2012 access to EH3 to install EH3-AD4 it was observed that the LS level in EH3-AD1 was dropping 
while the MO level was rising as shown in Fig.~\ref{fig-AD4}a. 
This feature was also accompanied by changes in the MO absorbance vs wavelength 
measured by the MO clarity monitor in EH3-AD1.
This trend continued for many months before reaching new equilibrium values in 2014.
Taking the  difference in AD liquid levels removes most of the temperature variation effects seen in Fig.~\ref{fig-AD4}(a).
Figure~\ref{fig-AD4}(b) shows the difference between the LS and MO overflow tanks levels for the EH3-AD1-3. 
Unlike the near constant levels seen in EH3-AD2 and EH3-AD3, 
EH3-AD1 shows a clear decline starting during the August access. 

This behavior can be understood as the 
consequence of a small crack at the bottom of the OAV that allows the LS to leak into the MO volume.
The crack must be at or near the bottom of the OAV  since the leak continued even when the MO level was 
several cm higher than the LS level.
Since LS is $\approx$~0.9\% denser than the MO, the differential pressure between the LS and MO volumes changes with
height.  Near the top of the OAV the MO is at higher pressure since the overflow tank levels are higher than the LS.
The situation was reversed at the bottom of the OAV at the start of the leak. 
As the MO levels rose and LS levels dropped  the leak slowed as the differential pressure 
between the LS and MO volumes decreased. The leak stopped when the LS and MO pressures across the
crack equalized. At the bottom of the OAV the pressure due to the denser LS  
is offset by the 4.8 cm higher MO level. 

LS is expected to dissolve completely in the MO and to diffuse  through the entire MO volume.  
Since the LS absorbance  at low wavelengths ($\leq410$ nm) is much higher than MO, 
 the absorbance of the MO in EH3-AD1 is expected to change as a function of time.
Above ~410 nm, the MO and LS absorbance curves have a very similar 
dependence on wavelength.
The absorbance measured  by the MO clarity monitor at 390 and 420 nm divided by the 
absorbance at  430 nm is plotted in Fig.~\ref{fig-AD4}(c).
In the wavelength region where the MO and LS have nearly identical absorbance, 
the ratio of the  420/430 nm  data is flat. 
In the region where the absorbance of the LS and MO differ significantly,  
the ratio of the  390/430 nm  absorbances exhibits  a similar time dependence as the LS level, supporting the 
interpretation of  a LS leak,  as an explanation of the changing overflow tank levels.

An estimated 50 liters of LS has leaked into  the 42800 liters of  MO (0.12\%) in EH3-AD1.
Several studies were made  to check if the AD performance was changed in any measurable way by this leak. 
Since almost all of the energy from Gd captures is contained within the GdLS and LS volumes, 
no changes to any of the oscillation analyses have been observed or are expected.  
The average  energy  of neutron captures on hydrogen particularly
for captures near the edge of the LS should in principle increase, however, the effect is well below the systematic errors
of the detector to detector energy response variation (0.2\%).  
An analysis of cosmogenic muon distributions in the ADs has 
seen increased light yield of  tracks traversing part of  MO volume in EH3-AD1 compared to other ADs. 
The  light yield increases with time as expected from the predicted LS contamination.

\section{Conclusion}
\label{sec-conclusion}

The Daya Bay experiment is unique in using multiple antineutrino detectors at the near and far 
experimental hall sites. 
Two antineutrino detectors  are deployed in each of the two near detector halls to monitor the 6 Daya Bay nuclear reactor cores.
Four antineutrino detectors  are deployed  in the far experimental hall to measure the antineutrino oscillation.
The interchangeable  antineutrino detectors  were built  and  assembled above ground using standardized procedures
to minimize  efficiency differences between detectors. Construction of the underground  tunnels and halls in parallel
with detector assembly minimized total project time and resulted in the first $\geq$ 5 standard deviation measurement of
sin$^22\theta_{13}$. 
Side-by-side comparisons between detectors at  the near  sites cross check  calculations of  relative 
detector efficiencies to better than 0.2\%.
The large target mass provided by the four detectors at the far hall enables  an unmatched statistical precision,
yielding the world's most precise measurements of sin$^22\theta_{13}$ and the  effective mass splitting $\Delta m_{ee}^2$,
now, and for the forseeable future.

\section{Acknowledgements }


The Daya Bay Experiment is supported in part by 
the Ministry of Science and Technology of China,
the United States Department of Energy,
the Chinese Academy of Sciences,
the CAS Center for Excellence in Particle Physics,
the National Natural Science Foundation of China,
the Guangdong provincial government,
the Shenzhen municipal government,
the China General Nuclear Power Group,
the Research Grants Council of the Hong Kong Special Administrative Region of China,
the MOST fund support from Taiwan,
the U.S. National Science Foundation,
Yale University,
the Ministry of Education, Youth and Sports of the Czech Republic,
the Joint Institute of Nuclear Research in Dubna, Russia,
the NSFC-RFBR joint research program,
the National Commission for Scientific and Technological Research of Chile.
We acknowledge Yellow River Engineering Consulting Co., Ltd.\ and China Railway 15th Bureau Group Co., Ltd.\ for building the underground laboratory.
We are grateful for the ongoing cooperation from the China Guangdong Nuclear Power Group and China Light~\&~Power Company.

\vspace{4mm}
\noindent
References:


\bibliographystyle{model1a-num-names}
\bibliography{<your-bib-database>}



\end{document}